%% file: main.tex
\documentclass[acmsmall,screen]{acmart}
\input{preamble}

\AtBeginDocument{%
  \providecommand\BibTeX{{%
    \normalfont B\kern-0.5em{\scshape i\kern-0.25em b}\kern-0.8em\TeX}}}

\setcopyright{acmlicensed}
\copyrightyear{2018}
\acmYear{2018}

\begin{document}

\title{A Usage-Aware Sequent Calculus for Differential Dynamic Logic}


\author{Myra Dotzel}
\affiliation{%
  \institution{Carnegie Mellon University}
  \country{USA}
}
\email{mdotzel@andrew.cmu.edu}

\author{Stefan Mitsch}
\affiliation{%
  \institution{DePaul University}
  \country{USA}
}

\author{Andr\'e Platzer}
\affiliation{%
 \institution{Karlsruhe Institute of Technology}
 \country{Germany}
}

\renewcommand{\shortauthors}{Dotzel et al.}

\begin{abstract}
    Ensuring that safety-critical applications behave as intended is an important yet challenging task. 
    Modeling languages like differential dynamic logic (\dL) have proof calculi capable of proving guarantees for such applications. 
    However, \dL programmers may unintentionally over-specify assumptions and program statements, which results in overly constrained models that yield weak or vacuous guarantees. 
    In hybrid systems models, such constraints are ubiquitous; they may appear as assumptions, conditions on control switches, and evolution domain constraints in systems of differential equations which makes it nontrivial to systematically detect which ones are over-specified. 
    Existing approaches are limited, either lacking formal correctness guarantees or the granularity to detect all kinds of bugs arising in a given formula. 
    
    As a remedy, we present a novel proof-based technique that detects which constraints in a \dL formula are vacuous or over-specified and suggests ways in which these components could be mutated while preserving correctness proofs.  %
    When properties follow entirely from constraints uninfluenced by program statements, this analysis spots outright flaws in models. Otherwise, it helps make models more flexible by identifying specific ways in which they may be generalized. 
    The resulting analysis is thorough, catching bugs at a fine-grained level and proposing mutations that could be applied in combination. 
    We prove soundness and completeness with respect to \dL to ensure the correctness of suggested mutations and general applicability of our technique. 
\end{abstract}



\received{20 February 2007}
\received[revised]{12 March 2009}
\received[accepted]{5 June 2009}

\maketitle
\section{Introduction}

Hybrid systems verification establishes correctness properties of safety-critical cyber-physical systems models like collision avoidance for aircraft~\cite{squires18}, autonomous vehicles~\cite{pek20}, and trains~\cite{platzer09}. 
Programmers who verify such properties first write a model that describes the physical behavior of their system. Then, they attempt to show that the desired properties hold after running the program under a certain set of assumptions. To build a proof, a programmer may apply a series of rules to dissect the model into a series of smaller proof obligations until elementary facts are reached. 

However, the results proven are only useful if the models adequately capture the physical dynamics of the system~\cite{DBLP:journals/fmsd/MitschP16,selvaraj2022}. 
As most cyber-physical systems models aspire to replicate realistic behavior, they are inherently difficult to write and analyze, combining  descriptions of control algorithms with descriptions of the physical dynamics of a given application~\cite{alur2015,lee2015introduction, Platzer10, Platzer18}. Modeling languages like differential dynamic logic (\dL)~\cite{DBLP:journals/jar/Platzer08,DBLP:journals/jar/Platzer17} represent control algorithms by a discrete program and physical dynamics by a system of differential equations and evolution domain constraints on which the dynamics apply. 
While uncertainty about the fidelity of the control model can be overcome by verified compilation~\cite{DBLP:conf/pldi/BohrerTMMP18}, the physics model requires a combination of offline and online verification at runtime~\cite{DBLP:journals/fmsd/MitschP16}. 

%
Due to the difficulty of writing hybrid systems models, \dL formulas are prone to two classes of modeling errors: vacuity and over-specification~\cite{selvaraj2022}. 
The first can be overcome by vacuity detection which determines vacuous components in temporal logic formulas~\cite{Kupferman2003} for cyber-physical systems~\cite{Dokhanchi2017}. 
To address the second, prior work has used techniques like mutation testing to determine over-specified components of cyber-physical systems models and suggest alternatives~\cite{10132190}. 
However, these techniques are limited to finding a single bug, whereas in practice there may be many, and are infeasible to implement naively. In particular, proofs involving real arithmetic, which are common in \dL, are doubly-exponential in the number of variables~\cite{davenport88} which makes the combinatorial exploration of proof alternatives inherent to mutation testing expensive~\cite{grebing2020usability}. 
To our knowledge, there are also no formal guarantees that ensure the correctness and general applicability of these tools. 

Recent work applies weakest precondition and strongest postcondition techniques to find bugs in \dL formulas~\cite{Chong2023}. 
However, bugs may occur anywhere in a \dL formula, not just in the pre- and post-condition. In \dL formulas, \emph{constraints} about parameters, initial states, conditionals in control algorithms, and evolution domain constraints -- which occur spread throughout a formula -- all regulate the conditions under which a safety property applies. 
%
%
For example, a safety proof about the effects of a hybrid systems controller is vacuous if safety already follows directly from evolution domain constraints which are actually constraints on the physics and not conclusions about the safety of the control actions. 
These challenges are not unique to \dL formulas; they plague hybrid automata too~\cite{alur1993,Henzinger2000}.
For this reason, detecting over-specification in \dL formulas is a challenging yet important task for strengthening safety guarantees of the hybrid systems these formulas are intended to model. 

\textbf{Our key observation in this work is that carefully inspecting hybrid systems correctness proofs provides insight on the fidelity~\cite{Platzer18} (and over-specification) of their models.} 
We develop \uapc a sequent calculus for \dL that tracks constraint usage in a proof. 
Augmenting the \dL sequent calculus with a dataflow analysis, \uapc computes proof-preserving \emph{mutations} of constraints in each proof step. 
The tracking of constraints and mutations serve complementary roles. Constraint tracking is useful while conducting the proof to determine where facts come from, and how they are linked to the original conjecture. Alternatively, mutations are most meaningful after the proof is finished where in a separate model improvement step a programmer selects from the tracked mutation alternatives. 
We prove that \uapc is both sound and complete, giving a programmer assurance that applying the computed mutations to their formula would preserve its validity, and that \uapc could analyze \emph{any} valid \dL formula. 

The techniques we introduce here are general enough that they could be extended to other proof systems, yet the significance and subtleties are special for hybrid systems. 
To demonstrate the utility of these techniques, we developed \uapc for the \dL proof calculus which handles safety-critical applications.

This paper makes the following contributions:
\begin{itemize}
    \item A novel core proof calculus \uapc that systematically tracks constraint usage in a proof. 
    \item \uapcplus, the sequential counterpart of \uapc that reuses information from previously-analyzed proof branches to guide the analysis.
    \item $\uapc_\texttt{diag}$, and extension of \uapc that diagnoses the usage of cuts in a proof.
    \item Soundness and completeness of all calculi with respect to \dL, verifying the analyses to be correct and applicable to any valid \dL formula.
\end{itemize}

\paragraph{Outline.}

In Sec.~\ref{sec:background}, we recall differential dynamic logic (\dL)~\cite{DBLP:journals/jar/Platzer08}. Through an example, we study how the \dL proof rules operate and speculate which constraints in the original conjecture are over-specified by eyeballing the proof in an ad-hoc fashion. In Sec.~\ref{sec:preliminaries} and~\ref{sec:UAPC}, we develop the \uapc proof calculus which is capable of detecting over-specified constraints \emph{automatically}. In Sec.~\ref{sec:metatheory} we provide the soundness and completeness properties of \uapc and sketch their proofs. In Sec.~\ref{sec:example}, we revisit our original example, this time analyzing with \uapc. We conclude with discussions of related (Sec.~\ref{sec:related}) and future work (Sec.~\ref{sec:conclusion}). 

%
\section{Background and Motivation}\label{sec:background}

In this section, we review \dL. We first give its syntax (Sec.~\ref{sec:dL}). Then we show an example of a \dL formula (Sec.~\ref{sec:parachute_model}) and an example safety proof of the formula (Sec.~\ref{sec:parachute_proof}). 

\subsection{Differential Dynamic Logic (\dL)}\label{sec:dL}
\dL is a logic for specifying hybrid programs~\cite{DBLP:journals/jar/Platzer08}. Programmers write \emph{\dL formulas} to express properties of hybrid systems. \emph{\dL programs} consist of statements and formulas, as shown below.
\begin{align*}
    \mathit{Term} \quad e, d &::= x \mid c \mid e + d \mid e \cdot d \mid e / d \mid \sqrt{e} \mid f(e_1, \dots ,e_k) \\
    \mathit{Stmt} \quad \alpha,\beta &::= x:=e \mid x:=* \mid x' = f(x), \dots, y' = f(y) \& Q \mid \, ?Q \mid \alpha; \beta \mid \alpha \cup \beta \mid \alpha^*\\
    \mathit{Fml} \quad P,Q  &::= \text{true} \mid \text{false} \mid e \sim d \mid p(e_1, \dots, e_k)\mid \lnot P \mid P \bowtie Q
    \mid \forall x \; P \mid \exists x \; P  \\   
    &\;\;\;\;\;\;\mid [\alpha] P \mid \langle\alpha\rangle P
\end{align*}
Terms $e, d$ in \dL are variables $x$, rational constants $c \in \mathbb{Q}$, polynomials, and functions of variable arguments. \dL statements $\alpha, \beta$ include deterministic and nondeterministic assignments of a term to a variable, differential equations ($x' = f(x), \dots, y' = f(y)$ with evolution domain contraint $Q$), test statements (?), program sequencing (;), nondeterminstic choice ($\cup$), and repetition (*). \dL formulas $P, Q$ include boolean primitives, binary relations $\sim \; \in \{=, \geq, >\}$, boolean predicates, negation, boolean connectives $\bowtie \; \in \{\land, \lor, \rightarrow, \leftrightarrow\}$, universal and existential quantification, and two kinds of modalities $\left[\cdot\right]$ (pronounced ``box") and $\langle\cdot\rangle$ (pronounced ``diamond"). The modality $\left[\alpha\right]P$ means that $P$ holds after \textit{all} runs of $\alpha$, whereas $\langle\alpha\rangle P$ means that $P$ holds after \textit{some} run of hybrid program $\alpha$.

To prove the correctness of a \dL formula, we reason about \emph{sequents} which take the form $\Gamma \vdash \Delta$ where $\Gamma$ and $\Delta$ are finite sets of \dL formulas and the assumptions in $\Gamma$ are used to prove $\Delta$. A sequent is \emph{valid} if $\bigwedge_{p \in \Gamma} p \rightarrow \bigvee_{q \in \Delta} q$ is true in all states. 

\subsection{Parachute Model}\label{sec:parachute_model}

For example, a programmer could write a \dL formula that models the aerodynamics of a skydiver's parachute (Fig.~\ref{fig:parachute}). Constraints on the formula occur spread throughout the formula, which we $\constraint{highlight}$ for emphasis. 
%
%
\begin{figure}[tbhp]
\begin{align*}
& \constraint{A} {\to} \left[\left(\left(?(\constraint{v - gT > -\sqrt{g/r_{op}}} \land \constraint{r = r_{cl}} \land \constraint{x > 100}) \cup \; r := r_{op}\right); t:= 0; \text{ODE} \& \constraint{Q}\right)^*\right] P\\
& \text{Parameters}~ g, r_{op}, v_{max}, T, x, v, r, t \\
& \constraint{A} \triangleq \constraint{g > 0} \land \constraint{r_{cl} = 0} \land \constraint{r_{op} > r_{cl}} \land \constraint{T > 0} \land \constraint{v_{max} < -\sqrt{g/r_{op}}} \land \constraint{x \geq 0} \land \constraint{-\sqrt{g/r_{op}} < v < 0} \land \constraint{r = r_{cl}} \\
& P \triangleq \constraint{x = 0} \to v \geq v_{max} \\
& \constraint{Q} \triangleq \constraint{t \leq T} \land \constraint{x \geq 0} \land \constraint{v < 0} \\
& \text{ODE} \triangleq x' = v, v' = -g + rv^2, t' = 1
\end{align*}
\caption{Model of parachute dynamics, adopted from~\cite{ITP2017}.}
\label{fig:parachute}
\end{figure}
%
%
The program runs under assumptions $\constraint{A}$. The overall program structure is a loop where $[(\cdot)^*] P$ expresses that $P$ holds after all iterations of the loop. 

At the beginning of each loop iteration, the program may choose to run either the branch $?(\constraint{v - gT > -\sqrt{g/r_{op}}} \land \constraint{r = r_{cl}} \land \constraint{x > 100})$ or the branch $r := r_{op}$. The former allows the skydiver to keep falling when safe, if the aerodynamic drag coefficient is $r=r_{cl}$ and the skydiver's position is a safe distance from the ground ($\constraint{x > 100}$). 
The latter represents the parachute opening which induces an aerodynamic drag coefficient by setting $r$ to the drag coefficient of the open parachute $r_{op}$. 
After this choice, time $t$ is reset to $0$ and the program runs the dynamics as allowed by the domain constraint $\constraint{Q}$. ODE models the dynamics of the skydiver's parachute in free fall but where velocity is affected by aerodynamic drag, while $t' = 1$ acts as a clock and $\constraint{t\leq T}$ imposes a time limit $T$. 
Domain constraint $\constraint{Q}$ enforces that the skydiver's position is above ground and that velocity is negative during evolution (parachute descends).

The postcondition $\constraint{x = 0} \to v \geq v_{max}$ expresses that when the skydiver hits the ground ($\constraint{x = 0}$), their velocity $v$ will be bounded by an acceptable touchdown velocity $v_{max}$ ($v \geq v_{max}$ due to the non-positive velocity where $v_{max}$ is considered ``maximal" w.r.t. magnitude). 

\begin{figure}
    \begin{proofleafboxexample}[(3h)]
    \begin{mathpar}
    \inferrule*[rightstyle=\normalfont,Right=\color{violet}M{:}\color{black}\(\protect{[:=]_1}\),width=\linewidth]{
        \inferrule*[rightstyle=\normalfont,Right=\color{violet}M{:}\color{black}\(\protect{[:=]_1}\),width=\linewidth]{
            \inferrule*[rightstyle=\normalfont,Right=\color{violet}M{:}\color{black}\(\protect{[:=]_2}\),width=\linewidth]{
                \inferrule*[rightstyle=\normalfont,Right=\color{violet}M{:}\color{black}QE,width=\linewidth]{
                    \ast
                }{
                    \examplelabel{i_5,i_6,i_7,i_8,u_1}{I(x_0,v_0)}, \examplelabel{j,k,l}{\mathcolorbox{yellow}{L(x_0,v_0)}}, \examplelabel{n}t= 0, \examplelabel{r}v = v_0 \vdashu{\Psi}{DA(\dots {\mid} \Psi)} \examplelabel{s}-g + rv^2 \geq -g
                }
            }{
                {\begin{aligned}
                &\examplelabel{i_5,i_6,i_7,i_8,u_1}{I(x_0,v_0)}, \examplelabel{j,k,l}{\mathcolorbox{yellow}{L(x_0,v_0)}}, \examplelabel{n}t= 0, \examplelabel{r}v = v_0 \vdashu{\Psi}{DA(\dots {\mid} \Psi) \cupm \{\any{\color{red}\{o_1\}}\color{blue}\}} \\
                &\left[\examplelabel{o_1}x' := v\right] \examplelabel{s}-g + rv^2 \geq -g
                \end{aligned}
                }
            }
        }{
            {\begin{aligned}
            & \examplelabel{i_5,i_6,i_7,i_8,u_1}{I(x_0,v_0)}, \examplelabel{j,k,l}{\mathcolorbox{yellow}{L(x_0,v_0)}}, \examplelabel{n}t= 0, \examplelabel{r}v = v_0 \vdashu{\Psi}{DA(\dots {\mid} \Psi) \cupm \{\any{\color{red}\{o_1\}}\color{blue}, \color{red}\{o_2\}_{\mId}\color{blue}\}} \\ 
            & \left[\examplelabel{o_2}v' := -g + rv^2\right]\left[\examplelabel{o_1}x' := v\right] \examplelabel{s}v' \geq -g
            \end{aligned}
            }
        }
    }{
        {\begin{aligned}
        & \examplelabel{i_5,i_6,i_7,i_8,u_1}{I(x_0,v_0)}, \examplelabel{j,k,l}{\mathcolorbox{yellow}{L(x_0,v_0)}}, \examplelabel{n}t= 0, \examplelabel{r}v = v_0 
        \vdashu{\Psi}{DA(\dots {\mid} \Psi) \cupm \{\any{\color{red}\{o_1\}}\color{blue}, \color{red}\{o_2\}_{\mId}\color{blue}, \color{red}\{o_3\}_{\mId}\color{blue}\}} \\ 
        & \left[\examplelabel{o_3}t' := 1\right]\left[\examplelabel{o_2}v' := -g + rv^2\right]\left[\examplelabel{o_1}x' := v\right] \examplelabel{s}v' \geq -gt'
        \end{aligned}
        }
    }
    \end{mathpar}
    \end{proofleafboxexample}
    \begin{proofmiddleboxexample}[(3a)]
    \begin{mathpar}
    \inferrule*[rightstyle=\normalfont,Right=\color{violet}M{:}\color{black}\(\protect{[}?\protect{]}\),width=\linewidth]{
        \inferrule*[rightstyle=\normalfont,Right=\color{violet}M{:}\color{black}{\([]\land\)},width=\linewidth]{
            \inferrule*[rightstyle=\normalfont,Right=\color{violet}M{:}\color{black}{\(\land\)}R,width=\linewidth]{
                (3c)\\
                {
                \inferrule*[rightstyle=\normalfont,Right=\color{violet}M{:}\color{black}dC,width=\linewidth]{
                    \inferrule*[rightstyle=\normalfont,right=\color{violet}M{:}\color{black}dI,width=\linewidth]{
                        (3g)
                        \\
                        (3h)
                    }{ 
                        {\begin{aligned}
                            &\examplelabel{i_5,i_6,i_7,i_8,u_1}{I(x,v)}, \examplelabel{j,k,l}{\mathcolorbox{yellow}{L(x,v)}}, \examplelabel{n}t= 0
                            \vdashu{\Psi}{\Theta'} \\ 
                            & \left[\examplelabel{\vec{o}}\text{ODE} \& \examplelabel{\vec{p}}Q\right]\examplelabel{s}v\geq v_0-gt
                        \end{aligned}
                        }
                    }
                    \\
                    (3f)
                    \\
                    \text{\color{red}$s$ fresh}
                }{
                    {\begin{aligned}
                    &\examplelabel{i_5,i_6,i_7,i_8,u_1}{I(x,v)}, \examplelabel{j,k,l}{\mathcolorbox{yellow}{L(x,v)}}, \examplelabel{n}t= 0 \vdashu{\Psi}{\Theta} \left[\examplelabel{\vec{o}}\text{ODE} \& \examplelabel{\vec{p}}Q\right] \examplelabel{i_7}v > -\sqrt{g/r_{op}}
                    \end{aligned}}
                }
                }
            }{
                {\begin{aligned}
                    &\examplelabel{i_5,i_6,i_7,i_8,u_1}{I(x,v)}, \examplelabel{j,k,l}{\mathcolorbox{yellow}{L(x,v)}}, \examplelabel{n}t= 0 \vdashu{\Psi}{\Theta} \\
                    &\left[\examplelabel{\vec{o}}\text{ODE} \& \examplelabel{\vec{p}}Q\right] (\examplelabel{i_5}x \geq 0 \land \examplelabel{i_6}v < 0 \land \examplelabel{u_1}x > -1 \land \examplelabel{i_8}r_\textit{cl}=0) 
                    \land \left[\examplelabel{\vec{o}}\text{ODE} \& \examplelabel{\vec{p}}Q\right] \examplelabel{i_7}v > -\sqrt{g/r_{op}}
                \end{aligned}
                }
            }
        }{
            \dots
        }
    }{
        \examplelabel{i_5,i_6,i_7,i_8,u_1}{I(x,v)} \vdashu{\Psi}{\Theta}
        \left[?\examplelabel{j,k,l}{\mathcolorbox{yellow}{L(x,v)}}\right] \left[\examplelabel{n}t:= 0; \examplelabel{\vec{o}}\text{ODE} \& \examplelabel{\vec{p}}Q\right] \examplelabel{i_5,i_6,i_7,i_8,u_1}{I(x,v)}
    }
    \end{mathpar}
    \end{proofmiddleboxexample}
    \begin{proofstartboxexample}
    \begin{mathpar}
    \inferrule*[rightstyle=\normalfont,Right=\color{violet}M{:}\color{black}\(\protect{{\to}R}\),width=\linewidth]{
            \inferrule*[rightstyle=\normalfont,Right=\color{violet}M{:}\color{black}\(\protect{\text{loop}}\),width=\linewidth]{
                (1) \\
                (2) \\
                (3) \\
                \color{red}u_1 \text{ fresh}\color{black}
            }{
                \examplelabel{\vec{i}}A \vdashu{\Psi}{\Omega
                }
                \left[\left(\{?(\examplelabel{j,k,l}{\mathcolorbox{yellow}{L(x,v)}}); \cup\; \examplelabel{m}r := r_{op};\}; \examplelabel{n}t:= 0; \examplelabel{\vec{o}}\text{ODE} \& \examplelabel{\vec{p}}Q\right)^*\right] 
                (\examplelabel{\vec{q}}P)
            }
        }{
            \vdashu{\Psi}{\Omega}
            \examplelabel{\vec{ {i}}}A \rightarrow \left[\left(\{?(\examplelabel{ {j,k,l}}{\mathcolorbox{yellow}{L(x,v)}}); \cup\; \examplelabel{   m}r := r_{op};\}; \examplelabel{   n}t:= 0; \examplelabel{\vec{  o}}\text{ODE} \& \examplelabel{\vec{  p}}Q\right)^*\right] 
                (\examplelabel{\vec{  q}}P)
        }
    \end{mathpar}
    \end{proofstartboxexample}
%
\text{where:} \quad    
    $\examplelabel{j,k,l}L(x,v) \triangleq \examplelabel{j}{\mathcolorbox{yellow}{v - gT > -\sqrt{g/r_{op}}}} \land \examplelabel{k}{r = r_{cl}} \land \examplelabel{l}{\mathcolorbox{yellow}{x > 100}}$ \\
    $\examplelabel{i_5,i_6,i_7,i_8,u_1}I(x,v) 
    \triangleq \examplelabel{i_5}x \geq 0 \land \examplelabel{i_6}v<0 \land \examplelabel{i_7}v>-\sqrt{g/r_{op}} \land \examplelabel{i_8}r_\textit{cl}=0 \land \examplelabel{u_1}{{x > -1}} \color{blue}$ \\
    $\color{blue}\Psi = {\color{red}\vec{o}}_{\mId}\color{blue} \qquad \color{blue}DA(\dots {\mid} \Psi) = \{\color{red}k_{\mId}\color{blue}, \color{red}s_{\mId}\color{blue}, {\color{red}\{i_8\}}_{\mW}\color{blue}, \any{\color{red}\{i_5\}}\color{blue}, \any{\color{red}\{i_6\}}\color{blue}, \any{\color{red}\{i_7\}}\color{blue}, \any{\color{red}j}\color{blue}, \any{\color{red}l}\color{blue}, \color{red}n_{\color{violet}\mR}\color{blue}, \any{\color{red}r}\color{blue}, \any{\color{red}\{u_1\}}, {\color{red}\vec{o}}_{\mId}\color{blue}\}$ \\
    $\color{blue}\Theta = \{{\color{red}\fuse{i_5}{p_2}}_{\color{violet}\mW}\color{blue},\any{\color{red}{\fuse{i_6}{p_3}}}\color{blue}, \any{\color{red}j}\color{blue}, \any{\color{red}l}\color{blue}, \any{\color{red}\{p_1\}}\color{blue}, {\color{red}\{i_7\}}_{\mId}\color{blue}, {\color{red}\{i_8\}}_\mId\color{blue}, {\color{red}k}_{\mId}\color{blue}, {\color{red}\{o_1\}}_{\mId}\color{blue}, {\color{red}\{o_2\}}_{\mId}\color{blue}, {\color{red}\{o_3\}}_{\mId}\color{blue}, {\color{red}n}_{\mId}\color{blue}, \any{\color{red}\{u_1\}}\color{blue}\}$ \\
    $\color{blue}\Omega = \{{\color{red}\fuse{i_5}{p_2}}_{\color{violet}\mId}\color{blue},{\color{red}\fuse{i_6}{p_3}}_\mId\color{blue},\color{red}k_{\mId}\color{blue}, 
    \any{\color{red}j}\color{blue}, \any{\color{red}l}\color{blue}, {\color{red}\{o_1\}}_{\mId}\color{blue}, \color{red}\{o_2\}_{\mId}\color{blue}, \color{red}\{o_3\}_{\mId}\color{blue}, \color{red}n_{\mId}\color{blue}, \color{red}\{i_7\}_{\mId}, \any{\color{red}m}\color{blue}, \any{\color{red}\{p_1\}}\color{blue},$ \\
    $\any{\color{red}\{i_1\}}\color{blue}, \any{\color{red}\{i_2\}}\color{blue}, \any{\color{red}\{i_3\}}\color{blue}, \any{\color{red}\{i_4\}}\color{blue}, {\color{red}\{i_8\}}_\mId\color{blue}, \color{red}\vec{q}_{\mId}\color{blue}\}$
\caption{Parachute example proof fragment in \uapc (with \color{red}red\color{black}, \color{blue}blue\color{black}, and \color{violet}violet \color{black} labels) or \dL (without). We write $\color{blue}DA(\dots {\mid} \Psi)$ for $\color{blue}DA(i_5,i_6,i_7,i_8,u_1,j,k,l,n,r,s \,{\mid} \Psi)$.}
\label{fig:parachute_ex}
\end{figure}

\subsection{Parachute Proof}\label{sec:parachute_proof}

The parachute formula in Fig.~\ref{fig:parachute} is valid~\cite{ITP2017}, meaning that it has a proof in the \dL proof calculus~\cite{Platzer12b}. 
We show a small chunk of an example \emph{proof tree} in boxes in Fig.~\ref{fig:parachute_ex} (full syntax-directed proof in Appendix~\ref{app:parachuteproof}). A zig-zag border at the top indicates that the proof continues in another snippet at the annotated subgoals, while a zig-zag border at the bottom indicates that the snippet continues from an open subgoal of another snippet. For now, we can ignore all the \color{red}red\color{black}, \color{blue}blue\color{black}, and \color{violet}violet \color{black} annotations and just focus on the \dL proof tree (written in black) and the $\mathcolorbox{yellow}{\text{yellow}}$ highlights. 

The proof begins in the bottom fragment of Fig.~\ref{fig:parachute_ex}, and should be read from conclusion (formula to prove valid) to premises (elementary facts that can be proven relatively easily). 
The first proof step, \color{violet}M:\color{black}$\rightarrow$R, moves the left-hand side of the implication $A$ into the set of assumptions on left of the turnstile. In the next step, \color{violet}M:\color{black}loop introduces a \emph{loop invariant} $I(x,v)$, which is an intermediate formula to help complete the proof. The application of the \color{violet}M:\color{black}loop rule spawns three proof branches, each corresponding to a smaller obligation to prove about the loop invariant: (1) that $I(x,v)$ holds initially, (2) that $I(x,v)$ can be used to prove the postcondition, and (3) that $I(x,v)$ holds after one loop iteration. 

We show the proof of obligation (3) in the middle fragment of Fig.~\ref{fig:parachute_ex} (branch (3a)). Here, the proof continues with \color{violet}M:\color{black}$[?]$ which converts the test $[? L(x,v)][t:=0;\text{ODE}\&Q]I(x,v)$ into an implication $L(x,v) \rightarrow [t:=0;\text{ODE}\&Q]I(x,v)$. After a few more steps, the proof reaches the step \color{violet}M:\color{black}$[]\land$ (derivable from the axiom \color{violet}M:\color{black}K) which breaks the formula into two separate conjuncts that should hold after running the same program. 
To prove these separately, the rule \color{violet}M:\color{black}$\land$R, forks the proof into two branches, each one proving different properties of ODE. 
The proof of the right conjunct follows by an application of the ule \color{violet}M:\color{black}dC which uses a \emph{differential cut} $v \geq v_0 - gt$ to instead prove two smaller obligations: that the cut holds after program execution (branch shown) and that the cut can be used to prove the postcondition when inserted into the ODE domain constraint (branch (3f), see Appendix). 
The step \color{violet}M:\color{black}dI proves the former obligation by showing that the domain constraint implies the postcondition (branch (3g), see Appendix) and turning differential equations into assignments.

Branch (3h) uses these assignments to avoid needing to solve the differential equations they represent~\cite{DBLP:conf/lics/PlatzerT18,DBLP:journals/jar/Platzer17}. Here, each \color{violet}M:\color{black}$[:=]_1$ step substitutes $t'$ and $v'$ with their respective terms in the assignments into the postcondition. Lastly, the leaf rule \color{violet}M:\color{black}QE for real arithmetic is applied, concluding this proof branch.
%





\textbf{By carefully inspecting the proof tree in Fig.~\ref{fig:parachute_ex}, we notice that the original model is \emph{over-specified}. }
\new{In particular, $\mathcolorbox{yellow}{\text{highlighted}}$ components $x > 100$ and $v - gT > - \sqrt{g/r_{op}}$ are not \emph{used} in the proof; they are neither necessary to prove the closing step \color{violet}M:\color{black}QE in (3h) nor are they involved in proving the rest of the branch. Therefore, they could be removed entirely from the original conjecture while producing the same (or a similar) proof. Thus, a formula without this information, or some generalized form of it, would still uphold its original correctness properties.} 

In the next section, we will cover the key ideas of \uapc, a proof calculus that will help us catch these bugs automatically. We will explain the annotations in the Fig.~\ref{fig:parachute_ex} proof tree, which is the translation of a \dL proof into \uapc, and formalize the removal and generalization of constraints as \emph{mutations} applied to \dL formulas at a fine-grained level. 

\section{Key Ideas of Usage-Aware Sequents}\label{sec:preliminaries}
%
Towards translating \dL proofs into \uapc, we embellish \dL sequent $\Gamma \vdash \Delta$ with annotations to obtain a \emph{usage-aware sequent} $\veclabel{{i}}{\Gamma} \vdashu{\Psi}{\Sigma} \veclabel{{j}}{\Delta}$ where the annotations $\color{red}\vec{\mathbf i}, \vec{\mathbf j}, \color{blue}\Psi, \Sigma$ facilitate the data-flow analysis of a proof. 

%
%
%
%

%

In this section, we define the key constructs of \uapc -- atoms and their identifying labels (Sec.~\ref{sec:atoms_labels}), mutations (Sec.~\ref{sec: union}), and sets that track labels and possible mutations (Sec.~\ref{sec:label_sets}). In Sec.~\ref{sec:applying_muts}, we introduce a mutation operator that simulates applying mutations post-analysis.



%
\subsection{Atoms and Labels}\label{sec:atoms_labels}

The fine-grained model components that \uapc analyzes are called \textit{atoms}. They are statements and formulas that cannot be further decomposed (defined below): 
%
    \begin{align*}
        \textit{Atomic Stmt} \quad \alpha_\text{atom} &::= x := e \mid x := * \mid x' = f(x) \\
        \textit{Atomic Fml} \quad P_\text{atom} &::= \text{true} \mid \text{false} \mid e \sim d \mid p(e_1 \dots e_k) 
    \end{align*}
%
Atoms include boolean primitives, binary operations, predicates, assignments, and differential equations (without their domain constraints, which are composed of more atoms). 
%
In Fig.~\ref{fig:parachute}, $x > 100$, $v - gT > - \sqrt{g/r_{op}}$, and $r:=r_{op}$ are atoms whereas $\text{ODE}$ is not due to the multiple equations. 

To identify atoms and help track their usage throughout a proof, we introduce \emph{labels} $\color{red}\mathbf{l_1,l_2}$. Labels are defined inductively below where $\color{red}l$ denotes a \emph{singleton label}, and $\color{red}\fuse{\mathbf l_1}{\mathbf l_2}$ denotes a \emph{fused label}. 
In the conclusion of a proof, labels are all singletons, while fused labels are formed during the analysis to track relationships between atoms, effectively cross-labeling them. The relationship enforced by fused labels will be discussed in Sec.~\ref{sec:label_sets}. 
In general, we write $\color{red}\fuse{i_1}{\fuse{\dots}{i_n}}$ to denote a fused label with $n$ singleton labels (which could accumulate in a proof). 
Fusing is commutative, meaning that $\color{red}\fuse{i_1}{\fuse{\dots}{i_n}}$ is equivalent to any fused permutation of $\color{red}i_1,\dots,i_n$, and fused labels with duplicate singletons degenerate, e.g., $\color{red}\fuse{l}{l}$ is equivalent to writing $\color{red}l$. 
We use a \emph{label vector} $\color{red}\vec{\mathbf l}$ to denote a nonempty collection of labels. 
We define labels and label vectors below:
\begin{align*}
    \textit{Label} \quad \color{red}\mathbf{l}\color{black} ::= & \, \color{red}l\color{black} \mid \color{red}\fuse{\mathbf l_1}{\mathbf l_2} \\
    \textit{Label Vector} \quad \color{red}\vec{\mathbf l} \color{black} ::= & \, \color{red}\mathbf{l} \color{black}\mid \color{red}\mathbf{l},\vec{\mathbf l}
\end{align*}
%

We assign each atom in a statement or formula an identifying label $\color{red}{\mathbf{l}}$, inductively defined below. We use label vectors $\color{red}\vec{\mathbf{l}}$ to denote the collection of labels in a compound statement $\alpha,\beta$ or formula $P,Q$.
%
%
%
\begin{align*}
    \textit{Labeled Stmt} \quad \veclabel{{\mathbf l}}{\alpha}, {\beta} ::=\, &\atomlabel{\mathbf l}{x := e} {\kern-0.20em}
    \mid \atomlabel{\mathbf l}{x := *} {\kern-0.20em}
    \mid {\atomlabel{\mathbf l_1}{x' = f(x)}, \dots, \atomlabel{\mathbf l_n}{y' = f(y)} \& \veclabel{{\mathbf l}}{Q}} {\kern-0.20em}  
    \mid \, ?\veclabel{{\mathbf l}}{Q} {\kern-0.20em} \mid {\veclabel{{\mathbf l}_1}{\alpha};\veclabel{{\mathbf l}_2}{\beta}} {\kern-0.20em} \\
    &\mid (\veclabel{{\mathbf l}}{\alpha})^* {\kern-0.20em} 
    \mid {\veclabel{{\mathbf l}_1}{\alpha}\cup\veclabel{{\mathbf l}_2}{\beta}}\\
  \textit{Labeled Fml} \quad \veclabel{{\mathbf l}}{P,Q} ::=\, &\atomlabel{\mathbf l}{\text{true}} \mid \atomlabel{\mathbf l}{\text{false}} \mid \atomlabel{\mathbf l}{e \sim d} 
                     \mid \atomlabel{\mathbf l}{p(e_1,\dots,e_k)}  
                     \mid \neg \veclabel{{\mathbf l}}{P} 
                     \mid {\veclabel{{\mathbf l}_1}{P} \bowtie \veclabel{{\mathbf l}_2}{Q}}
                     \mid  \forall x\, \veclabel{{\mathbf l}}{P} \\
                     &\mid  \exists x\, \veclabel{{\mathbf l}}{P}
                     \mid  {[\veclabel{{\mathbf l}_1}{\alpha}] \veclabel{{\mathbf l}_2}{P}}
                     \mid  {\langle\veclabel{{\mathbf l}_1}{\alpha}\rangle \veclabel{{\mathbf l}_2}{P}}
  \end{align*}
In the Fig.~\ref{fig:parachute_ex} definition of $I(x,v)$, the singleton labels $\color{red}i_5$, $\color{red}i_6$, $\color{red}i_7$, $\color{red}u_1$ identify the atoms $x \geq 0$, $v < 0$, $v > -\sqrt{g/r_{op}}$, $x > -1$, respectively.

\paragraph{Properties. }

Each atom in the usage-aware sequent $\veclabel{{i}}{\Gamma} \vdashu{\Psi}{\Sigma} \veclabel{{j}}{\Delta}$ has a label in $\color{red}\vec{\mathbf i}\color{black}, \color{red}\vec{\mathbf j}$ that the analysis uses to track its propagation throughout a proof. 
Property~\ref{obs:unique_labels} enforces that each atom in a \dL formula is assigned a \emph{singleton} and \emph{unique} label before the analysis. 
Labels in the conclusion of a proof are distinct from each other even if the same atom occurs multiple times in different places. The propagation of labels throughout a proof tree allows us to track atom usage, and may subsequently spawn \emph{fresh} labels if new atoms are introduced (to be discussed in Sec.~\ref{sec:UAPC}).

  \begin{definition}[\textbf{Uniqueness}]
        Let $\veclabel{{\mathbf i}}{\Gamma} \vdashu{\Psi}{\Sigma} \veclabel{{\mathbf j}}{\Delta}$. A singleton label $\color{red}l \color{black}\in \color{red}\vec{\mathbf i}\color{black}, \color{red}\vec{\mathbf j}\color{black}$ is \emph{unique} iff $\color{red}l \color{black}\neq \color{red}l'\color{black} \; \forall \color{red}l'\color{black} \in \color{red}\vec{\mathbf i}, \vec{\mathbf j}\color{black} \setminus \color{red}l$.
  \end{definition}
\begin{definition}[\textbf{Freshness}]\label{defn:fresh}
        A singleton label $\color{red}l$ is \emph{fresh} for a valid sequent $\veclabel{{\mathbf i}}{\Gamma} \vdashu{\Psi}{\Sigma} \veclabel{{\mathbf j}}{\Delta}$ iff $\color{red}l \color{black}\notin \color{blue}\Psi\color{black}, \color{blue}\Sigma\color{black}, \color{red}\vec{\mathbf i}\color{black}, \color{red}\vec{\mathbf j}$.
  \end{definition}
  \begin{observation}[\textbf{Labeling schema}]\label{obs:unique_labels}
  The labels $\color{red}\vec{\mathbf i}\color{black}, \color{red}\vec{\mathbf j}$ in the conclusion of a proof $\veclabel{{\mathbf i}}{\Gamma} \vdashu{\Psi}{\Sigma} \veclabel{{\mathbf j}}{\Delta}$ are singleton and unique. The labels of a proof tree contain either singletons that appear in the conclusion of the proof or that are fresh.
  \end{observation}

\subsection{Mutations}\label{sec: union}
%
The data-flow analysis uses labels to determine which atoms in a formula could be \emph{mutated} while preserving its proof. 
%
%
In this work, we consider three different \textit{mutation levels} --- $\mId$, $\mW$, and $\mR$ (pronounced identity, generalize, and remove).

These mutation levels (\emph{mutations}, for short) are functions that take atomic statements to atomic statements, and atomic formulas to atomic formulas. They indicate \emph{the extent} to which an atom can be modified, and \textbf{the goal of the analysis is to solve for these mutations for every atom in a \dL formula.} 
We define the mutation $\mathcal{M}$ in Fig.~\ref{fig:mutation_for_atoms}. 
\begin{figure}[tbh]
    \begin{align*}
    \mathcal{M} &: \textit{Atomic Stmt} \to \textit{Atomic Stmt},\textit{Atomic Fml} \to \textit{Atomic Fml} \\
    \mathcal{M}(\mathbf{b}) &\triangleq \mathbf{b} \qquad\qquad{\mathbf{b} \in \{\text{true}, \text{false}\}, \mathcal{M} \in \{\mId, \mW, \mR\} } \\
    \mathcal{M}(e \sim d) &\triangleq \begin{cases} e \sim d  \; &\quad{\text{if} \; \mathcal{M} = {\mId}}   \\
                        e \geq d^*  \; &\quad{\text{if} \; \mathcal{M} = {\mW}}   \\
                        \text{true}  \; &\quad{\text{if} \; \mathcal{M} = {\mR}}   \\ 
                        \end{cases}  \\
    \mathcal{M}(p(e_1 \dots e_k)) &\triangleq \begin{cases} p(e_1 \dots e_k)  \; &\quad{\text{if} \; \mathcal{M} \in \{\mId, \mW\}}   \\
                        \text{true}  \; &\quad{\text{if} \; \mathcal{M} = {\mR}}   \\ 
                        \end{cases}  \\
    \mathcal{M}(x := e) &\triangleq \begin{cases} x := e  \; &\quad{\text{if} \; \mathcal{M} = {\mId}}   \\
                        x := d^*  \; &\quad{\text{if} \; \mathcal{M} = {\mW}}   \\
                        ?\text{true}  \; &\quad{\text{if} \; \mathcal{M} = {\mR}}   \\ 
                        \end{cases}  \\
    \mathcal{M}(x := *) &\triangleq \begin{cases}
                            x := *  \; &\quad{\text{if} \; \mathcal{M} \in \{\mId, \mW\}}   \\
                            ?\text{true}  \; &\quad{\text{if} \; \mathcal{M} = {\mR}}
                            \end{cases}
    \end{align*}
    \caption{Definition of mutation $\mathcal{M}$}\label{fig:mutation_for_atoms}
\end{figure}
%
%
When applied, $\mId$ produces an atom identical to its input whereas $\mR$ removes an atom from a \dL formula while maintaining its validity. The mutation $\mR$ replaces formulas with ``\text{true}'' (uninformative constraint) and programs with ``?\text{true}'' (program without effect). Mutation $\mW$ generalizes $e \sim d$ ($e = d$, $e > d$, or $e \geq d$) to $e \geq d^*$ for a term $d^*$ such that the replacement of $e \sim d$ with $e \geq d^*$ in a \dL formula preserves its proof. The term $d^*$ is a fixed user-defined value with the same type as $d$. Because equalities (such as $x = e$) may come from assignments in a proof (to be discussed in Sec.~\ref{sec:UAPC}), we let the $\mW$ mutation of $x := e$ be $x := d^*$. Any mutation of the booleans true and false produces the same output because these are primitive. 

We wait to define $\color{violet}\mathcal{M}$ for differential equations because we treat their mutation together with their domain constraints which, as a whole, is not atomic. We instead consider them in  Sec.~\ref{sec:applying_muts} when discussing how to apply mutations to non-atomic formulas and statements. 


\paragraph{Properties. }

We observe that if an atom could be removed without affecting the validity of the original conjecture, then the atom could be generalized or remain identical, because removal implies that the atom is not used anywhere in the proof. 
If an atom can be generalized, then it can remain identical too but should not be removed. If an atom must remain identical, then it should not be generalized or removed. 
These observations characterize a semilattice over the mutations $\mId, \mW$, and $\mR$ (Defn.~\ref{defn:mutation_structure}).

\begin{definition}[Mutation structure]\label{defn:mutation_structure}
    $(\{\mId, \mW, \mR\}, \meet)$ forms a meet-semilattice where $\mId \sqsubseteq \mW \sqsubseteq \mR$.
    %
\end{definition}


%
The mutation $\mId$ admits the \emph{least} flexibility, whereas the mutation $\mR$ admits the \emph{most} flexibility. The mutation $\mW$ admits more flexibility than $\mId$, but less than $\mR$, so we have the ordering $\mId \sqsubseteq \mW \sqsubseteq \mR$. 
The meet operation (written $\meet$) computes the greatest lower bound between mutations, thereby selecting the more conservative mutation; it enforces that $\mR \meet \mW = \mW$, $\mW \meet \mId = \mId$, and $\mR \meet \mId = \mId$. Because meet is commutative and idempotent, we also have that $\color{violet}\mathcal{M}\color{black} \meet \color{violet}\mathcal{N}\color{black} = \color{violet}\mathcal{N}\color{black} \meet \color{violet}\mathcal{M}\color{black}$ and $\color{violet}\mathcal{M}\color{black} \meet \color{violet}\mathcal{M}\color{black} = \color{violet}\mathcal{M}\color{black}$ for all $\color{violet}\mathcal{M}\color{black}, \color{violet}\mathcal{N}\color{black} \in \{\mId, \mW, \mR\}$.

\subsection{Tracking Label Sets}\label{sec:label_sets}

To track mutations for each atom in a formula, we rely on \emph{mutation-tracking labels}, written $\color{red}{\mathbf l}_{\color{violet}\mathcal{M}}$ where $\color{violet}\mathcal{M}$ is the allowable mutation level for the atom labeled by $\color{red}{\mathbf l}$. 
To track the evolution of mutations for each label over the course of analyzing a proof, each usage-aware sequent $\veclabel{{\mathbf i}}{\Gamma} \vdashu{\Psi}{\Sigma} \veclabel{{\mathbf j}}{\Delta}$ in a proof computes sets of mutation-tracking labels $\color{blue}\Psi$, $\color{blue}\Sigma$ (\emph{label sets} for short), defined below. 
The \emph{input} set $\color{blue}\Psi$ (to be preserved) contains mutations that the programmer allows for each label, and the \emph{output} set $\color{blue}\Sigma$ (to be updated) contains mutations that the analysis allows for each label. 
\begin{align*}
    \textit{Label Set} \quad \color{blue}\Sigma\color{black} &::= \color{blue}\cdot\color{black} \mid \color{red}{\mathbf l}_{\color{violet}\mathcal{M}}\color{blue}, \Sigma 
\end{align*}

To add or remove from a label set, we define operators \emph{merge} ($\cupm$) and \emph{discard} ($\setminus$). 

\paragraph{Merge ($\cupm$). } Fig.~\ref{fig:merge} defines $\cupm$ over label sets $\color{blue}\Sigma_1$ and $\color{blue}\Sigma_2$, which has commutative and identity properties (Lines 1 and 2). 
When a label $l_\mathcal{M}$ in $\color{blue}\Sigma_1$ does not appear in $\color{blue}\Sigma_2$, then the remainder of $\color{blue}\Sigma_1$ and $\color{blue}\Sigma_2$ merge and $l_\mathcal{M}$ in $\color{blue}\Sigma_1$ is added to the resulting set by a normal set union (Line 3). 
Line 4 gives a similar definition for a fused label occurring in $\color{blue}\Sigma_1$ where all of its singletons do not occur in $\color{blue}\Sigma_2$. 

Otherwise, $\cupm$ merges mutations of the same label appearing in $\color{blue}\Sigma_1$ and $\color{blue}\Sigma_2$, choosing the more conservative mutation level (or meet) between the two occurrences. 
%
%

Line 5 merges singletons, Line 6 merges a fused label with a singleton, and Line 7 merges fused labels. The resulting fused label in Line 7 may contain duplicate occurrences of singletons, but these are harmless because fused duplicates of the same singleton carry the same semantic meaning as the singleton itself.


\begin{figure}[tbh]
    \begin{align*}
        &\cupm : \mathit{Label \, Set} \times \mathit{Label \, Set} \rightarrow \mathit{Label \, Set} \\
        \color{blue}\Sigma_1\color{black} &\cupm \color{blue}\Sigma_2\color{black} = \color{blue}\Sigma_2\color{black} \cupm \color{blue}\Sigma_1\color{black} \\
        \color{blue}\Sigma\color{black} &\cupm \color{blue}\cdot\color{black} = \color{blue}\Sigma\color{black} \\
        \color{blue}(\Sigma_1, l_\mathcal{M}) \color{black}&\cupm \color{blue}\Sigma_2 \color{black} =  \left(\color{blue}\Sigma_1 \color{black}\cupm \color{blue}\Sigma_2\right), \color{blue}l_{\mathcal{M}} \color{black} &(\text{if } \color{blue}l \color{black}\notin \color{blue}\Sigma_2\color{black})\\
        \color{blue}(\Sigma_1, \fuse{{i_1}}{ \fuse{\dots}{ {i_n}}}_\mathcal{M}) \color{black}&\cupm \color{blue}\Sigma_2 \color{black} =  \color{blue}(\Sigma \cupm \Sigma_2), \fuse{{i_1}}{ \fuse{\dots}{ {i_n}}}_{\mathcal{M}} \color{black} &(\text{if $\color{blue}\vec{i} \color{black}\notin \color{blue}\Sigma_2$})\\
        \color{blue}(\Sigma_1, l_\mathcal{M}) \color{black}&\cupm \color{blue}(\Sigma_2, l_\mathcal{N}) \color{black} =  \left(\color{blue}\Sigma_1 \color{black}\cupm \color{blue}\Sigma_2\right), \color{blue}l_{\mathcal{M} \meet \mathcal{N}} \\
        \color{blue}(\Sigma_1, \fuse{{i_1}}{ \fuse{\dots}{ {i_n}}}_\mathcal{M}) \color{black}&\cupm \color{blue}(\Sigma_2, {l}_\mathcal{N}) \color{black} =  \color{blue}(\Sigma_1, \fuse{{i_1}}{ \fuse{\dots}{ {i_n}}}_{\mathcal{M} \meet \mathcal{N}}) \color{black}\cupm \color{blue}\Sigma_2 \color{black} \color{black} &(\text{if $\color{blue}l \color{black}\in \color{blue}\{i_1, \dots , i_n\}$})\\
        \color{blue}(\Sigma_1, \fuse{{i_1}}{ \fuse{\dots}{ {i_m}}}_\mathcal{M}) \color{black}&\cupm \color{blue}(\Sigma_2, {\fuse{{j_1}}{ \fuse{\dots}{ {j_n}}}}_\mathcal{N}) \color{black} \\
        &=  \color{blue}\left(\Sigma_1 \color{black}\cupm \color{blue}\Sigma_2\right) \color{black}\cupm \color{blue}\{{\fuse{{i_1}}{ \fuse{\dots}{ \fuse{i_m}{ \fuse{j_1}{ {\fuse{\dots}{j_n}}}}}}}_{\mathcal{M} \meet \mathcal{N}}\} \color{black} &(\text{if $\color{blue}\vec{i} \color{black}\cap \color{blue}\vec{j} \color{black}\neq \color{blue}\emptyset$}) 
    \end{align*}
    \caption{Definition of merge operator $\cupm$}\label{fig:merge} 
\end{figure}
\paragraph{Discard ($\setminus$). } We define the discard operator for recursively removing singleton labels from a label set (Fig.~\ref{fig:difference}), where we write $\color{blue}\Sigma \color{black}\setminus \color{red}\vec{l} \color{black}$ as shorthand for $\color{blue}\Sigma \color{black}\setminus \color{blue} \color{red}\vec{l}_{\color{violet}\mathcal{M}}\color{black} \; (\forall \, \color{violet}\mathcal{M}\color{black})$. 
Line 1 defines discarding labels from an empty label set, which produces the empty set. 
Line 2 defines discarding a singleton label $\color{red}l$ from a label set that contains a matching singleton label, which is the usual set difference. 
Line 3 discards $\color{red}l_1$ from a fused label, which produces a fused label without $\color{red}l_1$. 
The commutativity of fused labels extends the definition on Line 3 to allow discarding a singleton from \emph{any} position in a fused label, not just the first. 
Line 4 defines discarding a label that does not appear in $\color{blue}\Sigma$, which is just $\color{blue}\Sigma$ with remaining $\color{red}\vec{l'}$ discarded.
Line 5 is the base case, defining the removal of no labels from $\color{blue}\Sigma$ to be just $\color{blue}\Sigma$.
\begin{figure}[tbh]
    \begin{align*}
        &\setminus : \textit{Label Set} \times \textit{Label Vector} \rightarrow \textit{Label Set} \\
        \color{blue}\cdot \color{black} &\setminus\, \color{red}\vec{l}  \color{black}= \color{blue}\cdot \\
        \color{blue}(\Sigma, {l}_{\mathcal{M}}) \color{black} &\setminus  \color{red}{l},\vec{l'} \color{black}= \color{blue}\Sigma  \color{black}\setminus \color{red}\vec{l'} \color{black} \\
        \color{blue}(\Sigma, {\fuse{l_1}{{\mathbf l_2}}}_{\mathcal{M}}) \color{black} &\setminus  \color{red}{l}_1,\vec{l'}\color{black}= \color{blue}(\Sigma, {\mathbf l_2}_{\mathcal{M}}) \color{black}\setminus \color{red}\vec{l'} \color{black} \\
        \color{blue}\Sigma \color{black} &\setminus \color{red}l,\vec{l'}  \color{black}= \color{blue}\Sigma \color{black} \setminus \color{red}\vec{l'} \color{black} \qquad\qquad{(\text{if} \, \color{red}l \color{black}\notin \color{blue}\Sigma\color{black})} \\
        \color{blue}\Sigma \color{black} &\setminus \color{red}\cdot \color{black} = \color{blue}\Sigma
    \end{align*}
    \caption{Definition of discard operator $\setminus$}
    \label{fig:difference}
\end{figure}
\paragraph{Properties. }

In $\veclabel{{\mathbf i}}\Gamma \vdashu{\Psi}{\Sigma} \veclabel{{\mathbf j}}\Delta$, sets $\color{blue}\Psi$ and $\color{blue}\Sigma$ are well-formed (Lemma~\ref{obs:well_formed_sets}), and the labels in input $\Psi$ also appear in output $\Sigma$ (Property~\ref{obs:superset_output}). 
A label set is well-formed if each label in it appears only once. Assuming that the input set $\color{blue}\Psi$ does not contain duplicate labels, we show that the output set $\color{blue}\Sigma$ also does not contain duplicates, which is preserved due to the semantics of merge and discard.

\begin{lemma}[\textbf{Well-formed sets}]\label{obs:well_formed_sets}
    If $\veclabel{{\mathbf i}}\Gamma \vdashu{\Psi}{\Sigma} \veclabel{{\mathbf j}}\Delta$ is valid and the labels in $\color{blue}\Psi\color{black}$ are unique, then for all $\color{blue}\Psi$ each label in $\color{red}\vec{\mathbf i},\vec{\mathbf j}$ appears exactly once in $\color{blue}\Sigma$. 
\end{lemma}

Property~\ref{obs:superset_output} enforces that the output set $\color{blue}\Sigma$ mentions at least every label $\color{red}\mathbf{l}_{\color{violet}\mathcal{M}}$ in $\color{blue}\Psi$ but with possibly stricter mutations ($\color{red}\mathbf{l}_{\color{violet}\mathcal{N}}$ where $\color{violet}\mathcal{N} \color{black}\sqsubseteq \color{violet}\mathcal{M}$). \new{In other words, from input to output, the analysis only contributes new information, tracking new labels as they are encountered and enforcing stricter mutations when needed.}

\begin{observation}[\textbf{Output subsumes input}]\label{obs:superset_output}
    If $\veclabel{{\mathbf i}}{\Gamma} \vdashu{\Psi}{\Sigma} \veclabel{{\mathbf j}}{\Delta}$ is valid, then $\forall \, {\color{red}\mathbf l}_{\color{violet}\mathcal{M}} \in \color{blue}\Psi$, $\exists \, \color{violet}\mathcal{N} \color{black}\sqsubseteq \color{violet}\mathcal{M}$ such that ${\color{red}\mathbf l}_{\color{violet}\mathcal{N}} \in \color{blue}\Sigma$. 
\end{observation}

Together, Lemma~\ref{obs:well_formed_sets} and Property~\ref{obs:superset_output} ensure that \textbf{\uapc yields information about every atom in a given formula, and that the computed mutations in the output set respect any constraints given in the input set.}

\subsection{Applying Mutations}\label{sec:applying_muts} 
The output of the analysis is a set mutation-tracking labels. In a separate model improvement step, a programmer may apply to the formula one tracked mutation for each label in the set. 
To simulate applying mutations, we define the \emph{mutation instance} ${\color{violet}\mu_{\Sigma}}$ that chooses a single mutation for each label in $\color{blue}\Sigma$ (Fig.~\ref{fig:mutationmapping}). 
\begin{figure}[htb]
    \begin{align*}
    {\color{violet}\mu_{\Sigma}} &: \mathit{Stmt} \to \mathit{Stmt},\mathit{Fml} \to \mathit{Fml} &\\
    {\color{violet}\mu_{\Sigma}}(\lnot P) &\triangleq \lnot {\color{violet}\mu_{\Sigma}}(P) &\\
    {\color{violet}\mu_{\Sigma}}(P \bowtie Q) &\triangleq {\color{violet}\mu_{\Sigma}}(P) \bowtie {\color{violet}\mu_{\Sigma}}(Q) &\text{where}~{\bowtie}\in\{\land,\lor,\rightarrow,\leftrightarrow\}\\
    {\color{violet}\mu_{\Sigma}}(\mathcal Q\, x \, P) &\triangleq \mathcal Q\, x \, {\color{violet}\mu_{\Sigma}}(P) &\text{where}~{\mathcal Q}\in\{\forall, \exists\}\\
    {\color{violet}\mu_{\Sigma}}(\lstrike\alpha\rstrike P) &\triangleq \lstrike{\color{violet}\mu_{\Sigma}}(\alpha)\rstrike {\color{violet}\mu_{\Sigma}}(P) &
    \text{where}~{\lstrike\cdot\rstrike}\in\{\left[\cdot\right], \langle\cdot\rangle\} \\
    {\color{violet}\mu_{\Sigma}}(?Q) &\triangleq \;?{\color{violet}\mu_{\Sigma}}(Q) &\\
    {\color{violet}\mu_{\Sigma}}(\alpha\triangleright\beta) &\triangleq {\color{violet}\mu_{\Sigma}}(\alpha) \triangleright {\color{violet}\mu_{\Sigma}}(\beta) &
    \text{where}~{\triangleright}\in\{\cup, ;\} \\
    {\color{violet}\mu_{\Sigma}}(\alpha^*) &\triangleq {\color{violet}\mu_{\Sigma}}(\alpha)^* & 
    \end{align*}
    \begin{align*}
    {\color{violet}\mu_{\Sigma}}(x' = f(x) \; \& \; Q) &\triangleq \begin{cases} x' = f(x) \; \& \; {\color{violet}\mu_{\Sigma}}(Q)  \; &\quad{\text{if} \; {\color{red}\mathbf{l}} : x' = f(x), \; {\color{red}\mathbf{l}}_{\color{violet}\mathcal{N}} \in {\color{blue}\Sigma}, \; {\color{violet}\mathcal{M}} \sqsubseteq {\color{violet}\mathcal{N}}, {\color{violet}\mathcal{M}} \in \color{violet}\{{\mId}, {\mW}\} }    \\
                        ? {\color{violet}\mu_{\Sigma}}(Q)  \; &\quad{\text{if} \; {\color{red}\mathbf{l}} : x' = f(x), \; {\color{red}\mathbf{l}}_{\color{violet}\mathcal{N}} \in {\color{blue}\Sigma}, \; {\color{violet}\mathcal{M}} \sqsubseteq {\color{violet}\mathcal{N}}, {\color{violet}\mathcal{M}} = {\color{violet}\mR}}   \\ 
                        \end{cases} \\ 
    {\color{violet}\mu_{\Sigma}}(\alpha_\text{atom}) &\triangleq \begin{cases} {\color{violet}\mathcal{M}}(\alpha_\text{atom})  \; &\quad{\text{if} \; {\color{red}\mathbf{l}}:\alpha_\text{atom}, \, {\color{red}\mathbf{l}}_{\color{violet}\mathcal{N}} \in {\color{blue}\Sigma}, \, {\color{violet}\mathcal{M}} \sqsubseteq {\color{violet}\mathcal{N}}}   \\
                        \alpha_\text{atom}  \; &\quad{\text{else
                        }}    \\
        \end{cases} \\
        {\color{violet}\mu_{\Sigma}}(P_\text{atom}) &\triangleq \begin{cases} {\color{violet}\mathcal{M}}(P_\text{atom})  \; &\quad{\text{if} \; {\color{red}\mathbf{l}}:P_\text{atom}, \, {\color{red}\mathbf{l}}_{\color{violet}\mathcal{N}} \in {\color{blue}\Sigma}, \, {\color{violet}\mathcal{M}} \sqsubseteq {\color{violet}\mathcal{N}}}   \\
                        P_\text{atom}  \; &\quad{\text{else
                        }}
        \end{cases} 
    \end{align*}
\caption{Definition of mutation instance ${\color{violet}\mu_{\Sigma}}$}\label{fig:mutationmapping}
\end{figure}
${\color{violet}\mu_{\Sigma}}$ maps statements to statements, and formulas to formulas, performing a recursive traversal until it reaches the atoms (Lines 1-7). 
The exception is differential equations (Line 8) which we handle in combination with their domain constraints. For differential equations, the mutations $\mId$ and $\mW$ are treated the same. The $\mR$ mutation replaces the entire differential equation and its domain constraint $Q$ with $?Q$, because differential equations are statements. 

At the atomic level (Lines 9-10), ${\color{violet}\mu_{\Sigma}}$ applies a mutation $\color{violet}\mathcal{M}$ to an atomic statement ($\alpha_\text{atom}$) or atomic formula ($P_\text{atom}$) labeled by $\color{red}\mathbf l$ if $\color{red}\mathbf l_{\color{violet}\mathcal{N}}\color{black} \in \color{blue}\Sigma$ and the applied mutation $\color{violet}\mathcal{M}$ is bounded by the computed mutation $\color{violet}\mathcal{N}$ ($\color{violet}\mathcal{M} \sqsubseteq \mathcal{N}$). 
Otherwise, if the atom's label is not in $\color{blue}\Sigma$, ${\color{violet}\mu_{\Sigma}}$ does not apply a mutation to the atom, skipping over it. 
This case only triggers when the input provided contains a label irrelevant to the formula being analyzed, because Lemma~\ref{obs:well_formed_sets} ensures that every label in a formula is analyzed and appears exactly once in the output set $\color{blue}\Sigma$. 
%
The fact that labels appear exactly once in a given output set enforces that ${\color{violet}\mu_{\Sigma}}$ applies exactly one mutation per label. 

We write ${\color{violet}\mu_{\Sigma}}(\Gamma \vdash \Delta)$ as shorthand for ${\color{violet}\mu_{\Sigma}}(P_1), \dots, {\color{violet}\mu_{\Sigma}}(P_n) \vdash {\color{violet}\mu_{\Sigma}}(Q_1), \dots, {\color{violet}\mu_{\Sigma}}(Q_m)$ where $P_i \in \Gamma$ for $1 \leq i \leq n$ and $Q_j \in \Delta$ for $1 \leq j \leq m$. 
Statements in Fig.~\ref{fig:mutationmapping} are accompanied by the side condition $\vars{{\color{violet}\mu_{\Sigma}}(\alpha)} \subseteq \vars{\alpha}$ 
which enforces that mutations do not introduce new variables (which will be important for defining axioms in Sec.~\ref{sec:axioms}). 
Generalizations to systems of differential equations follow accordingly.

Fused labels $\color{red}\fuse{{l_1}}{\fuse{{\dots}}{{l_n}}}_{\color{violet}\mathcal{M}}$ place an additional constraint on mutation. 
%
%
Because the fused labels represent cross-labeled atoms, the mutation instance can choose any mutation allowed by $\color{violet}\mathcal{M}$ for $\color{red}l_1, \dots, l_n$ \emph{as long as they are the same}; any mutation of the atom labeled by $\color{red}l_1$ must match the mutation of the atom labeled by $\color{red}l_2$, and so on.
%
Defn.~\ref{defn:fusing} formalizes this property by restricting the mutation operator to apply the same mutation $\color{violet}m$ to the atoms labeled by singletons $\color{red}l_1,\dots,l_n$ occurring in fused label $\color{red}\fuse{{l_1}}{\fuse{{\dots}}{{l_n}}}$.


\begin{definition}[Fusing]\label{defn:fusing}
    $\color{violet}\mu_{\Sigma, \fuse{{l_1}}{\fuse{{\dots}}{{l_n}}}_{\color{violet}\mathcal{M}}}(\Gamma \vdash P)$ is valid iff ${\color{violet}\mu_{\Sigma}}(\color{violet}\mu_{{\color{blue}{l_1}_{\color{violet}{m}}, \dots, {l_n}_{\color{violet}{m}}\color{blue}}}(\Gamma \vdash P))$ is valid for all $\color{violet}m \color{black}\sqsubseteq \color{violet}\mathcal{M}$.
\end{definition}

We write, for example, $\color{violet}\mu_{\{{i}_{\color{violet}{m}}, {j}_{\color{violet}{n}}\color{blue}\}}$ to denote a \emph{concrete} mutation instance instantiated with concrete mutations ${\color{violet}{m}}$, ${\color{violet}{n}}$ that obey the definition of ${\color{violet}\mu_{\Sigma}}$. 
For example, if $\color{blue}\Sigma \color{black}= \color{blue}\{i_{\mId}\color{black}, \any{j}\color{blue}\}$, one concrete mutation instance could be $\color{violet}\mu_{\{i_{\mId}\color{black}, j_{\mW}\color{blue}\}}$, but not $\color{violet}\mu_{\{j_{\mW}, j_{\mR}\color{blue}\}}$ because multiple mutations for $\color{red}j$ are instantiated and a mutation for $\color{red}i$ is not included, violating Lemma~\ref{obs:well_formed_sets}. 
%

%

%

\paragraph{Properties. }

While \uapc strives to compute the most flexible mutations when possible, Property~\ref{obs: id_mut} enforces that $\mId$ is always a safe choice for any label in a formula. 
Additionally, the same mutations can be applied to a valid formula in any order, all resulting in the same mutated formula (Lemma~\ref{lem:union_muts}).

\begin{observation}[\textbf{Identity admissibility}]\label{obs: id_mut}
    If $\veclabel{{i}}{\Gamma} \vdashu{\Psi}{\Sigma} \veclabel{{j}}{\Delta}$ is valid, then $\color{violet}\color{violet}\mu_{\{\vec{\mathbf i}_{\mId}, \vec{\mathbf j}_{\mId}\}}\color{black}(\Gamma \vdash \Delta)$ is valid.
\end{observation}

\begin{lemma}[\textbf{Composition of mutations}]\label{lem:union_muts}
    If $\color{blue}\Sigma$ and $\color{blue}\Omega$ are sets s.t. $\forall \color{red}l \color{black}\in \color{blue}\Sigma$ $\color{red}l \color{black}\notin \color{blue}\Omega$
    , then $\color{violet}\color{violet}\mu_{\Sigma \cup \Omega}\color{black}(\Gamma \vdash \Delta) \equiv \color{violet}{\color{violet}\mu_{\Sigma}}\color{black}(\color{violet}\color{violet}\mu_{\Omega}\color{black}(\Gamma \vdash \Delta)) \equiv \color{violet}\color{violet}\mu_{\Omega}\color{black}(\color{violet}{\color{violet}\mu_{\Sigma}}\color{black}(\Gamma \vdash \Delta))$.
\end{lemma}


\subsection{Takeaways}

In this section, we introduced usage-aware sequents. 
They extend \dL sequents with labels and label sets to help track mutations of atoms. 
For example, consider the following valid sequent from Fig.~\ref{fig:parachute_ex}: 
\[
        \examplelabel{i_5,i_6,i_7,i_8,u_1}I(x_0,v_0), \examplelabel{j,k,l}L(x_0,v_0), \examplelabel{n}t = 0, \examplelabel{r}v = v_0 \vdashu{\Psi}{DA(\dots {\mid} \Psi)} \examplelabel{s}-g + rv^2 \geq -g
\]
where 

\begin{itemize}
    \item $\examplelabel{i_5,i_6,i_7,i_8,u_1}I(x_0,v_0) \triangleq \examplelabel{i_5}x_0 \geq 0 \land \examplelabel{i_6}v_0<0 \land \examplelabel{i_7}v_0>-\sqrt{g/r_{op}} \land \examplelabel{i_8}r_\textit{cl}=0 \land \examplelabel{u_1}{{x > -1}} \color{blue}$
    \item $\examplelabel{j,k,l}L(x,v) \triangleq \examplelabel{j}{{v - gT > -\sqrt{g/r_{op}}}} \land \examplelabel{k}{r = r_{cl}} \land \examplelabel{l}{{x > 100}}$
    \item $\color{blue}\Psi \color{black}= {\color{red}\vec{o}}_{\mId}$ 
    \item $\color{blue}DA(\dots {\mid} \Psi) \color{black}= \color{blue}\{\color{red}k_{\mId}\color{blue}, \color{red}s_{\mId}\color{blue}, {\color{red}\{i_8\}}_{\mW}\color{blue}, \any{\color{red}\{i_5\}}\color{blue}, \any{\color{red}\{i_6\}}\color{blue}, \any{\color{red}\{i_7\}}\color{blue}, \any{\color{red}j}\color{blue}, \any{\color{red}l}\color{blue}, \color{red}n_{\color{violet}\mR}\color{blue}, \any{\color{red}r}\color{blue}, \any{\color{red}\{u_1\}}\color{blue}, {\color{red}\vec{o}}_{\mId}\color{blue}\}$.
\end{itemize}

While we define $\color{blue}DA(\dots {\mid} \Psi)$ in the next section, we can already see that the output set is a function of the input set $\color{blue}\Psi$.  
The computed output set indicates that atoms labeled by $\color{red}k$, $\color{red}s$, and ${\color{red}\vec{o}}$ should remain unchanged (\mId), $r_{cl} = 0$ (labeled by $\color{red}i_8$) could be generalized to $r_{cl} \geq 0$ (\mW, supposing that $d^*$ is instantiated with the same constant $0$), and all other atoms could be removed ($\mR$) while preserving validity. 
For example, we could choose a concrete mutation instance $\color{violet}\color{violet}\mu_{DA(\dots {\mid} \Psi)}$ instantiated with exactly these mutations such that when applied: 
\begin{align*}
    &\color{violet}\color{violet}\mu_{DA(\dots {\mid} \Psi)}\left(\color{black}I(x_0,v_0), L(x_0,v_0), t = 0, v = v_0 \vdash -g + rv^2 \geq -g\color{violet}\right)\color{black} \\
    \equiv &\left(\color{violet}\textbf{true}\color{black}, \color{violet}\textbf{true}\color{black}, \color{violet}\textbf{true}\color{black}, \color{violet}\textbf{true}\color{black}, r = r_\textit{cl}, \color{violet}\textbf{true}\color{black}, \color{violet}\textbf{true}\color{black}, r_\textit{cl} \geq 0 \vdash -g + rv^2 \geq -g\right)
\end{align*}
The valid mutated sequent above is logically equivalent to $r = r_\textit{cl}, r_\textit{cl} \geq 0 \vdash -g + rv^2 \geq -g$, which verifies our intuition that $r$ is non-negative is the only information needed to prove that $-g + rv^2 \geq -g$. 
The fact that all other atoms in this formula were unused makes us question whether all atoms of $L(x_0,v_0)$ and the loop invariant $I(x,v)$ were necessary, but so far our analysis has only been local. In the next section, we introduce a proof-based mechanism for computing mutations of the original conjecture by computing mutations across proof steps. 

%
\section{Usage-Aware Proof Calculus (\uapc)}\label{sec:UAPC}

%
In this section, we give the proof rules of \uapc which define the interaction between usage-aware sequents in a proof. 
We annotate the core \dL proof rules~\cite{cheatsheet} with labels and label sets, and define how they propagate information in each proof step. 
We show a few rules to explain the mechanics of \uapc and give the full calculus in Appendix~\ref{app:uapccalculus}.

\subsection{Inductive Proof Rules}

\begin{figure}[t!b!h]
    \begin{mathpar}
        \quad{\text{\color{violet}M:\color{black}$\lor$R}} \;\;
        \inferrule
        { \veclabel{{k}}{\Gamma} \vdashu{\Psi}{\Sigma} \veclabel{{l}}{\Delta}, \veclabel{{i}}{P}, \veclabel{{j}}{Q} }
        { \veclabel{{k}}{\Gamma} \vdashu{\Psi}{\Sigma} \veclabel{{i}}{P} \lor \veclabel{{j}}{Q}, \veclabel{{l}}{\Delta} }
\and
        \quad{\text{\color{violet}M:\color{black}$\to$R}} \;\;
        \inferrule
        { \veclabel{{k}}{\Gamma}, \veclabel{{i}}{P} \vdashu{\Psi}{\Sigma} \veclabel{{j}}{Q},\veclabel{{l}}{\Delta} }
        { \veclabel{{k}}{\Gamma} \vdashu{\Psi}{\Sigma} \veclabel{{i}}{P} \to \veclabel{{j}}{Q},\veclabel{{l}}{\Delta} }
\and
        \quad{\text{\color{violet}M:\color{black}$\land$R}} \;\;
        \inferrule
        { \veclabel{{k}}{\Gamma} \vdashu{\Psi}{\Sigma} \veclabel{{i}}{P}, \veclabel{{l}}{\Delta}
        \\ \veclabel{{k}}{\Gamma} \vdashu{\Psi}{\Omega} \veclabel{{j}}{Q}, \veclabel{{l}}{\Delta} }
        { \veclabel{{k}}{\Gamma} \vdashu{\Psi}{\Sigma \cupm \Omega} \veclabel{{i}}{P} \land \veclabel{{j}}{Q}, \veclabel{{l}}{\Delta} }
\and
        \quad{\text{\color{violet}M:\color{black}$\land$L}} \;\;
        \inferrule
        { \veclabel{{k}}{\Gamma},\veclabel{{i}}{P},\veclabel{{j}}{Q} \vdashu{\Psi}{\Sigma} \veclabel{{l}}{\Delta} }
        { \veclabel{{k}}{\Gamma},\veclabel{{i}}{P} \land \veclabel{{j}}{Q} \vdashu{\Psi}{\Sigma} \veclabel{{l}}{\Delta} }
\and
        \quad{\text{\color{violet}M:\color{black}WL}} \;\;
        \inferrule
        { \veclabel{{k}}{\Gamma} \vdashu{\Psi}{\Sigma} \veclabel{{l}}{\Delta} }
        { \veclabel{{i}}{P}, \veclabel{{k}}{\Gamma} \vdashu{\Psi}{\Sigma \cupm \atoms{\any{\vec{\mathbf i}}}} \veclabel{{l}}{\Delta} }
\and    
        \quad{\text{\color{violet}M:\color{black}$\forall$R}} \;\;
        \inferrule
        { \veclabel{{i}}{\Gamma} \vdashu{\Psi}{\Sigma} \veclabel{{j}}{p(y)}, \veclabel{{k}}{\Delta} }
        { \veclabel{{i}}{\Gamma} \vdashu{\Psi}{\Sigma} \forall x \; \veclabel{{j}}{p(x)}, \veclabel{{k}}{\Delta} }
        \quad{(y \notin \Gamma, \Delta, \forall x \; p(x))}
\and       
        \quad{\text{\color{violet}M:\color{black}$\exists$R}} \;\;
        \inferrule
        { \veclabel{{i}}{\Gamma} \vdashu{\Psi}{\Sigma} \veclabel{{j}}p(e), \veclabel{{k}}{\Delta} }
        { \veclabel{{i}}{\Gamma} \vdashu{\Psi}{\Sigma} \exists x \; \veclabel{{j}}{p(x)}, \veclabel{{k}}{\Delta} }
        \quad{(\text{arbitrary term $e$})}
\and
        \quad{\text{\color{violet}M:\color{black}loop}^{\color{red}\vec{l}\color{black}}} \;\;
        \inferrule
        { \veclabel{{k}}{\Gamma} \vdashu{\Psi}{\Sigma} \veclabel{{n}}{J}, \veclabel{{i}}{\Delta}
        \\ \veclabel{{n}}{J} \vdashu{\Psi}{\Omega} \veclabel{{j}}{P}
        \\ \veclabel{{n}}{J} \vdashu{\Psi}{\Theta} [\veclabel{{m}}{\alpha}] \veclabel{{n}}{J}
        }
        { \veclabel{{k}}{\Gamma} \vdashu{\Psi}{(\Sigma \cupm \Omega \cupm \Theta) \setminus {\vec{l}}} [\veclabel{{m}}{\alpha}^*]\veclabel{{j}}{P},\veclabel{{i}}{\Delta} }
        \quad{(\forall J_a \in \text{atoms}(J). \, \color{red}\mathbf{n_a} \color{black}= \phifunc{}{J_a}{(\veclabel{{k}}{\Gamma}, \veclabel{{i}}{\Delta})}{\color{red}\vec{l}\color{black}})} 
\and    
        \quad{\text{\color{violet}M:\color{black}cut}^{\color{red}\vec{l}\color{black}}} \;\;
        \inferrule
        { \veclabel{{i}}{\Gamma}, \veclabel{{j}}{C} \vdashu{\Psi}{\Sigma} \veclabel{{k}}{\Delta}
        \\ 
        \veclabel{{i}}{\Gamma} \vdashu{\Psi}{\Omega} \veclabel{{k}}{\Delta}, \veclabel{{j}}{C}
        }
        { \veclabel{{i}}{\Gamma} \vdashu{\Psi}{(\Sigma \cupm \Omega) \setminus {\vec{l}}} \veclabel{{k}}{\Delta} }
        \quad{(\forall C_a \in \text{atoms}(C). \, \color{red}\mathbf{j_a} \color{black}= \phifunc{}{C_a}{(\veclabel{{i}}{\Gamma}, \veclabel{{k}}{\Delta})}{\color{red}\vec{l}\color{black}})} 
\and        
        \quad{\text{\color{violet}M:\color{black}GVR}} \;\;
        \inferrule
        { \veclabel{{i}}{\Gamma_{\text{const}}} \vdashu{\Psi}{\Omega} \veclabel{{k}}{P}, \veclabel{{l}}{\Delta_{\text{const}}} }
        { \veclabel{{i}}{\Gamma} \vdashu{\Psi}{\Omega \cupm \{\any{\vec{\mathbf j}}\} } \left[\veclabel{{j}}\alpha\right] \veclabel{{k}}{P}, \veclabel{{l}}{\Delta}}
\and
        \quad{\text{\color{violet}M:\color{black}dI}} \;\;
        \inferrule
        { \veclabel{{l}}\Gamma, \veclabel{{i}}Q \vdashu{\Psi}{\Sigma} \veclabel{{k}}P, \veclabel{m}{\Delta} \\
        \veclabel{{i}}Q \vdashu{\Psi}{\Omega} [\atomlabel{j}x':=f(x)]\veclabel{{k}}(P)' }
        { \veclabel{{l}}\Gamma \vdashu{\Psi}{\Sigma \cupm \Omega} [\atomlabel{j}x'=f(x) \& \veclabel{{i}}Q]\veclabel{{k}}P, \veclabel{m}{\Delta} }
\and
        \quad{\text{\color{violet}M:\color{black}dC}} \;\;
        \inferrule
        { \veclabel{{i}}\Gamma \vdashu{\Psi}{\Sigma} [\atomlabel{j}x'=f(x)\&\veclabel{{k}}Q]\veclabel{{l}}C,\veclabel{{m}}\Delta 
        \\
        \veclabel{{i}}\Gamma \vdashu{\Psi}{\Omega} [\atomlabel{j}x'=f(x)\&(\veclabel{{k}}Q \land \veclabel{{l}}C)]\veclabel{{n}}P, \veclabel{{m}}\Delta }
        { \veclabel{{i}}\Gamma \vdashu{\Psi}{\Sigma \cupm \Omega} [\atomlabel{j}x'=f(x)\&\veclabel{{k}}Q]\veclabel{{n}}P, \veclabel{{m}}\Delta }
\and
        \quad{\text{\color{violet}M:\color{black}CER}^{\color{red}\vec{n}\color{black}}} \;\;
        \inferrule
        { \veclabel{{i}}{\Gamma} \vdashu{\Psi}{\Omega} \veclabel{ m}{C(\veclabel{j}{Q})}, \veclabel{{k}}{\Delta}
        \\ \vdashu{\Psi}{\Sigma} \veclabel{{j}}{Q} \leftrightarrow \veclabel{{l}}{P} }
        { \veclabel{{i}}{\Gamma} \vdashu{\Psi}{(\Omega \cupm \Sigma) \setminus {\vec{l}}} \veclabel{ m}{C(\veclabel{{l}}{P})}, \veclabel{{k}}{\Delta} }
        \quad{(\forall Q_a \in \text{atoms}(Q). \, \color{red}\mathbf{j_a} \color{black}= \phifunc{}{Q_a}{\veclabel{{l}}{P}}{\color{red}\vec{n}\color{black}})} 
    \end{mathpar}
    \caption{Representative \uapc inductive proof rules}
    \label{fig:uapcprop}
\end{figure}
%

Fig.~\ref{fig:uapcprop} shows a representative selection of \uapc inductive proof rules, which propagate the \emph{input set $\color{blue}\Psi\color{black}$ from conclusion to premise} and the \emph{output set $\color{blue}\Sigma\color{black}$ from premise to conclusion}. 

The rules \color{violet}M:\color{black}$\lor$R,  \color{violet}M:\color{black}$\rightarrow$R, \color{violet}M:\color{black}$\land$L, \color{violet}M:\color{black}$\forall$R, and \color{violet}M:\color{black}$\exists$R all preserve the same output sets from premise to conclusion because the labels in the conclusion match the ones in the premise, and no new information is learned in these steps. 
%
%
The rule \color{violet}M:\color{black}$\forall$R eliminates the quantifier $\forall x$ by binding a variable $y$ to every free instance of $x$ in the predicate $p(x)$. From conclusion to premise, the formula $p(x)$ transforms into $p(y)$, maintaining the same labels $\color{red}\vec{\mathbf j}$. The restriction $y \notin \Gamma, \Delta, \forall x \; p(x)$ ensures that the concretizing variable $y$ does not clash with other variables. Analogously, the rule \color{violet}M:\color{black}$\exists$R eliminates an existential quantifier from conclusion to premise by instantiating $x$ with an arbitrary term $e$. 

%

In the rule \color{violet}M:\color{black}$\land$R, the first premise has output set $\color{blue}\Sigma$ and the second has output set $\color{blue}\Omega$. The output set of the conclusion is the merge of these two $\color{blue}\Sigma \cupm \Omega$ because the conclusion mentions labels in both of the premises. The merge in the conclusion (defined in Fig.~\ref{fig:merge}) ensures that mutations computed in each of the premises agree.
Instead of proving the conclusion directly, the rules \color{violet}M:\color{black}cut and \color{violet}M:\color{black}loop introduce an auxiliary formula which they use to prove intermediary results in their premises. 
Notably, the rule \color{violet}M:\color{black}loop introduces a loop invariant $J$ and shows that it (1) holds under the assumptions $\Gamma$, (2) implies the postcondition, and (3) holds after one loop iteration provided that it held previously. 
The output set of the conclusion is obtained by merging the output sets of the three premises $\color{blue}\Sigma$, $\color{blue}\Omega$, and $\color{blue}\Theta$ sans the fresh labels $\color{red}\vec{l}$, which do not occur in the conclusion and hence do not inform the analysis of the original formula. 
%
%
%
The fresh labels are introduced by the loop invariant $\veclabel{{n}}{J}$ where $\color{red}\vec{\mathbf n}$ is a combination of pre-existing and new atoms. To compute $\color{red}\vec{\mathbf n}$, we define a function $\phi$ (in analogy to SSA~\cite{Cytron1991}) in Defn.~\ref{defn:phi}. The function $\phi$ decides whether each atom in $J$ is new or already appears elsewhere in the deriving sequent (in \color{violet}M:\color{black}loop, the first premise). If an atom is new, it is assigned a fresh label $\color{red}l_f$ from a predetermined set of fresh singleton labels $\color{red}\vec{l}\color{black}$. 
  If the atom is not new, then it preserves the same label used for other occurrences of that atom in the labeling context (in \color{violet}M:\color{black}loop, the other labels $\color{red}\vec{\mathbf k}, \vec{\mathbf i}$ in the deriving sequent). We use a fused label to cross-label in cases where there are duplicate atoms in the labeling context. 
  We write $\text{atoms}(J)$ to extract the atoms of formula $J$, and $\text{atoms}(\veclabel{{j}}{J})$ to extract the labeled atoms of $\veclabel{{j}}{J}$, and similarly for statements. 
  

  \begin{definition}[$\phi$ function] \label{defn:phi}
%
  \begin{align*}
        &\phi: \textit{Atom} \times \{\textit{Labeled Atom}\} \times \textit{Label Vector} \rightarrow \textit{Labeled Atom} \\
        &\phifunc{}{\text{atom}}{\text{lbl ctx}}{\text{fresh labels}}
        = \begin{cases} 
              \color{red}\fuse{j_1}{\fuse{\dots}{j_n}}\color{black} 
              & \textit{if } \exists \; \color{red}j_1,\dots,j_n \color{black}\textit{ s.t. }  \color{red}j_1\color{black} {:} \text{atom}, \dots, \color{red}j_n\color{black} {:} \text{atom} \in \text{lbl ctx}\\
              \color{red}l_f\color{black} 
              & \textit{ else, for some $\color{red}l_f \color{black}\in$ }\text{fresh labels}
           \end{cases}
      \end{align*}
  \end{definition}

  %
  
  %

For example, in the \color{violet}M:\color{black}loop proof step of Fig.~\ref{fig:parachute_ex}, $\phi($\_$,\examplelabel{\vec{i}}A,\color{red}\vec{u}\color{black})$ computes a label for each of the four atoms in $I(x,v)$. 
Since $x \geq 0$ appears in $A$, we reuse its label $\color{red}i_5$ ($\phi(x \geq 0,\examplelabel{\vec{i}}A,\color{red}\vec{u}\color{black}) = \color{red}i_5$). Similarly, the labels of $v < 0$ and $v > -\sqrt{g/r_{op}}$, which also appear in $A$, are $\color{red}i_6$ and $\color{red}i_7$. The atom $x > -1$, however, is new as it does not appear in $A$. So it deserves a fresh label ($\phi(x > -1,\examplelabel{\vec{i}}A,\vec{u}) = \color{red}u_1$). 

%

%

The rule \color{violet}M:\color{black}WL proves its conclusion without the assumption $\veclabel{{\mathbf i}}{P}$ in the premise. The output set $\color{blue}\Sigma \color{black}\cupm \any {\color{red}\vec{\mathbf i}}$ indicates that removing $P$ is allowed. 
Similarly, the rule \color{violet}M:\color{black}GVR has output set $\color{blue}\Omega \color{black}\cupm \any{\color{red}\vec{\mathbf j}}\color{blue}$ which indicates that $\veclabel{{\mathbf j}}{\alpha}$ is not needed to prove the conclusion. In \color{violet}M:\color{black}GVR, $\Gamma_\text{const}$ and $\Delta_\text{const}$ are $\Gamma$ and $\Delta$ restricted to formulas whose free variables do not intersect with the bound variables of $\alpha$. We label both $\Gamma$ and $\Gamma_\text{const}$ with $\color{red}\vec{\mathbf i}$ because the labels of $\Gamma_\text{const}$ draw from the same labels $\color{red}\vec{\mathbf i}$ in the conclusion. 

The rules \color{violet}M:\color{black}dI and \color{violet}M:\color{black}dC give two ways of using differential equations in a proof. The differential invariant rule \color{violet}M:\color{black}dI treats the differential equation $x' = f(x)$ as an assignment (second premise) given that the domain constraint implies the postcondition (first premise). In the second premise, the postcondition $P$ (once dependent on $x$) transforms into $P'$ (now dependent on $x'$) and moves the domain constraint (which is an ``assumption" about the physics) to the assumptions. The rule \color{violet}M:\color{black}dC is the analog of \color{violet}M:\color{black}cut to differential equations, injecting a cut into the proof to help with proving the conclusion. 
The first premise proves that the cut $C$ holds in all states after running the differential equation.
The second premise assumes $C$ in the domain constraint of the different equation to prove the post-condition. 

The contextual equivalence rule \color{violet}M:\color{black}CER applies equivalence axioms (discussed next in Sec.~\ref{sec:axioms}) to replace formulas with their equivalents in any context $C(\cdot)$. As with \color{violet}M:\color{black}loop, \color{violet}M:\color{black}CER analyzes one proof branch at a time. 
The second premise gives an equivalence between $P$ and $Q$. We use $\phi$ to compute labels for $Q$ because some of the atoms in $Q$ come from $P$ while others are new. The first premise plugs in $Q$ for all free occurrences of $P$ in $C(\cdot)$. 

%
\subsection{Axioms}\label{sec:axioms} 
We use contextual equivalence to apply axioms in a proof. 
We first discuss a selection of representative axioms in Fig.~\ref{fig:uapcaxioms} and then show how to use them in a proof.

\begin{figure}[htb]
    \begin{align*}
        \quad{\text{\color{violet}M:\color{black}$[?]$}} \;\; &\vdashu{\Psi}{\{\any{\color{blue}\vec{\mathbf j}}, \any{\color{blue}\vec{\mathbf k}}\color{blue}\} \cupm \Psi} [?\veclabel{{j}}{Q}]\veclabel{{k}}{P} \leftrightarrow \left(\veclabel{{j}}{Q} \to \veclabel{{k}}{P}\right) & \\
        \quad{\text{\color{violet}M:\color{black}$[{;}]$}} \;\; &\vdashu{\Psi}{\{\any{\color{blue}\vec{\mathbf i}}, \any{\color{blue}\vec{\mathbf j}}, \any{\color{blue}\vec{\mathbf k}}\color{blue}\} \cupm \Psi} [\veclabel{{i}}{\alpha};\veclabel{{j}}{\beta}]\veclabel{{k}}{P} \leftrightarrow [\veclabel{{i}}{\alpha}][\veclabel{{j}}{\beta}]\veclabel{{k}}{P} & \\
        \quad{\text{\color{violet}M:\color{black}$\langle\cdot\rangle$}} \;\; &\vdashu{\Psi}{\{\any{\color{blue}\vec{\mathbf j}}, \any{\color{blue}\vec{\mathbf k}}\color{blue}\} \cupm \Psi} \neg \left[\veclabel{{j}}{\alpha}\right]\neg \veclabel{{k}}{P} \leftrightarrow \langle \veclabel{{j}}{\alpha} \rangle \veclabel{{k}}{P} & \\
        %
        %
        \quad{\text{\color{violet}M:\color{black}V}} \;\; &\vdashu{\Psi}{\{\any{\color{blue}\vec{\mathbf j}}, \any{\color{blue}\vec{\mathbf k}}\color{blue}\} \cupm \Psi} \veclabel{{k}}{p} \rightarrow \left[\veclabel{{j}}{\alpha}\right]\veclabel{{k}}{p} \qquad\qquad\qquad {(\text{$FV(p) \cap BV(\alpha) = \emptyset$})} \\
        \quad{\text{\color{violet}M:\color{black}$[:=]_1$}} \;\; &\vdashu{\Psi}{\{{\color{blue}\mathbf j}_{\mId}, \any{\color{blue}\vec{\mathbf k}}\color{blue}\} \cupm \Psi} [\atomlabel{j}{x:=e}\;] \veclabel{{k}}{p(x)} \leftrightarrow \veclabel{{k}}{p(e)} \qquad{(x \in FV(p(x)))} \\  
        \quad{\text{\color{violet}M:\color{black}$[:=]_2$}} \;\; &\vdashu{\Psi}{{\{\any{\color{blue}\mathbf j}, \any{\color{blue}\vec{\mathbf k}}\color{blue}\} \cupm \Psi}} [\atomlabel{j}{x:=e}\;] \veclabel{{k}}{p} \leftrightarrow \veclabel{{k}}{p} \qquad\qquad {(x \notin FV(p))} \\
        \quad{\text{\color{violet}M:\color{black}$[\cup]$}} \;\; &\vdashu{\Psi}{\{\any{\color{blue}\vec{\mathbf i}}, \any{\color{blue}\vec{\mathbf j}}, \any{\overrightarrow{\fuse{{\mathbf k}}{{\mathbf l}}}}\color{blue}\} \cupm \Psi} [\veclabel{{i}}{\alpha} \cup \veclabel{{j}}{\beta}] {\color{red}\overrightarrow{\fuse{\mathbf{k}}{\mathbf{l}}}:}{P} \leftrightarrow \left([\veclabel{{i}}{\alpha}]\veclabel{{k}}{P} \land [\veclabel{{j}}{\beta}]\veclabel{{l}}{P}\right) & \\
        %
        %
    \end{align*}
    \caption{Representative \uapc axioms}
    \label{fig:uapcaxioms}
\end{figure}
%

The axiom \color{violet}M:\color{black}[?] converts between box tests ($[?Q]P$) and implications ($Q \rightarrow P$). 
The mutation $\mR$ means that $Q$ and $P$ could be uniformly removed from both sides while preserving the proof. 
We use the same labels on the left and right to ensure that mutations are applied uniformly to both sides. 
Axioms \color{violet}M:\color{black}[;] and \color{violet}M:\color{black}$\langle\cdot\rangle$ are similar in their labeling schemes. \color{violet}M:\color{black}[;] gives the sequencing rule for hybrid programs enveloped in box modalities, splitting $[\alpha;\beta]P$ into $[\alpha][\beta]P$, and vice versa. \color{violet}M:\color{black}$\langle\cdot\rangle$ translates between box and diamond modalities by duality. 

Axiom \color{violet}M:\color{black}V reasons about hybrid programs $\alpha$ that do not bind variables in the postcondition $p$. If $p$ holds, then $[\alpha]p$ holds given that the bound variables in $\alpha$ do not clash with the free variables of $p$, which is preserved even after mutation by definition because the variables of mutated $\alpha$ is a subset of the variables already in $\alpha$. 
The axioms M:$[:=]_1$ and M:$[:=]_2$ handle assignments in a proof by substituting free occurrences of $x$ in the postcondition $p$ with $e$ (M:$[:=]_1$), or disregarding it when there are no free occurrences of $x$ in $p$ to substitute (M:$[:=]_2$). The former enforces that $x:=e$ remains identical ($\color{blue}j_\mId$) under mutation because it is used to prove $p$, whereas the latter allows $x:=e$ to be removed ($\color{blue}j_\mR$) because it is not needed to prove $p$. The labels of $p$ ($\color{red}\vec{\mathbf k}$) are the same on both sides, and they can be uniformly removed on both sides while preserving the proof.

Lastly, we explain how to apply axioms in a proof by using the axiom \color{violet}M:\color{black}$\left[\cup\right]$ as an example. Axiom \color{violet}M:\color{black}$\left[\cup\right]$ reasons about nondeterministic choice. Since \color{violet}M:\color{black}$\left[\cup\right]$ is bidirectional, it may be applied in two different ways (forward $\rightarrow$ or backward $\leftarrow$). We explore both possibilities in Fig.~\ref{fig:union_axiom_ex}. When applied in the backward direction (Fig.~\ref{fig:union_axiom_ex} (B)), the conclusion of the proof containing $[\veclabel{{i}}{\alpha} \cup \veclabel{{j}}{\beta}] {\color{red}\overrightarrow{\fuse{{\mathbf k}}{{\mathbf k}}}:}{P}$ is transformed into the premise containing $[\veclabel{{i}}{\alpha}] \veclabel{{k}}{P} \land [\veclabel{{j}}{\beta}] \veclabel{{k}}{P}$. 
Alternatively, when applied in the forward direction (Fig.~\ref{fig:union_axiom_ex} (A)), the conclusion contains $[\veclabel{{i}}{\alpha}]\veclabel{{k}}{P} \land [\veclabel{{j}}{\beta}]\veclabel{{l}}{P}$. 
However, the premise has only one occurrence of $P$. To ensure proper label tracking, we assign a fused label $\color{red}\overrightarrow{\fuse{{\mathbf k}}{{\mathbf l}}}$ to preserve both $\color{red}\vec{\mathbf k}$ and $\color{red}\vec{\mathbf l}$. 
The label $\color{red}\overrightarrow{\fuse{{\mathbf k}}{{\mathbf l}}}$ means that each atom in $P$ is assigned label $\color{red}\fuse{{\mathbf k}}{{\mathbf l}}$ for the appropriate $\color{red}\mathbf k$ and $\color{red}\mathbf l$. 
In \color{violet}M:\color{black}$[\cup]$, the mutation $\mR$ on ${\color{red}\overrightarrow{\fuse{\mathbf{k}}{\mathbf{l}}}}$ indicates that we could uniformly remove all occurrences of $P$ (labeled by $\color{red}\vec{\mathbf k}$ and $\color{red}\vec{\mathbf l}$) on both sides of the axiom while preserving the proof. 

\begin{figure}
    \centering
    \begin{mathpar}
        \qquad{(\text{A})}\;\;
            \inferrule
            { {\begin{aligned}
                &\veclabel{{n}}{\Gamma} \vdashu{\Psi}{\Omega} \veclabel{ m}{C({\mathcolorbox{pink}{[\veclabel{{i}}{\alpha} \cup \veclabel{{j}}{\beta}] {\color{red}\overrightarrow{\fuse{\mathbf{k}}{\mathbf{l}}}:}{P}})}}, \veclabel{{o}}{\Delta}
            \\ &\cdot \; \vdashu{\Psi}{\{\any{\vec{\mathbf i}}, \any{\vec{\mathbf j}}, \any{\overrightarrow{\fuse{{\mathbf k}}{{\mathbf l}}}}\} \cupm \Psi} [\veclabel{{i}}{\alpha} \cup \veclabel{{j}}{\beta}] {\color{red}\overrightarrow{\fuse{\mathbf{k}}{\mathbf{l}}}:}{P} 
            \leftrightarrow 
            \left([\veclabel{{i}}{\alpha}]\veclabel{{k}}{P} \land [\veclabel{{j}}{\beta}]\veclabel{{l}}{P}\right) 
            \end{aligned}} }
            { \veclabel{{n}}{\Gamma} \vdashu{\Psi}{\Omega \cupm \{\any{\vec{\mathbf i}}, \any{\vec{\mathbf j}}, \any{\overrightarrow{\fuse{{\mathbf k}}{{\mathbf l}}}}\} \cupm \Psi} \veclabel{ m}{C({\mathcolorbox{pink}{[\veclabel{{i}}{\alpha}]\veclabel{{k}}{P} \land [\veclabel{{j}}{\beta}]\veclabel{{l}}{P}}})}, \veclabel{{o}}{\Delta} }
        \and
        \qquad{(\text{B})}\;\;
            \inferrule
            { {\begin{aligned}
                &\veclabel{{n}}{\Gamma} \vdashu{\Psi}{\Omega} \veclabel{ m}{C({\mathcolorbox{pink}{[\veclabel{{i}}{\alpha}]\veclabel{{k}}{P} \land [\veclabel{{j}}{\beta}]\veclabel{{k}}{P}}})}, \veclabel{{o}}{\Delta}
            \\ &\cdot \; \vdashu{\Psi}{\{\any{\vec{\mathbf i}}, \any{\vec{\mathbf j}}, \overrightarrow{\any{\fuse{\mathbf{k}}{\mathbf{k}}}}\} \cupm \Psi} [\veclabel{{i}}{\alpha} \cup \veclabel{{j}}{\beta}] {\color{red}\overrightarrow{\fuse{\mathbf{k}}{\mathbf{k}}}:}{P} 
            \leftrightarrow 
            \left([\veclabel{{i}}{\alpha}]\veclabel{{k}}{P} \land [\veclabel{{j}}{\beta}]\veclabel{{k}}{P}\right) 
            \end{aligned}} }
            { \veclabel{{n}}{\Gamma} \vdashu{\Psi}{\Omega \cupm \{\any{\vec{\mathbf i}}, \any{\vec{\mathbf j}}, \any{\overrightarrow{\fuse{{\mathbf k}}{{\mathbf k}}}}\} \cupm \Psi} \veclabel{ m}{C({\mathcolorbox{pink}{[\veclabel{{i}}{\alpha} \cup \veclabel{{j}}{\beta}] {\color{red}\overrightarrow{\fuse{\mathbf{k}}{\mathbf{k}}}:}{P}}})}, \veclabel{{o}}{\Delta} }
    \end{mathpar}
    \caption{Using axiom \color{violet}M:\color{black}$[\cup]$ with contextual equivalence rule \color{violet}M:\color{black}CER in two different styles}
    \label{fig:union_axiom_ex}
    \end{figure}


%
\subsection{Leaf Rules}

To close a proof, we apply leaf rules (for example, the \color{violet}M:\color{black}QE step in Fig.~\ref{fig:parachute_ex}). 
Leaf rules do not have any premises and provide the base cases of the proof; when given a concrete input set, each leaf rule returns a concrete output set that is a function of the input set and labels present in that proof step. 
We give all of the \uapc leaf rules in Fig.~\ref{fig:uapcleafrules}. 

\begin{figure}[tb]
    \begin{mathpar}
        \quad{\text{\color{violet}M:\color{black}id}} \;\;
        \inferrule
        {  }
        { 
            \veclabel{{i}}{P}, 
            \veclabel{{k}}{\Gamma} 
            \vdashu{\Psi}{\{\overrightarrow{\fuse{{\mathbf i}}{{\mathbf j}}}_{\mR}, {\vec{\mathbf k}}_{\mR}, {\vec{\mathbf l}}_{\mR}\} \cupm \Psi } 
            \veclabel{{j}}{P}, 
            \veclabel{{l}}{\Delta} 
        }
        \quad{\text{\color{violet}M:\color{black}QE}} \;\;
        \inferrule
        {  }
        { \veclabel{{i}}{\Gamma} \vdashu{\Psi}{DA(\vec{\mathbf i}, \vec{\mathbf j} {\mid} \Psi)} \veclabel{{j}}{\Delta} }
        \and
        \quad{\text{\color{violet}M:\color{black}$\top$R}} \;\;
        \inferrule
        {  }
        { \veclabel{{k}}{\Gamma} \vdashu{\Psi}{\{\mathbf i_{\mId}, {\vec{\mathbf k}}_{\mR}, {\vec{\mathbf j}}_{\mR}\} \cupm \Psi} \atomlabel{i}{\text{true}}, \veclabel{{j}}{\Delta} }
        \and
        \quad{\text{\color{violet}M:\color{black}$\mathbb{R}$}} \;\;
        \inferrule
        {  }
        { \veclabel{{i}}{\Gamma} \vdashu{\Psi}{DA(\vec{\mathbf i}, \vec{\mathbf j} {\mid} \Psi)} \veclabel{{j}}{\Delta} }
        \and
        \quad{\text{\color{violet}M:\color{black}$\bot$L}} \;\;
        \inferrule
        {  }
        { \atomlabel{i}{\text{false}}, \veclabel{{k}}{\Gamma} \vdashu{\Psi}{\{\mathbf i_{\mId}, {\vec{\mathbf k}}_{\mR}, {\vec{\mathbf j}}_{\mR}\} \cupm \Psi} \veclabel{{j}}{\Delta} }
        \and
        \quad{\text{\color{violet}M:\color{black}auto}} \;\;
        \inferrule
        {  }
        { \veclabel{{i}}{\Gamma} \vdashu{\Psi}{DA(\vec{\mathbf i}, \vec{\mathbf j} {\mid} \Psi)} \veclabel{{j}}{\Delta} }
    \end{mathpar}
    \caption{\uapc leaf rules}\label{fig:uapcleafrules}
\end{figure}

The rule \color{violet}M:\color{black}id closes a proof if one of the assumptions is syntactically equal to a succedent $P$. 
Given an input set $\color{blue}\Psi$, \color{violet}M:\color{black}id computes the output set $\color{blue}\{\any{\overrightarrow{\fuse{{\mathbf i}}{{\mathbf j}}}}\color{blue}, \any{\vec{\mathbf k}}\color{blue}, \any{\vec{\mathbf l}}\color{blue}\} \cupm \Psi$. 
The fused label $\color{blue}\any{\overrightarrow{\fuse{{\mathbf i}}{{\mathbf j}}}}$ indicates that $\color{red}\vec{\mathbf i}$ and $\color{red}\vec{\mathbf j}$ are allowed to (uniformly) take on at most the mutation $\mR$ because the rule only requires that the antecedent $P$ matches the succedent $P$, which would hold even if both instances were mutated to $\mathit{true}$. 
$\Gamma$ and $\Delta$ are each mutable under $\mR$ as they could be removed while preserving the proof. 

The rules \color{violet}M:\color{black}$\mathbb{R}$ and \color{violet}M:\color{black}auto take an input set $\color{blue}\Psi$ and produce an output set $\color{blue}DA(\vec{\mathbf i}, \vec{\mathbf j} {\mid} \Psi)$, defined according to Defn.~\ref{defn:DA}. When applied, the mutations in $\color{blue}DA(\vec{\mathbf i}, \vec{\mathbf j} {\mid} \Psi)$ yield a valid sequent. 
Additionally, each label $\color{red}\vec{\mathbf i}, \vec{\mathbf j}$ appears in $\color{blue}DA(\vec{\mathbf i}, \vec{\mathbf j} {\mid} \Psi)$ exactly once, which is necessary to preserve the property that \uapc analyzes every atom in a formula. 
We leave the concrete instantiation of $\color{blue}DA(\vec{\mathbf i}, \vec{\mathbf j} {\mid} \Psi)$ to future implementations of \uapc. 

\begin{definition}[\textbf{Dynamic Analysis}]\label{defn:DA}
$\color{blue}DA(\vec{\mathbf i}, \vec{\mathbf j} {\mid} \Psi)$ is a set of mutation-tracking labels such that: 
\begin{itemize}
    \item If $\veclabel{{i}}\Gamma \vdashu{\Psi}{{DA(\vec{\mathbf i}, \vec{\mathbf j} {\mid} \Psi)}} \veclabel{{j}}\Delta$ is valid, then $\color{violet}\mu_{DA(\vec{\mathbf i}, \vec{\mathbf j} {\mid} \Psi)}\color{black}(\Gamma \vdash \Delta)$ is valid for any $\color{violet}\mu_{DA(\vec{\mathbf i}, \vec{\mathbf j} {\mid} \Psi)}$. 
    \item Each label $\color{red}\vec{\mathbf i}, \vec{\mathbf j}$ appears exactly once in $\color{blue}DA(\vec{\mathbf i}, \vec{\mathbf j} {\mid} \Psi)$.
\end{itemize}
\end{definition}



\subsection{Takeaways}

The \uapc proof rules propagate label sets through a proof to learn usage information for each label. 
Given a concrete input set, leaf rules return a concrete output set with information about every label in that step. 
Rules that have multiple premises \emph{merge} their output sets into the conclusion output set. Rules that use \emph{$\phi$} discard fresh labels from their output sets because they are not mentioned in the conclusion, and hence do not give usage information about the starting formula. 
If these fresh labels were not discarded, however, they would yield usage diagnostics of cuts in a proof. 
Appendix~\ref{apx:proof_diag} formulates such a calculus. 

\uapc is flexible in allowing different ways of instantiating the constraint tracking. 
The core calculus explores proof branches in parallel and merges their usage information in each step. 
In Appendix~\ref{app:uapcpluscalculus}, we formulate $\uapcplus$ as the sequential counterpart of \uapc which threads label sets through branches, thereby guiding the analysis of each proof branch with the output of the prior. 
Due to its sequential analysis, proof rules in \uapcplus have the advantage of being defined without merge, which could reduce computational overhead in future implementations of \uapc.

\section{Key Theorems}\label{sec:metatheory} 
We prove soundness and completeness of \uapc with respect to \dL. We show representative proof cases for both theorems and motivate key lemmas when needed in the proofs. 
Full proofs of all results are in Appendix~\ref{app:metatheory}. 

\subsection{\uapc is correct.}

To show the correctness of mutations computed by \uapc, we prove that it is \emph{sound} with respect to \dL. Soundness means that if a formula has a proof in \uapc, then all of its computed mutations also have a proof in the \dL sequent calculus (Theorem~\ref{thm:soundness}).

\begin{theorem}[\textbf{Soundness}]\label{thm:soundness}
If $\veclabel{{i}}{\Gamma} \vdashu{\Psi}{\Sigma} \veclabel{{j}}{\Delta}$ is valid, then ${\color{violet}\mu_{\Sigma}}(\Gamma \vdash \Delta)$ is valid for all ${\color{violet}\mu_{\Sigma}}$.
\end{theorem}

    The proof is by structural induction on the proof of $\veclabel{{i}}{\Gamma} \vdashu{\Psi}{\Sigma} \veclabel{{j}}{\Delta}$. 
    We highlight a couple of representative cases here, M:id (representing the base cases) and M:cut (representing the inductive cases). We use boxes to distinguish key lemmas from the rest of the proof. 
    
    \begin{description}
        \item[Case M:id. ] Suppose the last rule in the proof tree is M:id:
    
        \begin{mathpar}
            \quad{\text{M:id}} \;\;
            \inferrule
            {  }
            { 
                \veclabel{{i}}{P}, 
                \veclabel{{k}}{\Gamma} 
                \vdashu{\Psi}{\{\overrightarrow{\fuse{{\mathbf i}}{{\mathbf j}}}_{\mR}, {\vec{\mathbf k}}_{\mR}, {\vec{\mathbf l}}_{\mR}\} \cupm \Psi } 
                \veclabel{{j}}{P}, 
                \veclabel{{l}}{\Delta} 
            }
        \end{mathpar}

        Let ${\color{blue}\Sigma}$ represent $\color{blue}{\{\overrightarrow{\fuse{{\mathbf i}}{{\mathbf j}}}_{\mR}, {\vec{\mathbf k}}_{\mR}, {\vec{\mathbf l}}_{\mR}\}}$. 
        \textbf{Our goal is to show that $\color{violet}\mu_{\Sigma \cupm \Psi}\color{black}(P, \Gamma \vdash P, \Delta)$ is valid for any $\color{violet}\mu_{\Sigma \cupm \Psi}$. }
        We first show that ${\color{violet}\mu_{\Sigma}}(P, \Gamma \vdash P, \Delta)$ is valid for any ${\color{violet}\mu_{\Sigma}}$. Then, we extend the result to $\color{violet}\mu_{\Sigma \cupm \Psi}$. \\
        
        Our strategy is to \emph{eliminate what we don't need}, which are $\Gamma$ and $\Delta$ for \color{violet}M:\color{black}id. 
        Fix an arbitrary ${\color{violet}\mu_{\Sigma}}$. Observe that if we can show that ${\color{violet}\mu_{\Sigma}}(P \vdash P)$ is valid, then by the \dL rule \textit{W}LR, we can show that ${\color{violet}\mu_{\Sigma}}(P, \Gamma \vdash P, \Delta)$ is valid:
        
        \begin{mathpar}
            \quad{\text{\textit{W}LR}} \;\;
            \inferrule{
                {\color{violet}\mu_{\Sigma}}(P \vdash P)
            }{
                {\color{violet}\mu_{\Sigma}}(P, \Gamma \vdash P, \Delta)
            }
        \end{mathpar}
    
        \textbf{We need to show that ${\color{violet}\mu_{\Sigma}}(P \vdash P)$ is valid. } 
        
        By Defn.~\ref{defn:fusing} and the definition of $\mu$ in  Fig.~\ref{fig:mutationmapping}, the mutation $\color{violet}m$ selected for both labels $\color{red}\vec{\mathbf i}$ and $\color{red}\vec{\mathbf j}$ must satisfy $\color{violet}m \color{black}\sqsubseteq \color{violet}\mR$. We have one case for each possible $\color{violet}m$ (three in total). We elide mutations for $\color{red}\vec{\mathbf k}$ and $\color{red}\vec{\mathbf l}$ from $\mu$ to focus the argument because $\Gamma$ and $\Delta$ do not appear in $P \vdash P$. \\
        
        \underline{Subcase 1: ${\color{violet}m} = \mId$.} By definition of $\mu$, we have
        
        \[{\color{violet}\mu_{\Sigma}}\color{black}(P \vdash P) \equiv \left(\color{violet}\mu_{\{\vec{\mathbf i}_{\mId}, \vec{\mathbf j}_{\mId}\}}\color{black}(P) \color{black}\vdash \color{violet}\mu_{\{\vec{\mathbf i}_{\mId}, \vec{\mathbf j}_{\mId}\}}\color{black}(P)\right) \equiv \left(P \vdash P\right)\]

        \underline{Subcase 2: ${\color{violet}m} = \mW$.} By definition of $\mu$, we have
        
        \[{\color{violet}\mu_{\Sigma}}\color{black}(P \vdash P) \equiv \left(\color{violet}\mu_{\{\vec{\mathbf i}_{\mW}, \vec{\mathbf j}_{\mW}\}}\color{black}(P) \vdash \color{violet}\mu_{\{\vec{\mathbf i}_{\mW}, \vec{\mathbf j}_{\mW}\}}\color{black}(P)\right) \equiv \left(\tilde{P} \vdash \tilde{P}\right)\]

        \underline{Subcase 3: ${\color{violet}m} = \mR$.} By definition of $\mu$, we have
        
        \[{\color{violet}\mu_{\Sigma}}(P \vdash P) \equiv \left(\color{violet}\mu_{\{\vec{\mathbf i}_{\mR}, \vec{\mathbf j}_{\mR}\}}\color{black}(P) \vdash \color{violet}\mu_{\{\vec{\mathbf i}_{\mR}, \vec{\mathbf j}_{\mR}\}}\color{black}(P)\right) \equiv \left(true \vdash true\right)\]

        By \dL rule id, each sequent is valid. 
        \textbf{We have just shown that ${\color{violet}\mu_{\Sigma}}(P \vdash P)$ is valid.} 
        
        We can now apply \textit{W}LR to learn that \textbf{${\color{violet}\mu_{\Sigma}}(\Gamma, P \vdash P, \Delta)$ is valid.} 
        Since ${\color{violet}\mu_{\Sigma}}$ was arbitrary, the result holds for \textit{any} ${\color{violet}\mu_{\Sigma}}\color{black}(P, \Gamma \vdash P, \Delta)$. \\

        To finish the proof, we need to show that $\color{violet}\mu_{\Sigma \cupm \Psi}\color{black}(\Gamma, P \vdash P, \Delta)$ is valid for all $\color{violet}\mu_{\Sigma \cupm \Psi}$ (Lemma~\ref{lem: subset_muts}). 
        The key intuition is that \emph{introducing more mutational constraints limits the overall pool of mutations for a given atom} due to the semantics of merge $\cupm$. In short, the set of labels may grow, but the mutations can only become more conservative. \\

        \noindent\fbox{\begin{minipage}{\linewidth}
        \begin{lemma}[\textbf{Monotonicity}]\label{lem: subset_muts}
        If $\color{violet}\mu_{\Sigma_1}\color{black}(\Gamma \vdash \Delta)$ is valid for all $\color{violet}\mu_{\Sigma_1}$, then $\color{violet}\mu_{\Sigma_1 \cupm \Sigma_2}\color{black}(\Gamma \vdash \Delta)$ is valid for all $\color{violet}\mu_{\Sigma_1 \cupm \Sigma_2}$.
        \end{lemma}
        \end{minipage}} \\

        \textbf{By Lemma~\ref{lem: subset_muts}, we have:} $\mathbf{\color{violet}\mu_{{\{\overrightarrow{\fuse{{\mathbf i}}{{\mathbf j}}}_{\mR}, {\vec{\mathbf k}}_{\mR}, {\vec{\mathbf l}}_{\mR}\}} \cupm \Psi}\color{black}(P,\Gamma \vdash P,\Delta) \textbf{ is valid for all }\color{violet}\mu_{{\{\overrightarrow{\fuse{{\mathbf i}}{{\mathbf j}}}_{\mR}, {\vec{\mathbf k}}_{\mR}, {\vec{\mathbf l}}_{\mR}\}} \cupm \Psi}}$. \\
        

        \item[Case M:cut. ]  Suppose the last rule in the proof tree is M:cut:
    
        \begin{mathpar}
        \quad{\text{\color{violet}M:\color{black}cut}^{\color{red}\vec{l}\color{black}}} \;\;
        \inferrule
        { \veclabel{{i}}{\Gamma}, \veclabel{{j}}{C} \vdashu{\Psi}{\Sigma} \veclabel{{k}}{\Delta}
        \\ 
        \veclabel{{i}}{\Gamma} \vdashu{\Psi}{\Omega} \veclabel{{k}}{\Delta}, \veclabel{{j}}{C}
        }
        { \veclabel{{i}}{\Gamma} \vdashu{\Psi}{(\Sigma \cupm \Omega) \setminus {\vec{l}}} \veclabel{{k}}{\Delta} }
        \quad{(\forall C_a \in \text{atoms}(C). \, \color{red}j_a \color{black}= \phifunc{}{C_a}{(\veclabel{{i}}{\Gamma}, \veclabel{{k}}{\Delta}\color{black})}{\color{red}\vec{l}\color{black}})}
        \end{mathpar}

        \textbf{We want to show that $\color{violet}\mu_{(\Sigma \cupm \Omega) \setminus {\vec{l}}}\color{black}(\Gamma \vdash \Delta)$ is valid.} \\
        
        Because the last rule is M:cut, we know that there must exist \emph{smaller} proofs of 
    
        \begin{itemize}
            \item[(1)] $\veclabel{{i}}{\Gamma}, \veclabel{{j}}{C} \vdashu{\Psi}{\Sigma} \veclabel{{k}}{\Delta}$
            \item[(2)] $\veclabel{{i}}{\Gamma} \vdashu{\Psi}{\Omega} \veclabel{{k}}{\Delta}, \veclabel{{j}}{C}$ 
        \end{itemize}

        By application of the IH to (1) and (2), we have:
        
        \begin{itemize}
            \item[(3)] ${\color{violet}\mu_{\Sigma}}(\Gamma, C \vdash \Delta)$ is valid for all ${\color{violet}\mu_{\Sigma}}$.
            \item[(4)] ${\color{violet}\mu_{\Omega}}(\Gamma \vdash \Delta, C)$ is valid for all $\color{violet}\mu_{\Omega}$.
        \end{itemize}
        
        By Lemma~\ref{lem: subset_muts} applied to (3): 
        
        \[\color{violet}\mu_{\Sigma \cupm \Omega}\color{black}(\Gamma, C \vdash \Delta) \; \color{black}\text{ is valid for \emph{all} $\color{violet}\mu_{\Sigma \cupm \Omega}$.} \;\;\; (5)\]
    
         \emph{Fix an arbitrary $\color{violet}\mu_{\Sigma \cupm \Omega}$}. By Lemma~\ref{lem: subset_muts} applied to (4):
        
        \[\color{violet}\mu_{\Sigma \cupm \Omega}\color{black}(\Gamma \vdash \Delta, C) \; \text{ is valid for \emph{the same fixed} $\color{violet}\mu_{\Sigma \cupm \Omega}$}. \;\;\; (6)\]
        
        By the \dL proof rule cut applied to (5) and (6):
        
        \[\color{violet}\mu_{\Sigma \cupm \Omega}\color{black}(\Gamma \vdash \Delta) \; \text{ is valid for \emph{fixed} $\color{violet}\mu_{\Sigma \cupm \Omega}$}. \;\;\; (7)\]

        Because $\color{violet}\mu_{\Sigma \cupm \Omega}$ was arbitrary, we know that (7) holds \emph{for all} $\color{violet}\mu_{\Sigma \cupm \Omega}$.

        To finish the proof, we introduce Corollary~\ref{lem: minus_muts} which shows that $\color{violet}\mu_{(\Sigma \cupm \Omega) \setminus \vec{l}}\color{black}(\Gamma \vdash \Delta) \; \text{ is valid}$ for all $\color{violet}\mu_{(\Sigma \cupm \Omega) \setminus \vec{l}}$. Discarding labels $\color{red}\vec{l}$ is safe because they are fresh, meaning that they do not appear in $\Gamma \vdash \Delta$. Hence, their mutations could not affect $\Gamma \vdash \Delta$. 
        

        \noindent\fbox{\begin{minipage}{\linewidth}
        \begin{corollary}[\textbf{Fresh monotonicity}] \label{lem: minus_muts}
        If $\color{violet}\mu_{\Sigma_1}(\Gamma \vdash \Delta)$ is valid for all $\color{violet}\mu_{\Sigma_1}$ and $\color{red}\vec{l}$ is fresh, then $\color{violet}\mu_{\Sigma_1 \setminus {\vec{l}}}(\Gamma \vdash \Delta)$ is valid for all $\color{violet}\mu_{\Sigma_1 \setminus {\vec{l}}}$. 
        \end{corollary}
        \end{minipage}} \\
        
        Because $\color{red}\vec{l}$ is fresh, the desired result follows directly by Corollary~\ref{lem: minus_muts}:
        
        \[\color{violet}\mu_{(\Sigma \cupm \Omega) \setminus {\vec{l}}}\color{black}(\Gamma \vdash \Delta) \; \text{ is valid for \textit{all} $\color{violet}\mu_{(\Sigma \cupm \Omega) \setminus {\vec{l}}}$}.\]
    \end{description}

\subsection{\uapc can analyze any valid \dL formula.}

To show that \uapc can analyze any valid \dL formula, we prove that it is \emph{complete} with respect to \dL. Completeness means that if a formula has a proof in the \dL sequent calculus, then for all input sets $\color{blue}\Psi$ there exists an output set $\color{blue}\Sigma$ such that the formula has a proof in \uapc (Theorem~\ref{thm:completeness}). 

\begin{theorem}[\textbf{Completeness}]\label{thm:completeness}
If $\Gamma \vdash \Delta$ is valid, then $\forall \, \color{blue}\Psi$ $\exists \, \color{blue}\Sigma$ s.t. $\veclabel{{i}}{\Gamma} \vdashu{\Psi}{\Sigma} \veclabel{{j}}{\Delta}$ is valid.
\end{theorem}

The proof is by structural induction on the dL proof of $\Gamma \vdash \Delta$. We highlight a few interesting cases, $\land$R, [:=], and id.

\begin{description}
    \item[Case $\land$R. ] Suppose the last rule in the proof tree is $\land$R:
    
        \begin{mathpar}
            \quad{\text{$\land$R}} \;\;
            \inferrule
            { {\Gamma} \vdash {P}
            \\ {\Gamma} \vdash {Q} }
            { {\Gamma} \vdash {P} \land {Q} }
        \end{mathpar}
    
        Let $\color{blue}\Psi$ be an arbitrary input set. 
        By $\land$R, there exist \emph{smaller} proofs of

        \begin{itemize}
            \item[(1)] $\Gamma \vdash P$
            \item[(2)] $\Gamma \vdash Q$
        \end{itemize}
        By the IH applied to (1) and (2), there exist output sets $\color{blue}\Sigma$ and $\color{blue}\Omega$ such that 
        
        \begin{itemize}
            \item[(3)] $\veclabel{{k}}{\Gamma} \vdashu{\Psi}{\Sigma} \veclabel{{i}}{P}$ 
            \item[(4)] $\veclabel{{k}}{\Gamma} \vdashu{\Psi}{\Omega} \veclabel{{j}}{Q}$
        \end{itemize}
        
        By application of M:$\land$R to (3) and (4), we obtain the desired result with output set $\color{blue}\Sigma \cupm \Omega$:
        
        \begin{mathpar}
            \quad{\text{M:$\land$R}} \;\;
            \inferrule
            { \veclabel{{k}}{\Gamma} \vdashu{\Psi}{\Sigma} \veclabel{{i}}{P}
            \\ \veclabel{{k}}{\Gamma} \vdashu{\Psi}{\Omega} \veclabel{{j}}{Q} }
            { \veclabel{{k}}{\Gamma} \vdashu{\Psi}{\Sigma \cupm \Omega} \veclabel{{i}}{P} \land \veclabel{{j}}{Q} } \\
        \end{mathpar}
        %

        
    \item[Case {[:=]}. ] Suppose the last rule in the proof tree is $[:=]$:
    
        \begin{mathpar}
            \quad{[:=]} \;\; \cdot \vdash [{x:=e}] {p(x)} \leftrightarrow {p(e)}
        \end{mathpar}
    
        Let $\color{blue}\Psi$ be an arbitrary input set. The proof proceeds in two subcases: 

        \begin{itemize}
            \item[(1)] where $x$ appears in the postcondition (i.e. $x \in FV(p(x))$) and 
            \item[(2)] where $x$ does not appear in the postcondition (i.e. $x \notin FV(p)$).
        \end{itemize}
    
        \paragraph{Subcase $x \in FV(p(x))$} If $x \in FV(p(x))$, then we obtain the desired result by  \color{violet}M:\color{black}$[:=]_1$ with output set $\color{blue}\{\mathbf{j_{\mId}}, \vec{\mathbf k}_{\mR}\} \cupm \Psi$:
    
        \begin{mathpar}
            \quad{\text{\color{violet}M:\color{black}$[:=]_1$}} \;\; \vdashu{\Psi}{\{{\mathbf j}_{\mId}, \vec{\mathbf k}_{\mR}\} \cupm \Psi} [\atomlabel{j}{x:=e}] \;\veclabel{{k}}{p(x)} \leftrightarrow \veclabel{{k}}{p(e)} \;\; \quad{(x \in FV(p(x)))}
        \end{mathpar}
    
        \paragraph{Subcase $x \notin FV(p)$} If $x \notin FV(p)$, then we obtain the desired result by \color{violet}M:\color{black}$[:=]_2$ with output set $\color{blue}\{{\mathbf j}_{\mR}, \vec{\mathbf k}_{\mR}\} \cupm \Psi$:
    
        \begin{mathpar}
            \quad{\text{\color{violet}M:\color{black}$[:=]_2$}} \;\; \vdashu{\Psi}{{\{{\mathbf j}_{\mR}, \vec{\mathbf k}_{\mR}\} \cupm \Psi}} [\atomlabel{j}{x:=e}] \;\veclabel{{k}}{p} \leftrightarrow \veclabel{{k}}{p} \;\; \quad{(x \notin FV(p))}
        \end{mathpar}

    \item[Case id.] Suppose the last rule in the proof tree is id:
    
        \begin{mathpar}
            \quad{\text{id}} \;\;
            \inferrule
            {  }
            { {P}, {\Gamma} \vdash {P}, {\Delta} }
        \end{mathpar} 
    
        Let $\color{blue}\Psi$ be an arbitrary input set. Therefore, the desired result follows by the rule \color{violet}M:\color{black}id with output set $\color{blue}\{\any{\fuse{\vec{\mathbf i}}{\vec{\mathbf j}}}, \any {\vec{\mathbf k}}, \any {\vec{\mathbf l}}\} \cupm \Psi$:
        
        \begin{mathpar}
            \quad{\text{\color{violet}M:\color{black}id}} \;\;
            \inferrule
            {  }
            { \veclabel{{i}}{P}, \veclabel{{k}}{\Gamma} \vdashu{\Psi}{\{\any{\fuse{\vec{\mathbf i}}{\vec{\mathbf j}}}, \any {\vec{\mathbf k}}, \any {\vec{\mathbf l}}\} \cupm \Psi } \veclabel{{j}}{P}, \veclabel{{l}}{\Delta} }
        \end{mathpar}

\end{description}

All other cases follow a similar proof strategy; identify the last rule in the proof tree and use the assumed valid dL formula and an arbitrary input set $\color{blue}\Psi$ to obtain the desired result. \emph{In each case, we observe that for every dL proof rule, there exists a matching \uapc proof rule.}

\subsection{Takeaways}

\uapc is both sound and complete, meaning that mutations computed by \uapc can be applied to a valid \dL formula while preserving its proof, and that \uapc is general enough to analyze \emph{any} valid \dL formula. 
Our modular proofs of both results allow for straightforward extension to new and derived proof rules. 
The modular structure of our base cases (as seen in the \color{violet}M:\color{black}id soundness case) allow for straightforward extension to new mutations.

\section{Parachute Proof ($\uapc$ style)}\label{sec:example}
In this section, we revisit the parachute proof from Sec.~\ref{sec:background}, this time analyzing with \uapc. 
\new{The analysis begins at the root of the proof tree shown in the bottom fragment.} We suppose the programmer is pretty confident in their physical dynamics, so they provide the initial constraints $\color{blue}\Psi = {\color{red}\vec{o}}_{\mId}\color{blue}$ to enforce that the differential equations (labeled by $\color{red}\vec{o}$) should remain unchanged. The objective is to solve for the output set $\color{blue}\Omega$ in the conclusion, which \uapc accomplishes by analyzing the input set and synthesizing an output set in each proof step. 

\subsection{The Analysis}

\textbf{The input set $\color{blue}\Psi\color{black}$ propagates up through each proof step} until it reaches the leaf rules which use it to compute a concrete output set. For example, the \color{violet}M:\color{black}QE step of branch (3h) concludes that $r_{cl} = 0$, $r = r_{cl}$, and $-g + rv^2 \geq -g$ should remain unchanged under mutation ($\mId$). 

\textbf{Now, the output set $\color{blue}DA(\dots {\mid} \Psi)$ propagates down the proof tree.} 
In the \color{violet}M:\color{black}$[:=]_2$ step, \uapc determines that the assignment $x':=v$ is not used in this step and can be safely removed ($\mR$). 
The next two steps (both \color{violet}M:\color{black}$[:=]_2$) detect that assignments $v':=-g+rv^2$ and $t':=1$ are used and should remain unchanged ($\mId$). 

The output set continues to propagate down to (3a), where the outputs of (3g) and (3h) are merged by \color{violet}M:\color{black}dI, and then again with (3f) by \color{violet}M:\color{black}dC which also discards the fresh label $s$. 
The \color{violet}M:\color{black}$\land$R step merges the output sets of its premises forming the output set $\color{blue}\Theta$ in the conclusion. 
This output set is preserved for the remainder of (3a).

In the bottom snippet, the \color{violet}M:\color{black}loop step merges the output sets of (1), (2), and (3) and discards fresh label $\color{red}u_1$, producing the output set $\color{blue}\Omega$. 
The \color{violet}M:\color{black}$\rightarrow$R step preserves this output. \textbf{As this rule is the root of the proof tree, the analysis concludes and $\color{blue}\Omega$ is the final result:}

\begin{align*}
    \color{blue}\Omega = &\color{blue}\{{\color{red}\fuse{i_5}{p_2}}_{\color{violet}\mId}\color{blue},{\color{red}\fuse{i_6}{p_3}}_\mId\color{blue},\color{red}k_{\mId}\color{blue}, 
    \any{\color{red}j}\color{blue}, \any{\color{red}l}\color{blue}, {\color{red}\{o_1\}_{\mId}}\color{blue}, \color{red}\{o_2\}_{\mId}\color{blue}, \color{red}\{o_3\}_{\mId}\color{blue}, \color{red}n_{\mId}\color{blue}, \color{red}\{i_7\}_{\mId}, \any{\color{red}m}\color{blue}, \any{\color{red}\{p_1\}}\color{blue}, \any{\color{red}\{i_1\}}\color{blue}, \any{\color{red}\{i_2\}}\color{blue}, \any{\color{red}\{i_3\}}\color{blue}, \\
    &\any{\color{red}\{i_4\}}\color{blue}, {\color{red}\{i_8\}}_\mId\color{blue}, \color{red}\vec{q}_{\mId}\color{blue}\}
\end{align*}

\subsection{Model Improvement}

From $\color{blue}\Omega$, the programmer learns that the atoms labelled by $\color{red}k$, $\color{red}o_2$, $\color{red}o_3$, $\color{red}n$, $\color{red}i_7$, $\color{red}i_8$, and $\color{red}\vec{q}$ should all remain unchanged, with the additional constraint that $\color{red}i_5$ and $\color{red}p_2$, and $\color{red}i_6$ and $\color{red}p_3$ take on the same mutation (which degenerates to just remaining unchanged because $\mId$ is the only mutation allowed). 
The atom labeled by $\color{red}o_1$ (which corresponds to the differential equation $x' = v$) must also remain unchanged. Even though \uapc determined that $x' = v$ was not needed to close the proof, the input constraints enforced that it must remain unchanged, which is respected in the output set. 
All other atoms can take on any mutation. \textbf{In particular, we notice that $x > 100$ ($\color{red}l$) and $v - gT > - \sqrt{g/r_{op}}$ ($\color{red}j$), as expected, were unused in the proof, and \uapc verifies that these can be safely removed from the formula while preserving its validity.} 

If the programmer were to apply all these mutations at their extremes, they would learn that the following generalized formula is also valid:

\begin{figure}[htbp]
\begin{align*}
& \color{violet}\color{violet}\mu_\Omega\left(\color{black} {A} {\to} \left[\left(\left(?( {v - gT > -\sqrt{g/r_{op}}} \land  {r = r_{cl}} \land  {x > 100}) \cup \; r := r_{op}\right); t:= 0; \text{ODE} \&  {Q}\right)^*\right] P\color{violet}\right)\color{black}\\
\equiv
&  \color{violet}\color{violet}\mu_\Omega(\color{black}A\color{violet})\color{black} {\to} \left[\left(\left(?( \color{violet}\textbf{true}\color{black} \land  {r = r_{cl}} \land  \color{violet}\textbf{true}\color{black}) \cup \; ?\color{violet}\textbf{true}\color{black} \right); t:= 0; \color{violet}\color{violet}\mu_\Omega(\color{black}\text{ODE}\color{violet})\color{black} \&  \color{violet}\color{violet}\mu_\Omega(\color{black}Q\color{violet})\color{black}\right)^*\right] \color{violet}\color{violet}\mu_\Omega(\color{black}P\color{violet})\\
\\
&\text{where} \\
&\color{violet}\color{violet}\mu_\Omega(\color{black}A\color{violet})\color{black} \equiv  \color{violet}\textbf{true}\color{black} \land  \color{violet}\textbf{true}\color{black} \land  \color{violet}\textbf{true}\color{black} \land  \color{violet}\textbf{true}\color{black} \land  {v_{max} < -\sqrt{g/r_{op}}} \land  {x \geq 0} \land  {-\sqrt{g/r_{op}} < v < 0} \land  {r = r_{cl}} \\
& \color{violet}\color{violet}\mu_\Omega(\color{black}\text{ODE}\color{violet})\color{black} \equiv x' = v, v' = -g + rv^2, t' = 1 \\
&  \color{violet}\color{violet}\mu_\Omega(\color{black}Q\color{violet})\color{black} \equiv  {t \leq T} \land  {x \geq 0} \land  {v < 0} \\
& \color{violet}\color{violet}\mu_\Omega(\color{black}P\color{violet})\color{black} \equiv  {x = 0} \to v \geq v_{max}
\end{align*}
\caption{Model of parachute dynamics, \color{violet}\textbf{mutated}\color{black}.}
\end{figure}

The non-$\mId$ mutated atoms are highlighted in violet bold. By just looking at the number of \color{violet}\textbf{true}\color{black}'s, we see that there are many unused (and unnecessary) constraints in this formula. 



\subsection{Takeaways}

\textbf{\uapc makes studying large proofs of complex \dL formulas like this one straightforward.} At first glance, it is not immediately obvious that $v - gT > - \sqrt{g/r_{op}}$ could be removed from the original formula without affecting validity, yet through systematic analysis, \uapc is able to detect this. Additionally, the input set $\color{blue}\Psi$ could have instead been instantiated with a non-empty set of user-imposed mutations (for example, atom $\color{red}l$ should stay $\mId$) and \uapc would account for these constraints too. 

The proof-based technique allows us to identify subtle bugs. In this example, they are $\mathcolorbox{yellow}{\text{formulas}}$ that occur in a test statement and eventually propagate into the assumptions of the proof tree.

%
\section{Related Work}\label{sec:related}
%

\paragraph{Classifying modeling errors in CPSs.} 

\cite{fulton2019} describes using mutations to synthesize candidate models to be validated against the real-world-environment at runtime. 
\cite{selvaraj2022} stratifies modeling errors in \dL formulas into two categories: vacuity and over-specification. Our work \emph{diagnoses} both kinds. 


\paragraph{Finding modeling errors in CPSs.}

ModelPlex~\cite{DBLP:journals/fmsd/MitschP16} identifies modeling errors in \dL formulas at runtime by determining whether or not a model conforms to the real-world (or simulated) environment. Here, we identify modeling errors \emph{prior to runtime}.
CESAR~\cite{Kabra_2024} synthesizes missing hybrid program components according to a specification. Our work instead \emph{debugs an existing program by looking at its proof} to determine possible mutations, thereby enabling proof diagnostics. 
%

Static vacuity detection for temporal logical formulas finds program traces that vacuously satisfy the specification~\cite{Dokhanchi2017} and checks whether all subformulas of a specification affect its truth~\cite{Kupferman2003}. In \uapc, the mutation $\mR$ provides vacuity detection while the mutation $\mW$ identifies over-constrained program components, \emph{diagnosing a more subtle class of bugs not covered by vacuity detection}. 


Other work uses adaptive test case generation~\cite{BARTOCCI2022101254} and property-based mutation~\cite{10132190} testing to find modeling errors in signal temporal logic~\cite{STL2013} formulas. 
%
\cite{10132190} allows only one mutation to be applied and tested at a time, whereas \uapc allows \emph{multiple mutations to be applied at the same time} and uses fusing to reason about potential dependencies between them. 
Both approaches~\cite{BARTOCCI2022101254,10132190} operate directly on the formula whereas our approach operates on the formula's proof tree which we use to glean \emph{information about both the formula as well as potential proof diagnostics}. To our knowledge, none of these approaches are verified to be \emph{sound or complete}. 

Recent work applies weakest precondition and strongest postcondition techniques to detect over-specification in pre- and post-conditions of \dL formulas~\cite{Chong2023}. However, these techniques \emph{do not provide the granularity of bug detection that we are pursuing in this work.} \uapc is capable of diagnosing over-specification of atoms arising all over \dL formula, not just in the pre- and post-conditions.

%

\paragraph{Refinement for \dL.} 

Prior work uses similar proof-based techniques for finding refactorings of \dL programs~\cite{Mitsch2014}. Like \uapc, their work aims to find ways \dL formula could be updated while reusing or transforming the proof. 
However, each refactoring has to be justified semantically on its own, whereas \uapc embeds such refactorings (here, mutations) into its proof rules which are used to construct a proof that automatically justifies a set of possible refactorings. 

Differential refinement logic has been used to simplify \dL proofs and develop proofs of complex \dL formulas~\cite{Loos2016, Prebet2024}. They relate concrete \dL formulas to abstract ones which can be proved more easily. 
\uapc is a refinement of the \dL proof calculus, providing usage information that would not be so easily attained by working in \dL alone.
\uapc operates on a given \dL proof, generating mutations such that when applied, generate a similar proof to the original. Our extension $\uapc_\texttt{diag}$ diagnoses over-specification and vacuity of cuts in \dL proofs.

\paragraph{Linear logic.} 


Linear logic~\cite{10.1007/3-540-60983-0_5,chirimar_gunter_riecke_1996} characterizes the usage of atoms in formula \emph{quantitatively} whereas \uapc does so \emph{qualitatively}. 
The purposes of linear logic and \uapc are orthogonal. Linear logic ensures that resources in a proof are used exactly once whereas \uapc's qualitative analysis of atom usage allows it to detect bugs in formulas, which linear logic was not designed for. 
Atom usage in a \dL proof is non-linear in the sense of linear logic meaning that the same atom may be used in multiple places in a proof. To ensure the soundness of our mutation scheme, it is especially important to reason about the usage of \emph{all} occurrences of a given atom in a proof. 
Resource semantics~\cite{reed2009} labels formulas and uses a proof calculus to track the \emph{quantitative} usage of atoms in a proof. 
Proof normalization and cut elimination operate on proofs~\cite{COCKETT200188, pfenning2009} while \uapc operates on the conjecture itself, producing a mutated proof in the process. In Appendix~\ref{apx:proof_diag}, we formulate an extension of \uapc that yields usage information for cuts in a proof.


\subsection{Takeaways}

Subsuming vacuity detection~\cite{Dokhanchi2017,Kupferman2003,10.1145/3371078}, \uapc identifies over-constrained program components in addition to vacuous ones. Other approaches apply weakest precondition and strongest postcondition techniques~\cite{Chong2023}, yet they are limited to detecting bugs in a formula's precondition and postcondition, whereas \uapc analyzes every atom in a formula. 
To our knowledge, testing-based approaches ~\cite{BARTOCCI2022101254,10132190,STL2013} are not proved to be correct or complete whereas \uapc is. Unlike single fault analyses~\cite{10132190}, \uapc produces mutations for every atom in a formula making it capable of detecting multiple interconnected bugs. 
\uapc also provides proof diagnostics which arises naturally from its proof-based approach that could not be achieved by testing-based techniques. 

\section{Conclusion and Future Work}\label{sec:conclusion}

In this work, we developed \uapc, a proof-based technique for diagnosing vacuity and over-specification bugs in \dL formulas. 
The resulting system is capable of detecting all possible classes of bugs in \dL formulas~\cite{selvaraj2022} and helps programmers develop more generalized hybrid systems models with stronger safety guarantees. 
Our proof-based technique yields several benefits. First, it is capable of detecting bugs in any valid \dL formula, which is verified by our completeness property. The proof-based technique also lends itself to systematic analysis of every atom in a \dL formula, and determines combinations of mutations on these atoms. \uapc's soundness property guarantees that any combination of computed mutations preserve the validity of a \dL formula. We also formulated sound and complete extensions of \uapc to proof diagnostics, which could be used to find optimal \dL proofs, and sequential analysis, allowing future implementations to avoid merging sets.

\paragraph{Future work}

%
The techniques we introduce through \uapc are useful for studying model improvement and proof-preserving modifications. In this work, we used the mutations $\mId$, $\mW$, and $\mR$ as an example. However, extending to more fine-grained mutations is supported by our meet-semilattice formulation, which would involve defining the new mutations as functions and inserting them between $\mId$ and $\mR$ in the partial order. 
For example, we leave generalizing mutations for atomic differential equations to future work, which in this work we took to be the same as the identity mutation.

%
We leave implementations of \uapc, \uapcplus, and $\uapc_\texttt{diag}$ to future work, yet we anticipate them to be straightforward with the proposed formulation. Like all other high-level reasoning and tracking relation among proof steps, our calculi will be implemented outside of the KeYmaera X core~\cite{CADE2015}. The soundness and completeness properties proved in this work ensure that the implementations produce mutations and proof steps \textbf{guaranteed to succeed in the core}.

\textbf{\uapc is also capable of reducing complexity in its implementations.} A naive implementation of \color{violet}M:\color{black}$\mathbb{R}$ that combinatorially mutates each atom in a formula is infeasible because it involves exponential search on top of a doubly-exponential decision procedure~\cite{davenport88, grebing2020usability} which has been a pain point in related techniques~\cite{10132190}. 
In \uapc, the input set $\color{blue}\Psi\color{black}$ in $\color{blue}DA(\vec{\mathbf l} {\mid} \Psi)$ is an initial set of constraints that guides the analysis so that it only needs to test constraints as strict or stricter than the ones provided. 

\clearpage

\bibliographystyle{ACM-Reference-Format}
\bibliography{bibfile}

\clearpage

\appendix
\input{appendix_tr}

\end{document}

%% file: preamble.tex
\usepackage{amsmath}
\usepackage{amsfonts}
\usepackage{mathpartir}
\usepackage{xcolor}
\usepackage{color,soul}
\usepackage{cleveref}
\usepackage{comment}
\usepackage{float}
\usepackage{xspace}
\usepackage{todonotes}
\usepackage{tikz}
\usepackage{stmaryrd}
\usepackage[skins]{tcolorbox}
\usetikzlibrary{decorations.pathmorphing}

\DeclareFontFamily{OT1}{pzc}{}
\DeclareFontShape{OT1}{pzc}{m}{it}{<-> s * [1.10] pzcmi7t}{}
\DeclareMathAlphabet{\mathpzc}{OT1}{pzc}{m}{it}

\newtcolorbox{proofstartboxexample}[1][]{
    enhanced,
    oversize=0cm,
    center,
    frame code={%
        \draw[solid,line width=1pt,black,decoration={zigzag,amplitude=0.5mm},decorate] (frame.north west) -- (frame.north east);
        \draw[solid,line width=1pt,black] (frame.north east) -- (frame.south east);
        \draw[solid,line width=1pt,black] (frame.south east) -- (frame.south west);
        \draw[solid,line width=1pt,black] (frame.south west) -- (frame.north west);
    }
}

\newtcolorbox{proofleafboxexample}[1][]{
    enhanced,
    oversize=0cm,
    center,    
    frame code={%
        \draw[solid,line width=1pt,black] (frame.north west) -- (frame.north east);
        \draw[solid,line width=1pt,black] (frame.north east) -- (frame.south east);
        \draw[solid,line width=1pt,black,decoration={zigzag,amplitude=0.5mm},decorate] (frame.south east) -- (frame.south west);
        \draw[solid,line width=1pt,black] (frame.south west) -- (frame.north west);
    },
    title={#1},
    fonttitle=\bfseries,
    attach boxed title to top left={yshift=-8mm,yshifttext=-2mm,xshift=2mm}
}

\newtcolorbox{proofmiddleboxexample}[1][]{
    enhanced,
    oversize=0cm,
    center,    
    frame code={%
        \draw[solid,line width=1pt,black,decoration={zigzag,amplitude=0.5mm},decorate] (frame.north west) -- (frame.north east);
        \draw[solid,line width=1pt,black] (frame.north east) -- (frame.south east);
        \draw[solid,line width=1pt,black,decoration={zigzag,amplitude=0.5mm},decorate] (frame.south east) -- (frame.south west);
        \draw[solid,line width=1pt,black] (frame.south west) -- (frame.north west);
    },
    title={#1},
    fonttitle=\bfseries,
    attach boxed title to top left={yshift=-8mm,yshifttext=-2mm,xshift=2mm}
}

\newtcolorbox{proofstartbox}[1][]{
    enhanced,
    oversize=0cm,
    center,
    frame code={%
        \draw[solid,line width=1pt,black,decoration={zigzag,amplitude=0.5mm},decorate] (frame.north west) -- (frame.north east);
        \draw[solid,line width=1pt,black] (frame.north east) -- (frame.south east);
        \draw[solid,line width=1pt,black] (frame.south east) -- (frame.south west);
        \draw[solid,line width=1pt,black] (frame.south west) -- (frame.north west);
    }
}

\newtcolorbox{proofmiddlebox}[1][]{
    enhanced,
    oversize=0cm,
    center,    
    frame code={%
        \draw[solid,line width=1pt,black,decoration={zigzag,amplitude=0.5mm},decorate] (frame.north west) -- (frame.north east);
        \draw[solid,line width=1pt,black] (frame.north east) -- (frame.south east);
        \draw[solid,line width=1pt,black,decoration={zigzag,amplitude=0.5mm},decorate] (frame.south east) -- (frame.south west);
        \draw[solid,line width=1pt,black] (frame.south west) -- (frame.north west);
    },
    title={#1},
    fonttitle=\bfseries,
    attach boxed title to bottom center={yshift=2mm,yshifttext=1mm}
}

\newtcolorbox{proofleafbox}[1][]{
    enhanced,
    oversize=0cm,
    center,    
    frame code={%
        \draw[solid,line width=1pt,black] (frame.north west) -- (frame.north east);
        \draw[solid,line width=1pt,black] (frame.north east) -- (frame.south east);
        \draw[solid,line width=1pt,black,decoration={zigzag,amplitude=0.5mm},decorate] (frame.south east) -- (frame.south west);
        \draw[solid,line width=1pt,black] (frame.south west) -- (frame.north west);
    },
    title={#1},
    fonttitle=\bfseries,
    attach boxed title to bottom center={yshift=2mm,yshifttext=1mm}
}

\newcommand{\new}[1]{{{#1}}}

\newcommand{\mathcolorbox}[2]{\colorbox{#1}{$\displaystyle #2$}}
\newcommand{\vdashu}[2]{\ensuremath{\vdash\mkern-5mu\color{blue}^{#1}_{#2}}\color{black}\,}
\newcommand{\cupm}{\ensuremath{\uplus}}

\newcommand{\atomlabel}[2]{\ensuremath{\color{red}\mathbf{#1}{:}\color{black} #2}}
\newcommand{\veclabel}[2]{\ensuremath{\color{red}\vec{\mathbf{#1}}{:}\color{black} #2}}
\newcommand{\examplelabel}[2]{\ensuremath{\color{red}{#1}{:}\color{black} #2}}
\newcommand{\constraint}[1]{\mathbf{\color{orange}{#1}\color{black}}}

\newcommand{\atoms}[1]{\ensuremath{\{#1\}}}

\newcommand{\phifunc}[4]{\ensuremath{{\phi_{#1}(#2, #3, #4)}}}

\newcommand{\any}[1]{{#1}_{\mR}}
\newcommand{\fuse}[2]{#1 {\curlywedgeuparrow} #2}
\newcommand{\dL}{\text{\upshape\textsf{d{\kern-0.05em}L}}\xspace}
\newcommand{\uapc}{\texttt{UAPC}\xspace}
\newcommand{\uapcplus}{\texttt{UAPC$^+$}\xspace}

\newcommand{\vars}[1]{{\textit{Var}\,(#1)}\xspace}

\newcommand{\mId}[0]{\text{\color{violet}id\color{black}}}
\newcommand{\mW}[0]{\text{\color{violet}G\color{black}}}
\newcommand{\mR}[0]{\text{\color{violet}R\color{black}}}

\newcommand{\meet}[0]{\sqcap}

\newenvironment{sketch}[0]{\noindent \textit{Proof sketch.}}

\newtheorem{innercustomlem}{Lemma}

\newenvironment{customlem}[2]
  {\renewcommand\theinnercustomlem{#1 (#2).}\innercustomlem}
  {\endinnercustomlem}

\newtheorem{innercustomthm}{Theorem}

\newenvironment{customthm}[2]
  {\renewcommand\theinnercustomthm{#1 (#2).}\innercustomthm}
  {\endinnercustomthm}

  \newtheorem{innercustomcor}{Corollary}
  
  \newenvironment{customcor}[2]
    {\renewcommand\theinnercustomcor{#1 (#2).}\innercustomcor}
    {\endinnercustomcor}

\newtheorem{observation}{Property}

\newcommand{\lstrike}{\mathopen{\mathrel{\langle}\joinrel\mathrel{[}}}
\newcommand{\rstrike}{\mathclose{\mathrel{]}\joinrel\mathrel{\rangle}}}

%% file: appendix_tr.tex
\section{Parachute example} \label{app:parachuteproof}

\paragraph{Model.}

\[\vdash \mathbf{A} \to
\left[\left(\left(?(\mathbf{v - gT > -\sqrt{g/{r_{op}}} \land r = r_{cl} \land x > 100}); \cup\; r := r_{op};\right) t:= 0; \text{ODE} \& \mathbf{Q}\right)^*\right]
P\]
\begin{align*}
    \text{Parameters}&~ g, r_{op}, m, T, x, v, r, t \\
\examplelabel{\vec{i}}A &\triangleq \examplelabel{i_1}g > 0 \land \examplelabel{i_2}r_{op} > r_{cl} \land \examplelabel{i_3}T > 0 \land \examplelabel{i_4}v_\textit{max} < -\sqrt{g/{r_{op}}} \land \examplelabel{i_5}x \geq 0 \\
&\;\; \land \examplelabel{i_6}0 > v > -\sqrt{g/{r_{op}}} \land \examplelabel{i_7}r = r_{cl} \land \examplelabel{i_8}{r_{cl} = 0} \\
\examplelabel{j,k,l}L(x,v) &\triangleq \examplelabel{j}{v - gT > -\sqrt{g/{r_{op}}}} \land \examplelabel{k}{r = r_{cl}} \land \examplelabel{l}x > 100\\
\examplelabel{\vec{q}}P &\triangleq \examplelabel{q_1}x = 0 \to \examplelabel{q_2}v \geq v_\textit{max} \\
\examplelabel{\vec{o}}\text{ODE} &\triangleq \examplelabel{o_1}x' = v, \examplelabel{o_2}v' = -g + rv^2, \examplelabel{o_3}t' = 1 \\
\examplelabel{\vec{p}}Q &\triangleq \examplelabel{p_1}t \leq T,\examplelabel{p_2}x \geq 0, \examplelabel{p_3}v < 0 \\
\examplelabel{i_5,i_6,i_7,i_8,u_1}I(x,v) &\triangleq \examplelabel{i_5}x \geq 0 \land \examplelabel{i_6}v<0 \land \examplelabel{i_7}v>-\sqrt{g/{r_{op}}} \land \examplelabel{i_8}{r_{cl} = 0} \land \examplelabel{u_1}{{x > -1}} 
\end{align*}

\paragraph{Proof.}

The proof tree, using the rules from Section~\ref{sec:UAPC}, is presented in boxes: a zig-zag border at the top indicates that the proof continues in another snippet at the annotated subgoals, while a zig-zag border at the bottom indicates that the snippet continues from an open subgoal of another snippet. 
The input and output label sets for this example are the following:
\begin{align*}
    \Psi = &\, \vec{o}_{\mId}\\
    DA_1(\dots {\mid} \Psi) = &\, \{k_{\mId}, s_{\mId}, {\{i_8\}}_{\mW}, \any{\{i_5\}}, \any{\{i_6\}}, \any{\{i_7\}}, \any{j}, \any{l}, n_{\mR}, \any{r}, \any{\{u_1\}}, {\vec{o}}_{\mId}\} \\
    DA_2(\dots {\mid} \Psi) = &\, \{n_{\mId}, s_{\mId}, r_{\mW}, \any{\{i_5\}}, \any{\{i_6\}}, \any{\{i_7\}}, \any{\{i_8\}}, \any{j}, \any{k}, \any{l}, \any{\{u_1\}}\} \\
    DA_3(\dots {\mid} \Psi) = &\, \{j_{\mId}, n_{\mId}, r_{\mId}, \{p_1\}_{\mId}, \{i_5\}_{\mR}, \{i_6\}_{\mR}, \{i_7\}_{\mR}, \{i_8\}_{\mR}, \{u_1\}_{\mR}, k_{\mR}, l_{\mR}, \{p_2\}_{\mR}, \{p_3\}_{\mR}, \{s\}_{\mR}\} \\
    DA_4(\dots {\mid} \Psi) = &\, \{\any{\{p_2\}}, \any{\{p_3\}}, \any{\{p_1\}}\} \\
    DA_5(\dots {\mid} \Psi) = &\, \{\{i_7\}_{\mR}, \any{\{i_5\}}, \any{\{i_6\}}, \any{n}, \any{\{p_1\}}, \any{\{p_2\}}, \any{\{p_3\}}, \any{\{u_1\}}\} \\
    DA_6(\dots {\mid} \Psi) = &\, \{\any{\{u_1\}}, \{p_2\}_{\mId},\any{\{i_5\}},\any{\{i_6\}},\any{\{i_6\}}, \any{j}, \any{k}, \any{l}, \any{n}, \any{\{p_1\}}, \any{\{p_3\}}\} \\
    DA_7(\dots {\mid} \Psi) = &\, \{\any{\fuse{i_5}{p_2}},\any{\fuse{i_6}{p_3}}, \any{j}, \any{k}, \any{l}, \any{n}, \any{\{p_1\}}, \any{\{u_1\}}\} \\
    DA_8(\dots{\mid}\Psi) = &\, \{\{i_7\}_{\mId}, \{i_6\}_{\mR}, \{i_5\}_{\mR}, \any{n}, \{p_1\}_{\mR}, \{p_2\}_{\mR}, \{p_3\}_{\mR}, \any{\{u_1\}} \} \\
    DA_9(\dots{\mid}\Psi) = &\, \{\{i_5\}_{\mW}, \any{\{i_6\}}, \any{\{i_7\}}, {\{u_1\}}_{\mId}, \any{\{i_1\}}, \any{\{i_2\}}, \any{\{i_3\}}, \any{\{i_4\}}, \any{\{i_8\}}\} \\
    DA_{10}(\dots {\mid} \Psi) = &\, \{\any{\{i_5\}}, {\{i_7\}}_{\mId}, \any{\{i_6\}}, \vec{q}_{\mId}, \any{\{u_1\}}\} \\
    \Sigma = &\, DA_4(... \mid \Psi) \cupm \{\any{\{n\}}, \any{\{i_5\}}, \any{\{i_6\}}, \any{\{i_7\}}, \any{\{o_1\}}, \any{\{o_2\}}, \any{\{o_3\}}, \any{\{u_1\}}\} \cupm DA_5(\dots \mid \Psi) \\
    &\cupm DA_8(\dots{\mid}\Psi) \cupm \{o_1\}_{\mR}\cupm \{o_2\}_{\mId}\cupm \{o_3\}_{\mR} \cupm \any{m} \cupm DA_6(\dots \mid \Psi) \cupm DA_7(\dots \mid \Psi) \\
    &\cupm \any{\vec{o}} \cupm DA_1(\dots {\mid} \Psi) \cupm \{{o_1}_{\mR},{o_2}_{\mId}, {o_3}_{\mId}\} \cupm DA_2(\dots {\mid} \Psi) \cupm DA_3(\dots {\mid} \Psi) \cupm \vec{o}_{\mR}  \cupm n_{\mR} \\
    \color{black}\Omega = &\, \color{black}\{{\color{black}\fuse{i_5}{p_2}}_{\color{violet}\mId}\color{black},{\color{black}\fuse{i_6}{p_3}}_\mId\color{black},\color{black}k_{\mId}\color{black}, 
    \any{\color{black}j}\color{black}, \any{\color{black}l}\color{black}, {\color{black}\{o_1\}_{\mId}}\color{black}, \color{black}\{o_2\}_{\mId}\color{black}, \color{black}\{o_3\}_{\mId}\color{black}, \color{black}n_{\mId}\color{black}, \color{black}\{i_7\}_{\mId}, \any{\color{black}m}\color{black}, \any{\color{black}\{p_1\}}\color{black}, \any{\color{black}\{i_1\}}\color{black}, \any{\color{black}\{i_2\}}\color{black}, \any{\color{black}\{i_3\}}\color{black}, \\
    &\any{\color{black}\{i_4\}}\color{black}, {\color{black}\{i_8\}}_\mId\color{black}, \color{black}\vec{q}_{\mId}\color{black}\}
\end{align*}

\begin{proofstartbox}
\begin{mathpar}
\inferrule*[rightstyle=\normalfont,Right=M{:}\(\protect{{\to}R}\),width=\linewidth]{
    \inferrule*[rightstyle=\normalfont,Right=M{:}\(\protect{\text{loop}}\),width=\linewidth]{
        (1) \\
        (2) \\
        (3)
    }{
        \examplelabel{\vec{i}}A \vdash^\Psi_{\Psi \cupm \Omega
        }
        \left[\left(\{?(\examplelabel{j,k,l}{L}); \cup\; \examplelabel{m}r := r_{op};\} \examplelabel{n}t:= 0; \examplelabel{\vec{o}}\text{ODE} \& \examplelabel{\vec{p}}Q\right)^*\right] 
        (\examplelabel{\vec{q}}P)
    }
}{
    {\begin{aligned}
    &\vdash^\Psi_{\Sigma \cupm \Psi \cupm \{\{i_5\}_{\mR}, \{i_6\}_{\mR}, \{i_7\}_{\mR}, \any{\{i_1\}}, \any{\{i_2\}}, \any{\{i_3\}}, \any{\{i_4\}}, \any{\{i_8\}}\} \cupm DA_9(\dots{\mid}\Psi) \cupm DA_{10}(\dots {\mid} \Psi)} \\
    &\examplelabel{\vec{i}}A \rightarrow \left[\left(\{?(\examplelabel{j,k,l}{L}); \cup\; \examplelabel{m}r := r_{op};\} \examplelabel{n}t:= 0; \examplelabel{\vec{o}}\text{ODE} \& \examplelabel{\vec{p}}Q\right)^*\right] 
        (\examplelabel{\vec{q}}P)x
    \end{aligned}}
}
\end{mathpar}
\end{proofstartbox}

\begin{proofmiddlebox}[(1)]
\begin{mathpar}
\inferrule*[rightstyle=\normalfont,Right=M{:}\(\protect{\text{$\land$R}}\),width=\linewidth]{ 
    (1a) \\
    (1b)
}{
    {\begin{aligned}
    &\examplelabel{\vec{i}}A \vdash^\Psi_{\Psi \cupm \{\{i_5\}_{\mR}, \{i_6\}_{\mR}, \{i_7\}_{\mR}, \any{\{i_1\}}, \any{\{i_2\}}, \any{\{i_3\}}, \any{\{i_4\}}, \any{\{i_8\}}\} \cupm DA_9(\dots{\mid}\Psi)} \\ &\examplelabel{i_5}x \geq 0 \land \examplelabel{i_6}v < 0 \land \examplelabel{i_7}v > -\sqrt{g/{r_{op}}}\land \examplelabel{u_1}{{x > -1}}
    \end{aligned}
    }
}
\end{mathpar}
\end{proofmiddlebox}

\begin{proofleafbox}[(1a)]
    \begin{mathpar}
        \inferrule*[rightstyle=\normalfont,Right=M{:}\(\protect{\text{id}}\),width=\linewidth]{ 
        \ast 
        }{
            {\begin{aligned}
            &\examplelabel{\vec{i}}A \vdash^\Psi_{\Psi \cupm \{\{i_5\}_{\mR}, \{i_6\}_{\mR}, \{i_7\}_{\mR}, \any{\{i_1\}}, \any{\{i_2\}}, \any{\{i_3\}}, \any{\{i_4\}}, \any{\{i_8\}}\}} \\ &\examplelabel{i_5}x \geq 0 \land \examplelabel{i_6}v < 0 \land \examplelabel{i_7}v > -\sqrt{g/{r_{op}}}
            \end{aligned}
            }
        }
    \end{mathpar}
\end{proofleafbox}

\begin{proofleafbox}[(1b)]
    \begin{mathpar}
        \inferrule*[rightstyle=\normalfont,Right=M{:}\(\protect{\mathbb{R}}\),width=\linewidth]{ 
        \ast 
        }{
            {\begin{aligned}
            &\examplelabel{\vec{i}}A \vdash^\Psi_{DA_9(\dots{\mid}\Psi)} \examplelabel{u_1}{{x > -1}}
            \end{aligned}
            }
        } 
    \end{mathpar}
\end{proofleafbox}
\begin{proofleafbox}[(2)]
\begin{mathpar}
\inferrule*[rightstyle=\normalfont,Right=M{:}\(\protect{\text{QE}}\),width=\linewidth]{ \ast }{
    \examplelabel{i_5, i_6, i_7, u_1}I \vdash^\Psi_{DA_{10}(\dots {\mid} \Psi)} \examplelabel{q_1}x = 0 \to \examplelabel{q_2}v \geq v_\textit{max}\\
}
\end{mathpar}
\end{proofleafbox}

\begin{proofmiddlebox}[(3)]
\begin{mathpar}
\inferrule*[rightstyle=\normalfont,Right=M{:}\(\protect{[;]}\),width=\linewidth]{ 
    \inferrule*[rightstyle=\normalfont,Right=M{:}\(\protect{[\cup]}\),width=\linewidth]{
        \inferrule*[rightstyle=\normalfont,Right=M{:}\(\protect{\land R}\),width=\linewidth]{
            (3a) \\
            (3b)
        }{
            {\begin{aligned}
            &\examplelabel{i_5,i_6,i_7,i_8,u_1}I \vdash^\Psi_{\Sigma} \\
            &\left[?(\examplelabel{j}{v - gT > -\sqrt{g/{r_{op}}}} \land \examplelabel{k}{r = r_{cl}} \land \examplelabel{l}x > 100);\right]\left[\examplelabel{n}t:= 0; \examplelabel{\vec{o}}\text{ODE} \& \examplelabel{\vec{p}}Q\right] \examplelabel{i_5,i_6,i_7,i_8,u_1}I \\ 
            &\land \left[\examplelabel{m}r := r_{op};\right]\left[\examplelabel{n}t:= 0; \examplelabel{\vec{o}}\text{ODE} \& \examplelabel{\vec{p}}Q\right] \examplelabel{i_5,i_6,i_7,i_8,u_1}I\\ 
            \end{aligned}}
        }
    }{
        {\begin{aligned}
        &\examplelabel{i_5,i_6,i_7,i_8,u_1}I \vdash^\Psi_{\Sigma}\\ 
        &\left[?(\examplelabel{j}{v - gT > -\sqrt{g/{r_{op}}}} \land \examplelabel{k}{r = r_{cl}} \land \examplelabel{l}x > 100); \cup\; \examplelabel{m}r := r_{op};\right]\left[\examplelabel{n}t:= 0; \examplelabel{\vec{o}}\text{ODE} \& \examplelabel{\vec{p}}Q\right] \\
        &\examplelabel{i_5,i_6,i_7,i_8,u_1}I\\ 
        \end{aligned}
        }
    }
}{
    {\begin{aligned}
    &\examplelabel{i_5,i_6,i_7,i_8,u_1}I \vdash^\Psi_{\Sigma} \\
    &\left[\left(?(\examplelabel{j}{v - gT > -\sqrt{g/{r_{op}}}} \land \examplelabel{k}{r = r_{cl}} \land \examplelabel{l}x > 100); \cup\; \examplelabel{m}r := r_{op};\right) \examplelabel{n}t:= 0; \examplelabel{\vec{o}}\text{ODE} \& \examplelabel{\vec{p}}Q\right]\\
    &\examplelabel{i_5,i_6,i_7,i_8,u_1}I\\
    \end{aligned}}
}
\end{mathpar}
\end{proofmiddlebox}

\begin{proofmiddlebox}[(3a)]
\begin{mathpar}
\inferrule*[rightstyle=\normalfont,Right=M{:}\(\protect{[?]}\),width=\linewidth]{
    \inferrule*[rightstyle=\normalfont,Right=M{:}\(\protect{[{\to}R]}\),width=\linewidth]{
        \inferrule*[rightstyle=\normalfont,Right=M{:}\(\protect{[\land L]}\),width=\linewidth]{
            \inferrule*[rightstyle=\normalfont,Right=M{:}\(\protect{[;]}\),width=\linewidth]{
                \inferrule*[rightstyle=\normalfont,Right=M{:}\(\protect{[:=]}\),width=\linewidth]{
                    \inferrule*[rightstyle=\normalfont,Right=M{:}\(\protect{[]\land}\),width=\linewidth]{
                        \inferrule*[rightstyle=\normalfont,Right=M{:}\(\protect{\land R}\),width=\linewidth]{
                            (3c) \\
                            (3d)
                        }{
                            {\begin{aligned}
                            &\examplelabel{i_5,i_6,i_7,i_8,u_1}I(x,v), \examplelabel{j,k,l}L, \examplelabel{n}t= 0 \\
                            &\vdash^\Psi_{DA_6(\dots \mid \Psi) \cupm DA_7(\dots \mid \Psi) \cupm \any{\vec{o}} \cupm DA_1(\dots {\mid} \Psi) \cupm \{{o_1}_{\mR},{o_2}_{\mId}, {o_3}_{\mId}\} \cupm DA_2(\dots {\mid} \Psi) \cupm DA_3(\dots {\mid} \Psi) \cupm \vec{o}_{\mR}} \\ 
                            &\left[\examplelabel{\vec{o}}\text{ODE} \& \examplelabel{\vec{p}}Q\right] (\examplelabel{i_5}x \geq 0 \land \examplelabel{i_6}v < 0\land \examplelabel{u_1}x > -1)
                            \land \left[\examplelabel{\vec{o}}\text{ODE} \& \examplelabel{\vec{p}}Q\right] \examplelabel{i_7}v > -\sqrt{g/{r_{op}}}
                            \end{aligned}}
                        }
                    }{
                        {\begin{aligned}
                        &\examplelabel{i_5,i_6,i_7,i_8,u_1}I(x,v), \examplelabel{j,k,l}L, \examplelabel{n}t= 0 \\
                        &\vdash^\Psi_{DA_6(\dots \mid \Psi) \cupm DA_7(\dots \mid \Psi) \cupm \any{\vec{o}} \cupm DA_1(\dots {\mid} \Psi) \cupm \{{o_1}_{\mR},{o_2}_{\mId}, {o_3}_{\mId}\} \cupm DA_2(\dots {\mid} \Psi) \cupm DA_3(\dots {\mid} \Psi) \cupm \vec{o}_{\mR}} \\ 
                        &\left[\examplelabel{\vec{o}}\text{ODE} \& \examplelabel{\vec{p}}Q\right] \examplelabel{i_5,i_6,i_7,i_8,u_1}I(x,v)
                        \end{aligned}}
                    }
                }{
                    {\begin{aligned}
                    &\examplelabel{i_5,i_6,i_7,i_8,u_1}I(x,v), \examplelabel{j,k,l}L \\
                    &\vdash^\Psi_{DA_6(\dots \mid \Psi) \cupm DA_7(\dots \mid \Psi) \cupm \any{\vec{o}} \cupm DA_1(\dots {\mid} \Psi) \cupm \{{o_1}_{\mR},{o_2}_{\mId}, {o_3}_{\mId}\} \cupm DA_2(\dots {\mid} \Psi) \cupm DA_3(\dots {\mid} \Psi) \cupm \vec{o}_{\mR} \cupm n_{\mR}} \\
                    &\left[\examplelabel{n}t:= 0\right]\left[\examplelabel{\vec{o}}\text{ODE} \& \examplelabel{\vec{p}}Q\right] \examplelabel{i_5,i_6,i_7,i_8,u_1}I(x,v)
                    \end{aligned}}
                }
            }{
                {\begin{aligned}
                &\examplelabel{i_5,i_6,i_7,i_8,u_1}I(x,v), \examplelabel{j,k,l}L \\
                &\vdash^\Psi_{DA_6(\dots \mid \Psi) \cupm DA_7(\dots \mid \Psi) \cupm \any{\vec{o}} \cupm DA_1(\dots {\mid} \Psi) \cupm \{{o_1}_{\mR},{o_2}_{\mId}, {o_3}_{\mId}\} \cupm DA_2(\dots {\mid} \Psi) \cupm DA_3(\dots {\mid} \Psi) \cupm \vec{o}_{\mR} \cupm n_{\mR}} \\
                &\left[\examplelabel{n}t:= 0 ; \examplelabel{\vec{o}}\text{ODE} \& \examplelabel{\vec{p}}Q\right] \examplelabel{i_5,i_6,i_7,i_8,u_1}I(x,v)
                \end{aligned}}
            }
        }{
            {\begin{aligned}
            &\examplelabel{i_5,i_6,i_7,i_8,u_1}I(x,v) \land \examplelabel{j,k,l}L \\ 
            &\vdash^\Psi_{DA_6(\dots \mid \Psi) \cupm DA_7(\dots \mid \Psi) \cupm \any{\vec{o}} \cupm DA_1(\dots {\mid} \Psi) \cupm \{{o_1}_{\mR},{o_2}_{\mId}, {o_3}_{\mId}\} \cupm DA_2(\dots {\mid} \Psi) \cupm DA_3(\dots {\mid} \Psi) \cupm \vec{o}_{\mR} \cupm n_{\mR}} \\
            &\left[\examplelabel{n}t:= 0; \examplelabel{\vec{o}}\text{ODE} \& \examplelabel{\vec{p}}Q\right] \examplelabel{i_5,i_6,i_7,i_8,u_1}I(x,v)
            \end{aligned}}
        }
    }{
    {\begin{aligned}
    &\examplelabel{i_5,i_6,i_7,i_8,u_1}I(x,v) \\
    &\vdash^\Psi_{DA_6(\dots \mid \Psi) \cupm DA_7(\dots \mid \Psi) \cupm \any{\vec{o}} \cupm DA_1(\dots {\mid} \Psi) \cupm \{{o_1}_{\mR},{o_2}_{\mId}, {o_3}_{\mId}\} \cupm DA_2(\dots {\mid} \Psi) \cupm DA_3(\dots {\mid} \Psi) \cupm \vec{o}_{\mR} \cupm n_{\mR}} \\ 
    &\examplelabel{j,k,l}L \to \left[\examplelabel{n}t:= 0; \examplelabel{\vec{o}}\text{ODE} \& \examplelabel{\vec{p}}Q\right] \examplelabel{i_5,i_6,i_7,i_8,u_1}I(x,v) 
    \end{aligned}}
    }
}{
    {\begin{aligned}
    &\examplelabel{i_5,i_6,i_7,i_8,u_1}I(x,v) \\
    &\vdash^\Psi_{DA_6(\dots \mid \Psi) \cupm DA_7(\dots \mid \Psi) \cupm \any{\vec{o}} \cupm DA_1(\dots {\mid} \Psi) \cupm \{{o_1}_{\mR},{o_2}_{\mId}, {o_3}_{\mId}\} \cupm DA_2(\dots {\mid} \Psi) \cupm DA_3(\dots {\mid} \Psi) \cupm \vec{o}_{\mR}  \cupm n_{\mR}} \\
    &\left[?\examplelabel{j,k,l}L\right] \left[\examplelabel{n}t:= 0; \examplelabel{\vec{o}}\text{ODE} \& \examplelabel{\vec{p}}Q\right] \examplelabel{i_5,i_6,i_7,i_8,u_1}I(x,v)
    \end{aligned}}
}
\end{mathpar}
\end{proofmiddlebox}

\begin{proofmiddlebox}[(3c)]
\begin{mathpar}
        \inferrule*[rightstyle=\normalfont,Right=M{:}\(\protect{\text{dW}}\),width=\linewidth]{
            \inferrule*[rightstyle=\normalfont,Right=M{:}\(\protect{\land R}\),width=\linewidth]{
                (3m) \\
                (3n)
            }{
                \examplelabel{i_5,i_6,i_7,i_8,u_1}I(x_0,v_0), \examplelabel{j}{v_0 - gT > -\sqrt{g/{r_{op}}}} \land \examplelabel{k}{r = r_{cl}} \land \examplelabel{l}x_0 > 100, \examplelabel{n}t_0= 0, \examplelabel{\vec{p}}Q \\ \vdash^\Psi_{DA_6(\dots \mid \Psi) \cupm DA_7(\dots \mid \Psi)} \examplelabel{i_5}x \geq 0 \land \examplelabel{i_6}v < 0\land \examplelabel{u_1}x > -1
            }
        }{
            {\begin{aligned}
            &\examplelabel{i_5,i_6,i_7,i_8,u_1}I(x,v), \examplelabel{j}{v - gT > -\sqrt{g/{r_{op}}}} \land \examplelabel{k}{r = r_{cl}} \land \examplelabel{l}x > 100, \examplelabel{n}t= 0 \\
            &\vdash^\Psi_{DA_6(\dots \mid \Psi) \cupm DA_7(\dots \mid \Psi) \cupm \any{\vec{o}}} \\
            &\left[\examplelabel{\vec{o}}\text{ODE} \& \examplelabel{\vec{p}}Q\right] (\examplelabel{i_5}x \geq 0 \land \examplelabel{i_6}v < 0\land \examplelabel{u_1}x > -1)
            \end{aligned}}
        }
\end{mathpar}
\end{proofmiddlebox}

\begin{proofleafbox}[(3m)]
    \begin{mathpar}
        \inferrule*[rightstyle=\normalfont,Right=M{:}\(\protect{\text{id}}\),width=\linewidth]{
                \ast
                }{
                    {\begin{aligned}
                    &\examplelabel{i_5,i_6,i_7,i_8,u_1}I(x_0,v_0), \examplelabel{j}{v_0 - gT > -\sqrt{g/{r_{op}}}} \land \examplelabel{k}{r = r_{cl}} \land \examplelabel{l}x_0 > 100, \examplelabel{n}t_0= 0, \examplelabel{\vec{p}}Q \\
                    &\vdash^\Psi_{DA_7(\dots \mid \Psi)} \examplelabel{i_5}x \geq 0 \land \examplelabel{i_6}v < 0
                    \end{aligned}}
                }
    \end{mathpar}
\end{proofleafbox}

\begin{proofleafbox}[(3n)]
    \begin{mathpar}
        \inferrule*[rightstyle=\normalfont,Right=M{:}\(\protect{\mathbb{R}}\),width=\linewidth]{
                \ast
                }{
                    {\begin{aligned}
                    &\examplelabel{i_5,i_6,i_7,i_8,u_1}I(x_0,v_0), \examplelabel{j}{v_0 - gT > -\sqrt{g/{r_{op}}}} \land \examplelabel{k}{r = r_{cl}} \land \examplelabel{l}x_0 > 100, \examplelabel{n}t_0= 0, \examplelabel{\vec{p}}Q \\
                    &\vdash^\Psi_{DA_6(\dots \mid \Psi)} \examplelabel{u_1}x > -1
                    \end{aligned}}
                }
    \end{mathpar}
\end{proofleafbox}

\begin{proofmiddlebox}[(3d)]
\begin{mathpar}
        \inferrule*[rightstyle=\normalfont,Right=M{:}\(\protect{\text{dC}}\),width=\linewidth]{
            (3e)
            \\
            (3f)
        }{
            {\begin{aligned}
            &\examplelabel{i_5,i_6,i_7,i_8,u_1}I(x,v), \examplelabel{j}{v - gT > -\sqrt{g/{r_{op}}}} \land \examplelabel{k}{r = r_{cl}} \land \examplelabel{l}x > 100, \examplelabel{n}t= 0 \\ 
            &\vdash^\Psi_{DA_1(\dots {\mid} \Psi) \cupm \{{o_1}_{\mR},{o_2}_{\mId}, {o_3}_{\mId}\} \cupm DA_2(\dots {\mid} \Psi) \cupm DA_3(\dots {\mid} \Psi) \cupm \vec{o}_{\mR}} \left[\examplelabel{\vec{o}}\text{ODE} \& \examplelabel{\vec{p}}Q\right] \examplelabel{i_7}v > -\sqrt{g/{r_{op}}}
            \end{aligned}}
        }
\end{mathpar}
\end{proofmiddlebox}

\begin{proofmiddlebox}[(3e)]
\begin{mathpar}
        \inferrule*[rightstyle=\normalfont,Right=M{:}\(\protect{\text{dI}}\),width=\linewidth]{
            (3g)
            \\
            (3h)
        }{
            {\begin{aligned}
            &\examplelabel{i_5,i_6,i_7,i_8,u_1}I(x_0,v_0), \examplelabel{j}{v_0 - gT > -\sqrt{g/{r_{op}}}} \land \examplelabel{k}{r = r_{cl}} \land \examplelabel{l}x_0 > 100, \examplelabel{n}t= 0, \examplelabel{r}v = v_0 \\
            &\vdash^\Psi_{DA_1(\dots {\mid} \Psi) \cupm \{{o_1}_{\mR},{o_2}_{\mId}, {o_3}_{\mId}\} \cupm DA_2(\dots {\mid} \Psi)} \left[\examplelabel{\vec{o}}\text{ODE} \& \examplelabel{\vec{p}}Q\right] \examplelabel{s}v \geq v_0 - gt
            \end{aligned}}
        }
\end{mathpar}
\end{proofmiddlebox}

\begin{proofleafbox}[(3g)]
\begin{mathpar}
        \inferrule*[rightstyle=\normalfont,Right=M{:}\(\protect{\text{QE}}\),width=\linewidth]{
            \ast
        }{
            {\begin{aligned}
            &\examplelabel{i_5,i_6,i_7,i_8,u_1}I(x_0,v_0), \examplelabel{j}{v_0 - gT > -\sqrt{g/{r_{op}}}} \land \examplelabel{k}{r = r_{cl}} \land \examplelabel{l}x_0 > 100, \examplelabel{n}t= 0, \examplelabel{r}v = v_0 \\ 
            &\vdash^\Psi_{DA_2(\dots {\mid} \Psi)} \examplelabel{s}v \geq v_0 - gt
            \end{aligned}}
        }
\end{mathpar}
\end{proofleafbox}

\begin{proofleafbox}[(3h)]
\begin{mathpar}
        \inferrule*[rightstyle=\normalfont,Right=M{:}\(\protect{[:=]}\),width=\linewidth]{
            \inferrule*[rightstyle=\normalfont,Right=M{:}\(\protect{[:=]}\),width=\linewidth]{
                \inferrule*[rightstyle=\normalfont,Right=M{:}\(\protect{[:=]}\),width=\linewidth]{
                    \inferrule*[rightstyle=\normalfont,Right=M{:}\(\protect{\text{QE}}\),width=\linewidth]{
                        \ast
                    }{
                        {\begin{aligned}
                        &\examplelabel{i_5,i_6,i_7,i_8,u_1}I(x_0,v_0), \examplelabel{j}{v_0 - gT > -\sqrt{g/{r_{op}}}} \land \examplelabel{k}{r = r_{cl}} \land \examplelabel{l}x_0 > 100, \examplelabel{n}t= 0, \examplelabel{r}v = v_0 \\
                        &\vdash^\Psi_{DA_1(\dots {\mid} \Psi)} \examplelabel{s}-g + rv^2 \geq -g
                        \end{aligned}}
                    }
                }{
                    {\begin{aligned}
                    &\examplelabel{i_5,i_6,i_7,i_8,u_1}I(x_0,v_0), \examplelabel{j}{v_0 - gT > -\sqrt{g/{r_{op}}}} \land \examplelabel{k}{r = r_{cl}} \land \examplelabel{l}x_0 > 100, \examplelabel{n}t= 0, \examplelabel{r}v = v_0 \\ &\vdash^\Psi_{DA_1(\dots {\mid} \Psi) \cupm \{{o_1}_{\mR}\}} \left[\examplelabel{o_1}x' := v\right] \examplelabel{s}-g + rv^2 \geq -g
                    \end{aligned}}
                }
            }{
                {\begin{aligned}
                &\examplelabel{i_5,i_6,i_7,i_8,u_1}I(x_0,v_0), \examplelabel{j}{v_0 - gT > -\sqrt{g/{r_{op}}}} \land \examplelabel{k}{r = r_{cl}} \land \examplelabel{l}x_0 > 100, \examplelabel{n}t= 0, \examplelabel{r}v = v_0 \\
                &\vdash^\Psi_{DA_1(\dots {\mid} \Psi) \cupm \{{o_1}_{\mR},{o_2}_{\mId}\}} \left[\examplelabel{o_2}v' := -g + rv^2\right]\left[\examplelabel{o_1}x' := v\right] \examplelabel{s}v' \geq -g
                \end{aligned}}
            }
        }{
            {\begin{aligned}
            &\examplelabel{i_5,i_6,i_7,i_8,u_1}I(x_0,v_0), \examplelabel{j}{v_0 - gT > -\sqrt{g/{r_{op}}}} \land \examplelabel{k}{r = r_{cl}} \land \examplelabel{l}x_0 > 100, \examplelabel{n}t= 0, \examplelabel{r}v = v_0 \\ 
            &\vdash^\Psi_{DA_1(\dots {\mid} \Psi) \cupm \{{o_1}_{\mR},{o_2}_{\mId}, {o_3}_{\mId}\}} \left[\examplelabel{o_3}t' := 1\right]\left[\examplelabel{o_2}v' := -g + rv^2\right]\left[\examplelabel{o_1}x' := v\right] \examplelabel{s}v' \geq -gt'
            \end{aligned}}
        }
\end{mathpar}
\end{proofleafbox}

\begin{proofmiddlebox}[(3b)]
\begin{mathpar}
\inferrule*[rightstyle=\normalfont,Right=M{:}\protect{\([:=]\)},width=\linewidth]{ 
    \inferrule*[rightstyle=\normalfont,Right=M{:}\protect{\([;]\)},width=\linewidth]{
        \inferrule*[rightstyle=\normalfont,Right=M{:}\protect{\([:=]\)},width=\linewidth]{
            \inferrule*[rightstyle=\normalfont,Right=M{:}\protect{\([]\land\)},width=\linewidth]{
                \inferrule*[rightstyle=\normalfont,Right=M{:}\(\land\)R,width=\linewidth]{
                    (3i)
                    \\
                    (3j)
                }{
                    {\begin{aligned}
                    & \examplelabel{i_5,i_6,i_7,i_8,u_1}I(x,v), \examplelabel{n}t = 0 \\
                    & \vdash^\Psi_{DA_4(... \mid \Psi), \any{\{n\}}, \any{\{i_5\}}, \any{\{i_6\}}, \any{\{i_7\}}, \any{\{o_1\}}, \any{\{o_2\}}, \any{\{o_3\}}, \any{\{u_1\}} \cupm DA_5(\dots \mid \Psi) \cupm DA_8(\dots{\mid}\Psi) \cupm \{o_1\}_{\mR}\cupm \{o_2\}_{\mId}\cupm \{o_3\}_{\mR}} 
                    \\
                    &\phantom{\land}~ \left[\examplelabel{o_1}x' = v, \examplelabel{o_2}v' = -g + rv^2, \examplelabel{o_3}t' = 1 \& \examplelabel{p_1}t \leq T,\examplelabel{p_2}x \geq 0, \examplelabel{p_3}v < 0\right](\examplelabel{i_5}x \geq 0 \land \examplelabel{i_6}v < 0) \\
                    & \land \left[\examplelabel{o_1}x' = v, \examplelabel{o_2}v' = -g + rv^2, \examplelabel{o_3}t' = 1 \& \examplelabel{p_1}t \leq T,\examplelabel{p_2}x \geq 0, \examplelabel{p_3}v < 0\right]\examplelabel{i_7}v > -\sqrt{g/{r_{op}}}
                    \end{aligned}}
                }
            }{
                {\begin{aligned}
                &\examplelabel{i_5,i_6,i_7,i_8,u_1}I(x,v), \examplelabel{n}t = 0 \\
                & \vdash^\Psi_{DA_4(... \mid \Psi), \any{\{n\}}, \any{\{i_5\}}, \any{\{i_6\}}, \any{\{i_7\}}, \any{\{o_1\}}, \any{\{o_2\}}, \any{\{o_3\}}, \any{\{u_1\}}\} \cupm DA_5(\dots \mid \Psi) \cupm DA_8(\dots{\mid}\Psi) \cupm \{o_1\}_{\mR}\cupm \{o_2\}_{\mId}\cupm \{o_3\}_{\mR}}\\ 
                & \left[\examplelabel{o_1}x' = v, \examplelabel{o_2}v' = -g + rv^2, \examplelabel{o_3}t' = 1 \& \examplelabel{p_1}t \leq T,\examplelabel{p_2}x \geq 0, \examplelabel{p_3}v < 0\right]\\
                &\examplelabel{i_5,i_6,i_7,i_8,u_1}I(x,v)
                \end{aligned}}
            }
        }{
            {\begin{aligned}
            & \examplelabel{i_5,i_6,i_7,i_8,u_1}I(x,v) \\
            & \vdash^\Psi_{DA_4(... \mid \Psi), \any{\{n\}}, \any{\{i_5\}}, \any{\{i_6\}}, \any{\{i_7\}}, \any{\{o_1\}}, \any{\{o_2\}}, \any{\{o_3\}}, \any{\{u_1\}}\} \cupm DA_5(\dots \mid \Psi) \cupm DA_8(\dots{\mid}\Psi) \cupm \{o_1\}_{\mR}\cupm \{o_2\}_{\mId}\cupm \{o_3\}_{\mR} }\\
            & \left[\examplelabel{n}t:= 0\right]\left[\examplelabel{o_1}x' = v, \examplelabel{o_2}v' = -g + rv^2, \examplelabel{o_3}t' = 1 \& \examplelabel{p_1}t \leq T,\examplelabel{p_2}x \geq 0, \examplelabel{p_3}v < 0\right]\\
            &\examplelabel{i_5,i_6,i_7,i_8,u_1}I(x,v)
            \end{aligned}}
        }
    }{
        {\begin{aligned}
        &\examplelabel{i_5,i_6,i_7,i_8,u_1}I(x,v) \\
        & \vdash^\Psi_{DA_4(... \mid \Psi), \any{\{n\}}, \any{\{i_5\}}, \any{\{i_6\}}, \any{\{i_7\}}, \any{\{o_1\}}, \any{\{o_2\}}, \any{\{o_3\}}, \any{\{u_1\}}\} \cupm DA_5(\dots \mid \Psi) \cupm DA_8(\dots{\mid}\Psi) \cupm \{o_1\}_{\mR}\cupm \{o_2\}_{\mId}\cupm \{o_3\}_{\mR} }\\ 
        & \left[\examplelabel{n}t:= 0 ; \{\examplelabel{o_1}x' = v, \examplelabel{o_2}v' = -g + rv^2, \examplelabel{o_3}t' = 1 \& \examplelabel{p_1}t \leq T,\examplelabel{p_2}x \geq 0, \examplelabel{p_3}v < 0\}\right] \\
        &\examplelabel{i_5,i_6,i_7,i_8,u_1}I(x,v)
        \end{aligned}}
    }
}{
{\begin{aligned}
    &\examplelabel{i_5,i_6,i_7,i_8,u_1}I(x,v) \\
    & \vdash^\Psi_{DA_4(... \mid \Psi), \any{\{n\}}, \any{\{i_5\}}, \any{\{i_6\}}, \any{\{i_7\}}, \any{\{o_1\}}, \any{\{o_2\}}, \any{\{o_3\}}, \any{\{u_1\}}\} \cupm DA_5(\dots \mid \Psi) \cupm DA_8(\dots{\mid}\Psi) \cupm \{o_1\}_{\mR}\cupm \{o_2\}_{\mId}\cupm \{o_3\}_{\mR} \cupm \any{m}} \\
    & \left[\examplelabel{m}r := 1\right]\\ 
    & \left[\examplelabel{n}t:= 0 ; \{\examplelabel{o_1}x' = v, \examplelabel{o_2}v' = -g + rv^2, \examplelabel{o_3}t' = 1 \& \examplelabel{p_1}t \leq T,\examplelabel{p_2}x \geq 0, \examplelabel{p_3}v < 0\}\right] \\
    &\examplelabel{i_5,i_6,i_7,i_8,u_1}I(x,v)
\end{aligned}
}
}
\end{mathpar}
\end{proofmiddlebox}

\begin{proofleafbox}[(3f)]
    \begin{mathpar}
            \inferrule*[rightstyle=\normalfont,Right=M{:}\(\protect{\text{dW}}\),width=\linewidth]{
                \inferrule*[rightstyle=\normalfont,Right=M{:}\(\protect{\text{QE}}\),width=\linewidth]{
                    \ast
                }{
                    {\begin{aligned}
                    &\examplelabel{i_5,i_6,i_7,i_8,u_1}I(x_0,v_0), \examplelabel{j}{v_0 - gT > -\sqrt{g/{r_{op}}}} \land \examplelabel{k}{r = r_{cl}} \land \examplelabel{l}x_0 > 100, \examplelabel{n}t_0= 0, \examplelabel{r}v = v_0, \\
                    &\examplelabel{\vec{p}}Q \land \examplelabel{s}v \geq v_0 - gt \vdash^\Psi_{DA_3(\dots {\mid} \Psi)} \examplelabel{i_7}v > -\sqrt{g/{r_{op}}}
                    \end{aligned}}
                }
            }{
                {\begin{aligned}
                &\examplelabel{i_5,i_6,i_7,i_8,u_1}I(x,v), \examplelabel{j}{v - gT > -\sqrt{g/{r_{op}}}} \land \examplelabel{k}{r = r_{cl}} \land \examplelabel{l}x > 100, \examplelabel{n}t= 0, \examplelabel{r}v = v_0 \\ 
                &\vdash^\Psi_{DA_3(\dots {\mid} \Psi) \cupm \vec{o}_{\mR}} \left[\examplelabel{\vec{o}}\text{ODE} \& \examplelabel{\vec{p}}Q \land \examplelabel{s}v \geq v_0 - gt\right] \examplelabel{i_7}v > -\sqrt{g/{r_{op}}}
                \end{aligned}}
            }
    \end{mathpar}
\end{proofleafbox}

\begin{proofleafbox}[(3i)]
\begin{mathpar}
\inferrule*[rightstyle=\normalfont,Right=M{:}\protect{\(\text{dW}\)},width=\linewidth]{
    \inferrule*[rightstyle=\normalfont,Right=M{:}\protect{\(\land L\)},width=\linewidth]{
        \inferrule*[rightstyle=\normalfont,Right=M{:}\protect{\(id\)},width=\linewidth]{
            \ast
        }{
             \examplelabel{p_1}t \leq T,\examplelabel{p_2}x \geq 0, \examplelabel{p_3}v < 0 \vdash^\Psi_{DA_4(... \mid \Psi)} \examplelabel{p_2}x \geq 0 \land \examplelabel{p_3}v < 0
        }
    }{
        \examplelabel{p_1}t \leq T,\examplelabel{p_2}x \geq 0, \examplelabel{p_3}v < 0 \vdash^\Psi_{DA_4(... \mid \Psi)} \examplelabel{p_2}x \geq 0 \land \examplelabel{p_3}v < 0
    }
}{
    {\begin{aligned}
        & \examplelabel{i_5,i_6,i_7,i_8,u_1}I(x,v), \examplelabel{n}t = 0 \vdash^\Psi_{DA_4(... \mid \Psi), \any{\{n\}}, \any{\{i_5\}}, \any{\{i_6\}}, \any{\{i_7\}}, \any{\{o_1\}}, \any{\{o_2\}}, \any{\{o_3\}}, \any{\{u_1\}}\}} \\ 
        & \left[\examplelabel{o_1}x' = v, \examplelabel{o_2}v' = -g + rv^2, \examplelabel{o_3}t' = 1 \& \examplelabel{p_1}t \leq T,\examplelabel{p_2}x \geq 0, \examplelabel{p_3}v < 0\right] \examplelabel{i_5}x \geq 0 \land \examplelabel{i_6}v < 0
    \end{aligned}
    }
}
\end{mathpar}
\end{proofleafbox}

\begin{proofmiddlebox}[(3j)]
\begin{mathpar}
\inferrule*[rightstyle=\normalfont,Right=M{:}\protect{\(\text{dG}\)},width=\linewidth]{
    \inferrule*[rightstyle=\normalfont,Right=M{:}\protect{\(\text{dI}\)},width=\linewidth]{
        (3k)
        \\
        (3l)
    }{
        {\begin{aligned}
        &\examplelabel{i_5,i_6,i_7,i_8,u_1}I(x_0,v_0), \examplelabel{n}t = 0 \vdash^\Psi_{DA_5(\dots \mid \Psi) \cupm DA_8(\dots{\mid}\Psi) \cupm \{o_1\}_{\mR}\cupm \{o_2\}_{\mId}\cupm \{o_3\}_{\mR}\cupm \{r_{\mId}\}} \\
        &\exists y \left[\examplelabel{o_1}x' = v, \examplelabel{o_2}v' = -g + v^2, \examplelabel{o_3}t' = 1, \examplelabel{r}y' = -\frac{1}{2}py\sqrt{v - g/{r_{op}}} ~\&~ \examplelabel{p_1}t \leq T,\examplelabel{p_2}x \geq 0, \examplelabel{p_3}v < 0\right]\\
        &\examplelabel{i_7}v > -\sqrt{g/{r_{op}}}
        \end{aligned}}
    }
}{
    {\begin{aligned}
    &\examplelabel{i_5,i_6,i_7,i_8,u_1}I(x,v), \examplelabel{n}t = 0 \\
    &\vdash^\Psi_{DA_5(\dots \mid \Psi) \cupm DA_8(\dots{\mid}\Psi) \cupm \{o_1\}_{\mR}\cupm \{o_2\}_{\mId}\cupm \{o_3\}_{\mR}\cupm \{r_{\mId}\} \setminus r} \\
    &\left[\examplelabel{o_1}x' = v, \examplelabel{o_2}v' = -g + v^2, \examplelabel{o_3}t' = 1 \& \examplelabel{p_1}t \leq T,\examplelabel{p_2}x \geq 0, \examplelabel{p_3}v < 0\right] \examplelabel{i_7}v > -\sqrt{g/{r_{op}}}
    \end{aligned}}
}
\end{mathpar}
\end{proofmiddlebox}

\begin{proofleafbox}[(3k)]
\begin{mathpar}
\inferrule*[rightstyle=\normalfont,Right=M{:}\protect{\(\text{id}\)},width=\linewidth]{
    \ast
}{
    {\begin{aligned}
    &\examplelabel{i_5,i_6,i_7,i_8,u_1}I(x_0,v_0), \examplelabel{n}t = 0, \examplelabel{p_1}t \leq T,\examplelabel{p_2}x \geq 0, \examplelabel{p_3}v < 0 \\
    &\vdash^\Psi_{DA_5(\dots \mid \Psi)}
    \examplelabel{i_7}v_0 > -\sqrt{g/{r_{op}}}
    \end{aligned}}
}
\end{mathpar}
\end{proofleafbox}

\begin{proofleafbox}[(3l)]
\begin{mathpar}
\inferrule*[rightstyle=\normalfont,Right=M{:}\protect{\([:=]\)},width=\linewidth]{
    \inferrule*[rightstyle=\normalfont,Right=M{:}\protect{\([:=]\)},width=\linewidth]{
        \inferrule*[rightstyle=\normalfont,Right=M{:}\protect{\([:=]\)},width=\linewidth]{
            \inferrule*[rightstyle=\normalfont,Right=M{:}\protect{\([:=]\)},width=\linewidth]{
                \inferrule*[rightstyle=\normalfont,Right=M{:}\protect{\(\text{QE}\)},width=\linewidth]{
                    \ast
                }{
                    {\begin{aligned}
                    &\examplelabel{i_5,i_6,i_7,i_8,u_1}I(x_0,v_0), \examplelabel{n}t = 0 \vdash^\Psi_{DA_8(\dots{\mid}\Psi)} \\
                    &\exists y (\examplelabel{p_1}t \leq T,\examplelabel{p_2}x \geq 0, \examplelabel{p_3}v < 0 \rightarrow \examplelabel{i_7}y^2*(v+\sqrt{g/{r_{op}}}) = 1 )
                    \land 
                    \forall \; v \; \forall y \;\\
                    &\examplelabel{i_7}2*y*(-\frac{1}{2}y\sqrt{v - g/{r_{op}}})*(v+(g/{r_{op}})^{(1/2)}) + y^2*(-g + v^2 + 0) = 0
                    \end{aligned}}
                }
            }{
                {\begin{aligned}
                &\examplelabel{i_5,i_6,i_7,i_8,u_1}I(x_0,v_0), \examplelabel{n}t = 0 \vdash^\Psi_{DA_8(\dots{\mid}\Psi) \cupm \{o_1\}_{\mR}} \\
                &\exists y (\examplelabel{p_1}t \leq T,\examplelabel{p_2}x \geq 0, \examplelabel{p_3}v < 0 \rightarrow \examplelabel{i_7}y^2*(v+\sqrt{g/{r_{op}}}) = 1 )
                \land \forall \; v \; \forall y \;
                \left[\examplelabel{o_1}x' := v\right] \\
                &\examplelabel{i_7}2*y*(-\frac{1}{2}y\sqrt{v - g/{r_{op}}})*(v+\sqrt{g/{r_{op}}}) + y^2*(-g + v^2 + 0) = 0
                \end{aligned}
                }
            }
        }{
            {\begin{aligned}
            &\examplelabel{i_5,i_6,i_7,i_8,u_1}I(x_0,v_0), \examplelabel{n}t = 0 \vdash^\Psi_{DA_8(\dots{\mid}\Psi) \cupm \{o_1\}_{\mR}\cupm \{o_2\}_{\mId}} \\
            &\exists y (\examplelabel{p_1}t \leq T,\examplelabel{p_2}x \geq 0, \examplelabel{p_3}v < 0 \rightarrow \examplelabel{i_7}y^2*(v+\sqrt{g/{r_{op}}}) = 1 )
            \land 
            \forall v \forall y 
            \left[\examplelabel{o_2}v' := -g + v^2\right] \\
            &\left[\examplelabel{o_1}x' := v\right]
            \examplelabel{i_7}2*y*(-\frac{1}{2}y\sqrt{v - g/{r_{op}}})*(v+\sqrt{g/{r_{op}}}) + y^2*(v' + 0) = 0
            \end{aligned}}
        }
    }{
        {\begin{aligned}
        &\examplelabel{i_5,i_6,i_7,i_8,u_1}I(x_0,v_0), \examplelabel{n}t = 0\\
        &\vdash^\Psi_{DA_8(\dots{\mid}\Psi) \cupm \{o_1\}_{\mR}\cupm \{o_2\}_{\mId}\cupm \{o_3\}_{\mR}} \\
        &\exists y (\examplelabel{p_1}t \leq T,\examplelabel{p_2}x \geq 0, \examplelabel{p_3}v < 0 \rightarrow \examplelabel{i_7}y^2*(v+\sqrt{g/{r_{op}}}) = 1 )
        \\
        &\land 
        \forall \; v \; \forall y \;
        \left[\examplelabel{o_3}t' := 1\right]
        \left[\examplelabel{o_2}v' := -g + v^2\right]\left[\examplelabel{o_1}x' := v\right] \\
        &\examplelabel{i_7}2*y*(-\frac{1}{2}y\sqrt{v - g/{r_{op}}})*(v+\sqrt{g/{r_{op}}}) + y^2*(v' + 0) = 0
        \end{aligned}}
    }
}{
    {\begin{aligned}
    &\examplelabel{i_5,i_6,i_7,i_8,u_1}I(x_0,v_0), \examplelabel{n}t = 0 \\
    &\vdash^\Psi_{DA_8(\dots{\mid}\Psi) \cupm \{o_1\}_{\mR}\cupm \{o_2\}_{\mId}\cupm \{o_3\}_{\mR}\cupm \{r_{\mId}\}} \\
    &\exists y (\examplelabel{p_1}t \leq T,\examplelabel{p_2}x \geq 0, \examplelabel{p_3}v < 0 \rightarrow \examplelabel{i_7}y^2*(v+\sqrt{g/{r_{op}}}) = 1 )
    \\
    &\land 
    \forall \; v \; \forall y \;
    \left[\examplelabel{r}y' := -\frac{1}{2}y\sqrt{v - g/{r_{op}}}\right]\left[\examplelabel{o_3}t' := 1\right]
    \left[\examplelabel{o_2}v' := -g + v^2\right]\left[\examplelabel{o_1}x' := v\right] \\
    &\examplelabel{i_7}2*y*y'*(v+\sqrt{g/{r_{op}}}) + y^2*(v' + 0) = 0
    \end{aligned}}
}
\end{mathpar}
\end{proofleafbox}

\section{\uapc Proof Calculus}\label{app:uapccalculus}

\subsection{\uapc Propositional and Quantifier Sequent Calculus Rules}

\begin{mathpar}
    \quad{\text{\color{violet}M:\color{black}$\land$R}} \;\;
    \inferrule
    { \veclabel{{k}}{\Gamma} \vdashu{\Psi}{\Sigma} \veclabel{{i}}{P}, \veclabel{{l}}{\Delta}
    \\ \veclabel{{k}}{\Gamma} \vdashu{\Psi}{\Omega} \veclabel{{j}}{Q}, \veclabel{{l}}{\Delta} }
    { \veclabel{{k}}{\Gamma} \vdashu{\Psi}{\Sigma \cupm \Omega} \veclabel{{i}}{P} \land \veclabel{{j}}{Q}, \veclabel{{l}}{\Delta} }
\and    
        \quad{\text{\color{violet}M:\color{black}cut}^{\color{red}\vec{l}\color{black}}} \;\;
        \inferrule
        { \veclabel{{i}}{\Gamma}, \veclabel{{j}}{C} \vdashu{\Psi}{\Sigma} \veclabel{{k}}{\Delta}
        \\ 
        \veclabel{{i}}{\Gamma} \vdashu{\Psi}{\Omega} \veclabel{{k}}{\Delta}, \veclabel{{j}}{C}
        }
        { \veclabel{{i}}{\Gamma} \vdashu{\Psi}{(\Sigma \cupm \Omega) \setminus {\vec{l}}} \veclabel{{k}}{\Delta} }
        \quad{(\forall C_a \in \text{atoms}(C). \, \color{red}j_a \color{black}= \phifunc{}{C_a}{(\veclabel{{i}}{\Gamma}, \veclabel{{k}}{\Delta})}{\color{red}\vec{l}\color{black}})} 
\and
    \quad{\text{\color{violet}M:\color{black}$\neg$R}} \;\;
    \inferrule
    { \veclabel{{k}}{\Gamma}, \veclabel{{i}}{P} \vdashu{\Psi}{\Sigma} \veclabel{{j}}{\Delta} }
    { \veclabel{{k}}{\Gamma} \vdashu{\Psi}{\Sigma} \neg \veclabel{{i}}{P}, \veclabel{{j}}{\Delta} }
\and
    \quad{\text{\color{violet}M:\color{black}$\lor$R}} \;\;
    \inferrule
    { \veclabel{{k}}{\Gamma} \vdashu{\Psi}{\Sigma} \veclabel{{l}}{\Delta}, \veclabel{{i}}{P}, \veclabel{{j}}{Q} }
    { \veclabel{{k}}{\Gamma} \vdashu{\Psi}{\Sigma} \veclabel{{i}}{P} \lor \veclabel{{j}}{Q}, \veclabel{{l}}{\Delta} }
\and
    \quad{\text{\color{violet}M:\color{black}$\neg$L}} \;\;
    \inferrule
    { \veclabel{{k}}{\Gamma} \vdashu{\Psi}{\Sigma} \veclabel{{j}}{\Delta}, \veclabel{{i}}{P} }
    { \veclabel{{k}}{\Gamma}, \neg \veclabel{{i}}{P} \vdashu{\Psi}{\Sigma} \veclabel{{j}}{\Delta} }
\and
    \quad{\text{\color{violet}M:\color{black}$\land$L}} \;\;
    \inferrule
    { \veclabel{{k}}{\Gamma},\veclabel{{i}}{P},\veclabel{{j}}{Q} \vdashu{\Psi}{\Sigma} \veclabel{{l}}{\Delta} }
    { \veclabel{{k}}{\Gamma},\veclabel{{i}}{P} \land \veclabel{{j}}{Q} \vdashu{\Psi}{\Sigma} \veclabel{{l}}{\Delta} }
\and
    \quad{\text{\color{violet}M:\color{black}$\lor$L}} \;\;
    \inferrule
    { \veclabel{{i}}{P}, \veclabel{{k}}{\Gamma} \vdashu{\Psi}{\Sigma} \veclabel{{l}}{\Delta}
    \\ \veclabel{{j}}{Q}, \veclabel{{k}}{\Gamma} \vdashu{\Psi}{\Omega} \veclabel{{l}}{\Delta} }
    { \veclabel{{i}}{P} \lor \veclabel{{j}}{Q}, \veclabel{{k}}{\Gamma} \vdashu{\Psi}{\Sigma \cupm \Omega} \veclabel{{l}}{\Delta} }
\and
    \quad{\text{\color{violet}M:\color{black}$\to$R}} \;\;
    \inferrule
    { \veclabel{{k}}{\Gamma}, \veclabel{{i}}{P} \vdashu{\Psi}{\Sigma} \veclabel{{j}}{Q},\veclabel{{l}}{\Delta} }
    { \veclabel{{k}}{\Gamma} \vdashu{\Psi}{\Sigma} \veclabel{{i}}{P} \to \veclabel{{j}}{Q},\veclabel{{l}}{\Delta} }
\and
    \quad{\text{\color{violet}M:\color{black}$\leftrightarrow$R}} \;\;
    \inferrule
    { \veclabel{{k}}{\Gamma}, \veclabel{{i}}{P} \vdashu{\Psi}{\Sigma} \veclabel{{l}}{\Delta}, \veclabel{{j}}{Q}
    \\ \veclabel{{k}}{\Gamma}, \veclabel{{j}}{Q} \vdashu{\Psi}{\Omega} \veclabel{{l}}{\Delta}, \veclabel{{i}}{P} }
    { \veclabel{{k}}{\Gamma} \vdashu{\Psi}{\Sigma \cupm \Omega} \veclabel{{i}}{P} \leftrightarrow \veclabel{{j}}{Q}, \veclabel{{l}}{\Delta} }
\and   
    \quad{\text{\color{violet}M:\color{black}$\to$L}} \;\;
    \inferrule
    { \veclabel{{k}}{\Gamma} \vdashu{\Psi}{\Sigma} \veclabel{{l}}{\Delta}, \veclabel{{i}}{P}
    \\ \veclabel{{j}}{Q}, \veclabel{{k}}{\Gamma} \vdashu{\Psi}{\Omega} \veclabel{{l}}{\Delta} }
    { \veclabel{{i}}{P} \to \veclabel{{j}}{Q}, \veclabel{{k}}{\Gamma} \vdashu{\Psi}{\Sigma \cupm \Omega} \veclabel{{l}}{\Delta} }
\and   
    \quad{\text{\color{violet}M:\color{black}$\leftrightarrow$L}} \;\;
    \inferrule
    { \veclabel{{i}}{P} \land \veclabel{{j}}{Q}, \veclabel{{k}}{\Gamma} \vdashu{\Psi}{\Sigma} \veclabel{{l}}{\Delta}
    \\ \neg \veclabel{{i}}{P} \land \neg \veclabel{{j}}{Q}, \veclabel{{k}}{\Gamma} \vdashu{\Psi}{\Omega} \veclabel{{l}}{\Delta} }
    { \veclabel{{i}}{P} \leftrightarrow \veclabel{{j}}{Q}, \veclabel{{k}}{\Gamma} \vdashu{\Psi}{\Sigma \cupm \Omega} \veclabel{{l}}{\Delta} }
\and
    \quad{\text{\color{violet}M:\color{black}WR}} \;\;
    \inferrule
    { \veclabel{{i}}{\Gamma} \vdashu{\Psi}{\Sigma} \veclabel{{k}}{\Delta} }
    { \veclabel{{i}}{\Gamma} \vdashu{\Psi}{\Sigma \cupm \atoms{\any {\vec{j}}}} \veclabel{{j}}{P}, \veclabel{{k}}{\Delta} }
\and    
    \quad{\text{\color{violet}M:\color{black}PR}} \;\;
    \inferrule
    { \veclabel{{k}}{\Gamma} \vdashu{\Psi}{\Sigma} \veclabel{{j}}{Q}, \veclabel{{i}}{P}, \veclabel{{l}}{\Delta} }
    { \veclabel{{k}}{\Gamma} \vdashu{\Psi}{\Sigma} \veclabel{{i}}{P}, \veclabel{{j}}{Q}, \veclabel{{l}}{\Delta} }
\and    
    \quad{\text{\color{violet}M:\color{black}WL}} \;\;
    \inferrule
    { \veclabel{{k}}{\Gamma} \vdashu{\Psi}{\Sigma} \veclabel{{l}}{\Delta} }
    { \veclabel{{i}}{P}, \veclabel{{k}}{\Gamma} \vdashu{\Psi}{\Sigma \cupm \atoms{\any{\vec{i}}}} \veclabel{{l}}{\Delta} }
\and    
    \quad{\text{\color{violet}M:\color{black}PL}} \;\;
    \inferrule
    { \veclabel{{j}}{Q}, \veclabel{{i}}{P}, \veclabel{{k}}{\Gamma} \vdashu{\Psi}{\Sigma} \veclabel{{l}}{\Delta} }
    { \veclabel{{i}}{P}, \veclabel{{j}}{Q}, \veclabel{{k}}{\Gamma} \vdashu{\Psi}{\Sigma} \veclabel{{l}}{\Delta} }
\and   
    \quad{\text{\color{violet}M:\color{black}$\forall$R}} \;\;
    \inferrule
    { \veclabel{{i}}{\Gamma} \vdashu{\Psi}{\Sigma} \veclabel{{j}}{p(y)}, \veclabel{{k}}{\Delta} }
    { \veclabel{{i}}{\Gamma} \vdashu{\Psi}{\Sigma} \forall x \; \veclabel{{j}}{p(x)}, \veclabel{{k}}{\Delta} }
    \quad{(y \notin \Gamma, \Delta, \forall x \; p(x))}
\and   
    \quad{\text{\color{violet}M:\color{black}$\forall$L}} \;\;
    \inferrule
    { \veclabel{{i}}{\Gamma}, \veclabel{{j}}{p(e)} \vdashu{\Psi}{\Sigma} \veclabel{{k}}{\Delta} }
    { \veclabel{{i}}{\Gamma}, \forall x \; \veclabel{{j}}{p(x)} \vdashu{\Psi}{\Sigma} \veclabel{{k}}{\Delta} }
    \quad{(\text{arbitrary term $e$})}
\and   
    \quad{\text{\color{violet}M:\color{black}$\exists$R}} \;\;
    \inferrule
    { \veclabel{{i}}{\Gamma} \vdashu{\Psi}{\Sigma} \veclabel{{j}}p(e), \veclabel{{k}}{\Delta} }
    { \veclabel{{i}}{\Gamma} \vdashu{\Psi}{\Sigma} \exists x \; \veclabel{{j}}{p(x)}, \veclabel{{k}}{\Delta} }
    \quad{(\text{arbitrary term $e$})}
\and   
    \quad{\text{\color{violet}M:\color{black}$\exists$L}} \;\;
    \inferrule
    { \veclabel{{i}}{\Gamma}, \veclabel{{j}}{p(y)} \vdashu{\Psi}{\Sigma} \veclabel{{k}}{\Delta} }
    { \veclabel{{i}}{\Gamma}, \exists x \; \veclabel{{j}}{p(x)} \vdashu{\Psi}{\Sigma} \veclabel{{k}}{\Delta} }
    \quad{(y \notin \Gamma, \Delta, \forall x \; p(x))}
\end{mathpar}

\subsection{\uapc Sequent Calculus Proof Rules}

    \begin{mathpar} 
    \quad{\text{\color{violet}M:\color{black}loop}^{\color{red}\vec{l}\color{black}}} \;\;
        \inferrule
        { \veclabel{{k}}{\Gamma} \vdashu{\Psi}{\Sigma} \veclabel{{n}}{J}, \veclabel{{i}}{\Delta}
        \\ \veclabel{{n}}{J} \vdashu{\Psi}{\Omega} \veclabel{{j}}{P}
        \\ \veclabel{{n}}{J} \vdashu{\Psi}{\Theta} [\veclabel{{m}}{\alpha}] \veclabel{{n}}{J}
        }
        { \veclabel{{k}}{\Gamma} \vdashu{\Psi}{(\Sigma \cupm \Omega \cupm \Theta) \setminus {\vec{l}}} [\veclabel{{m}}{\alpha}^*]\veclabel{{j}}{P},\veclabel{{i}}{\Delta} }
        \quad{(\forall J_a \in \text{atoms}(J). \, \color{red}n_a \color{black}= \phifunc{}{J_a}{(\veclabel{{k}}{\Gamma}, \veclabel{{i}}{\Delta})}{\color{red}\vec{l}\color{black}})} 
\and   
    \quad{\text{\color{violet}M:\color{black}MR}^{\color{red}\vec{k}\color{black}}} \;\;
    \inferrule
    { \veclabel{{i}}{\Gamma} \vdashu{\Psi}{\Sigma} \left[\veclabel{{j}}{\alpha}\right] \veclabel{{k}}{Q}, \veclabel{{l}}{\Delta} 
    \\ \veclabel{{k}}{Q} \vdashu{\Psi}{\Omega} \veclabel{{m}}{P} }
    { \veclabel{{i}}{\Gamma} \vdashu{\Psi}{(\Sigma \cupm \Omega) \setminus {\vec{k}}} \left[\veclabel{{j}}{\alpha}\right] \veclabel{{m}}{P}, \veclabel{{l}}{\Delta} }
\and    
    \quad{\text{\color{violet}M:\color{black}ML}^{\color{red}\vec{l}\color{black}}} \;\;
    \inferrule
    { \veclabel{{i}}{\Gamma}, \left[\veclabel{{j}}{\alpha}\right] \veclabel{{k}}{Q} \vdashu{\Psi}{\Sigma} \veclabel{{l}}{\Delta}
    \\ \veclabel{{m}}{P} \vdashu{\Psi}{\Omega} \veclabel{{k}}{Q} }
    { \veclabel{{i}}{\Gamma}, \left[\veclabel{{j}}{\alpha}\right] \veclabel{{m}}{P} \vdashu{\Psi}{(\Sigma \cupm \Omega) \setminus {\vec{k}}} \veclabel{{l}}{\Delta} }
\and    
    \quad{\text{\color{violet}M:\color{black}GVR}} \;\;
    \inferrule
    { \veclabel{{i}}{\Gamma_{\text{const}}} \vdashu{\Psi}{\Omega} \veclabel{{k}}{P}, \veclabel{{l}}{\Delta_{\text{const}}} }
    { \veclabel{{i}}{\Gamma} \vdashu{\Psi}{\Omega \cupm \{\any{\vec{j}}\}} \left[\veclabel{{j}}\alpha\right] \veclabel{{k}}{P}, \veclabel{{l}}{\Delta}}
\and    
    \quad{\text{\color{violet}M:\color{black}iG}^{\color{red}j\color{black}}} \;\;
    \inferrule
    { \veclabel{{i}}{\Gamma} \vdashu{\Psi}{\Omega} \left[\atomlabel{j}{y:=e}\right]\veclabel{{k}}{p}, \veclabel{{l}}{\Delta} }
    { \veclabel{{i}}{\Gamma} \vdashu{\Psi}{\Omega \setminus \{j\}} \veclabel{{k}}{p}, \veclabel{{l}}{\Delta} }
    \quad{(\text{$y$ new})}
\and    
%
    \quad{\text{\color{violet}M:\color{black}CER}^{\color{red}\vec{n}\color{black}}} \;\;
    \inferrule
    { \veclabel{{i}}{\Gamma} \vdashu{\Psi}{\Omega} \veclabel{ m}{C(\veclabel{j}{Q})}, \veclabel{{k}}{\Delta}
    \\ \vdashu{\Psi}{\Sigma} \veclabel{{j}}{Q} \leftrightarrow \veclabel{{l}}{P} }
    { \veclabel{{i}}{\Gamma} \vdashu{\Psi}{(\Omega \cupm \Sigma) \setminus {\vec{l}}} \veclabel{ m}{C(\veclabel{{l}}{P})}, \veclabel{{k}}{\Delta} }
    \quad{(\forall Q_a \in \text{atoms}(Q). \, \color{red}j_a \color{black}= \phifunc{}{Q_a}{\veclabel{{l}}{P}}{\color{red}\vec{n}\color{black}})} 
\and
    \quad{\text{\color{violet}M:\color{black}CEL}^{\color{red}\vec{n}\color{black}}} \;\;
    \inferrule
    { \veclabel{{i}}{\Gamma}, \veclabel{ m}{C(\veclabel{j}{Q})} \vdashu{\Psi}{\Omega} \veclabel{{k}}{\Delta}
    \\ \vdashu{\Psi}{\Sigma} \veclabel{{j}}{Q} \leftrightarrow \veclabel{{l}}{P} }
    { \veclabel{{i}}{\Gamma}, \veclabel{ m}{C(\veclabel{{l}}{P})} \vdashu{\Psi}{(\Omega \cupm \Sigma) \setminus {\vec{l}}} \veclabel{{k}}{\Delta} }
    \quad{(\forall Q_a \in \text{atoms}(Q). \, \color{red}j_a \color{black}= \phifunc{}{Q_a}{\veclabel{{l}}{P}}{\color{red}\vec{n}\color{black}})} 
\and
    \quad{\text{\color{violet}M:\color{black}CQR}^{\color{red}{l}\color{black}}} \;\;
    \inferrule
    { \veclabel{{i}}{\Gamma} \vdashu{\Psi}{\Omega} \veclabel{{j}}{p(k)}, \veclabel{{m}}{\Delta}
    \\ \vdashu{\Psi}{\Sigma} \atomlabel{{l}}{k = e}}
    { \veclabel{{i}}{\Gamma} \vdashu{\Psi}{\Omega} \veclabel{{j}}{p(e)}, \veclabel{{m}}{\Delta} }
\and
    \quad{\text{\color{violet}M:\color{black}CQL}^{\color{red}{l}\color{black}}} \;\;
    \inferrule
    { \veclabel{{i}}{\Gamma}, \veclabel{{j}}{p(k)} \vdashu{\Psi}{\Omega} \veclabel{{m}}{\Delta}
    \\ \vdashu{\Psi}{\Sigma} \atomlabel{{l}}{k = e} }
    { \veclabel{{i}}{\Gamma}, \veclabel{{j}}{p(e)} \vdashu{\Psi}{\Omega} \veclabel{{m}}{\Delta} }
\and
    \quad{\text{\color{violet}M:\color{black}CTR}} \;\;
    \inferrule
    { \vdashu{\Psi}{\Omega} \atomlabel{j}{e = k} }
    { \veclabel{{i}}{\Gamma} \vdashu{\Psi}{\Omega \cupm \{\any{\vec{i}}, \any{\vec{l}}\}} \atomlabel{j}{c(e) = c(k)}, \veclabel{{l}}{\Delta} }
\and    
    \quad{\text{\color{violet}M:\color{black}CTL}} \;\;
    \inferrule
    { \vdashu{\Psi}{\Omega} \atomlabel{j}{e = k} }
    { \veclabel{{i}}{\Gamma}, \atomlabel{j}{c(e) = c(k)} \vdashu{\Psi}{\Omega \cupm \{\any{\vec{i}}, \any{\vec{l}}\}} \veclabel{{l}}{\Delta} }
\and    
    \quad{\text{\color{violet}M:\color{black}US}} \;\;
    \inferrule
    { \veclabel{{i}}{\Gamma} \vdashu{\Psi}{\Omega} \veclabel{{j}}{\Delta} }
    { \sigma(\veclabel{{i}}{\Gamma}) \vdashu{\Psi}{\Omega} \sigma(\veclabel{{j}}{\Delta}) }
\and    
    \quad{\text{\color{violet}M:\color{black}UR}} \;\;
    \inferrule
    { \veclabel{{i}}{\Gamma} \frac{y}{x} \vdashu{\Psi}{\Omega} \veclabel{{j}}{\Delta} \frac{y}{x}}
    { \veclabel{{i}}{\Gamma} \vdashu{\Psi}{\Omega} \veclabel{{j}}{\Delta} }
\and    
    \quad{\text{\color{violet}M:\color{black}BRR}} \;\;
    \inferrule
    { \veclabel{{i}}{\Gamma} \vdashu{\Psi}{\Omega} \left[\atomlabel{j}{y:=e}\right] \veclabel{{k}}{\phi} \frac{y}{x}, \veclabel{{l}}\Delta }
    { \veclabel{{i}}{\Gamma} \vdashu{\Psi}{\Omega} \left[\atomlabel{j}{x:=e}\right] \veclabel{{k}}{\phi}, \veclabel{{l}}\Delta }
    \quad{(y, y', x' \notin FV(\phi))}
\and    
    \quad{\text{\color{violet}M:\color{black}BRL}} \;\;
    \inferrule
    { \veclabel{{i}}{\Gamma}, \left[\atomlabel{j}{y:=e}\right] \veclabel{{k}}{\phi} \frac{y}{x} \vdashu{\Psi}{\Omega} \veclabel{{l}}{\Delta} }
    { \veclabel{{i}}{\Gamma}, \left[\atomlabel{j}{x:=e}\right] \veclabel{{k}}{\phi} \vdashu{\Psi}{\Omega} \veclabel{{l}}{\Delta} }
    \quad{(y, y', x' \notin FV(\phi))}    
    \end{mathpar}

In rule \text{\color{violet}M:\color{black}UR}, \(\Gamma\frac{y}{x}\) is the result of renaming \(x\) to \(y\) in all formulas of \(\Gamma\). 
In rules \text{\color{violet}M:\color{black}BRR} and \text{\color{violet}M:\color{black}BRL}, \(\phi\frac{y}{x}\) is the result of renaming \(x\) to \(y\) in formula \(\phi\).

\subsection{\uapc Axioms}

    \begin{align*}  
    \quad{\text{\color{violet}M:\color{black}$[?]$}} \;\;\;\; &\vdashu{\Psi}{\{\vec{j}_{\mR}, \vec{k}_{\mR}\} \cupm \Psi} [?\veclabel{{j}}{Q}]\veclabel{{k}}{P} \leftrightarrow \left(\veclabel{{j}}{Q} \to \veclabel{{k}}{P}\right)\\
    \quad{\text{\color{violet}M:\color{black}$[{;}]$}} \;\;\;\; &\vdashu{\Psi}{\{\vec{i}_{\mR}, \vec{j}_{\mR}, \vec{k}_{\mR}\} \cupm \Psi} [\veclabel{{i}}{\alpha};\veclabel{{j}}{\beta}]\veclabel{{k}}{P} \leftrightarrow [\veclabel{{i}}{\alpha}][\veclabel{{j}}{\beta}]\veclabel{{k}}{P}\\
    \quad{\text{\color{violet}M:\color{black}$\langle\cdot\rangle$}} \;\;\;\; &\vdashu{\Psi}{\{\vec{j}_{\mR}, \vec{k}_{\mR}\} \cupm \Psi} \neg \left[\veclabel{{j}}{\alpha}\right]\neg \veclabel{{k}}{P} \leftrightarrow \langle \veclabel{{j}}{\alpha} \rangle \veclabel{{k}}{P}\\
    \quad{\text{\color{violet}M:\color{black}K}} \;\;\;\; &\vdashu{\Psi}{\{\any{\fuse{\vec{j}}{\vec{l}}}, \vec{k}_{\mR}, \vec{m}_{\mR}\} \cupm \Psi} \left[\atomlabel{\fuse{\vec{j}}{\vec{l}}}{\alpha}\right](\veclabel{{k}}{P} \to \veclabel{{m}}{Q}) \rightarrow \left(\left[\veclabel{{j}}{\alpha}\right] \veclabel{{k}}{P} \to \left[\veclabel{{l}}{\alpha}\right] \veclabel{{m}}{Q}\right) \\
    \quad{\text{\color{violet}M:\color{black}$[\cup]$}} \;\;\;\; &\vdashu{\Psi}{\{\vec{i}_{\mR}, \vec{j}_{\mR}, \fuse{\vec{k}}{\vec{l}}_{\mR}\} \cupm \Psi} [\veclabel{{i}}{\alpha} \cup \veclabel{{j}}{\beta}] \atomlabel{\fuse{\vec{k}}{\vec{l}}}{P} \leftrightarrow \left([\veclabel{{i}}{\alpha}]\veclabel{{k}}{P} \land [\veclabel{{j}}{\beta}]\veclabel{{l}}{P}\right)\\
    \quad{\text{\color{violet}M:\color{black}$[']$}} \;\;\;\; &\vdashu{\Psi}{\{\vec{m}_{\mId}, \vec{n}_{\mId}, \vec{k}_{\mR}, \vec{j}_{\mR}, \vec{l}_{\mR}\} \cupm \Psi} [\atomlabel{{k}}x' = f(x) \;\& \veclabel{{j}}q(x)] \veclabel{{l}}p(x) \\
    &\leftrightarrow \left(\forall \veclabel{{m}}{t {\geq} 0} \;((\forall \veclabel{{n}}{0 {\leq} s {\leq} t} \; \veclabel{{j}}q(x(s))) \to [\atomlabel{{k}}x:=x(t)]\veclabel{{l}}p(x))\right)  \quad{(\text{if } x'(t) = f(x(t)))}\\
    \quad{\text{\color{violet}M:\color{black}$[*]$}} \;\;\;\; &\vdashu{\Psi}{\{\any{\fuse{\vec{j}}{\vec{m}}}, \any{\fuse{\vec{k}}{\vec{l}}}\} \cupm \Psi} [\atomlabel{\fuse{\vec{k}}{\vec{l}}}{\alpha}^*]\atomlabel{\fuse{\vec{j}}{\vec{m}}}{P} \leftrightarrow \left(\veclabel{{j}}{P} \land [\veclabel{{k}}{\alpha}][\veclabel{{l}}{\alpha}^*]\veclabel{{m}}{P}\right) \\
    \quad{\text{\color{violet}M:\color{black}I}} \;\;\;\; &\vdashu{\Psi}{\{\any{\fuse{\fuse{\vec{k}}{\vec{l}}}{\vec{n}}}, \any{\fuse{\vec{j}}{\vec{m}}}\} \cupm \Psi} \left[\atomlabel{\fuse{\vec{j}}{\vec{m}}}{\alpha}^*\right]\atomlabel{\fuse{\fuse{\vec{k}}{\vec{l}}}{\vec{n}}}{P} \leftrightarrow \left(\veclabel{{k}}{P} \land \left[\veclabel{{j}}{\alpha}^*\right](\veclabel{{l}}{P} \to \left[\veclabel{{m}}{\alpha}\right]\veclabel{{n}}{P})\right)\\
    \quad{\text{\color{violet}M:\color{black}V}} \;\;\;\; &\vdashu{\Psi}{\{\vec{k}_{\mR}, \vec{j}_{\mR}\} \cupm \Psi} \veclabel{{k}}{p} \rightarrow \left[\veclabel{{j}}{\alpha}\right]\veclabel{{k}}{p} \quad{(\text{$FV(p) \cap BV(\alpha) = \emptyset$})}\\
    \quad{\text{\color{violet}M:\color{black}$[:=]_1$}} \;\;  &\vdashu{\Psi}{\{j_{\mId}, \vec{k}_{\mR}\} \cupm \Psi} [\atomlabel{j}{x:=e}\;] \veclabel{{k}}{p(x)} \leftrightarrow \veclabel{{k}}{p(e)} \;\;  \quad{x \in FV(p(x))} \\
    \quad{\text{\color{violet}M:\color{black}$[:=]_2$}} \;\;  &\vdashu{\Psi}{{\{j_{\mR}, \vec{k}_{\mR}\} \cupm \Psi}} [\atomlabel{j}{x:=e}\;] \veclabel{{k}}{p} \leftrightarrow \veclabel{{k}}{p} \;\;  \quad{x \notin FV(p)} \\
    \quad{\text{\color{violet}M:\color{black}$[{:}{*}]_1$}} \;\;  &\vdashu{\Psi}{{\{j_{\mId}, \vec{k}_{\mR}\} \cupm \Psi}} [\atomlabel{j}{x:=*}\;] \veclabel{{k}}{p(x)} \leftrightarrow \forall x \veclabel{{k}}{p(x)} \;\;  \quad{x \in FV(p(x))} \\
    \quad{\text{\color{violet}M:\color{black}$[{:}{*}]_2$}} \;\;  &\vdashu{\Psi}{{\{j_{\mR}, \vec{k}_{\mR}\} \cupm \Psi}} [\atomlabel{j}{x:=*}\;] \veclabel{{k}}{p} \leftrightarrow \veclabel{{k}}{p} \;\; \quad{x \notin FV(p)}
    \end{align*}

\input{Sections/first_order_axioms.tex}
\input{Sections/diff_eq_axioms.tex}
\input{Sections/differential_axioms.tex}
\input{Sections/derived_rules.tex}

\subsection{\uapc Leaf Rules}

    \begin{mathpar}
        \quad{\text{\color{violet}M:\color{black}id}} \;\;
        \inferrule
        {  }
        { \veclabel{{i}}{P}, \veclabel{{k}}{\Gamma} \vdashu{\Psi}{\{\any{\fuse{\vec{i}}{\vec{j}}}, \any {\vec{k}}, \any {\vec{l}}\} \cupm \Psi } \veclabel{{j}}{P}, \veclabel{{l}}{\Delta} }
\and
        \quad{\text{\color{violet}M:\color{black}QE}} \;\;
        \inferrule
        {  }
        { \veclabel{{i}}{\Gamma} \vdashu{\Psi}{DA(\vec{i}, \vec{j} \mid \Psi)} \veclabel{{j}}{\Delta} }
\and
        \quad{\text{\color{violet}M:\color{black}$\mathbb{R}$}} \;\;
        \inferrule
        {  }
        { \veclabel{{i}}{\Gamma} \vdashu{\Psi}{DA(\vec{i}, \vec{j} \mid \Psi)} \veclabel{{j}}{\Delta} }
\and        
        \quad{\text{\color{violet}M:\color{black}$\top$R}} \;\;
        \inferrule
        {  }
        { \veclabel{{k}}{\Gamma} \vdashu{\Psi}{\{i_{\mId}, \any {\vec{k}}, \any {\vec{j}}\} \cupm \Psi} \atomlabel{i}{\text{true}}, \veclabel{{j}}{\Delta} }
\and
        \quad{\text{\color{violet}M:\color{black}auto}} \;\;
        \inferrule
        {  }
        { \veclabel{{i}}{\Gamma} \vdashu{\Psi}{DA(\vec{i}, \vec{j} \mid \Psi)} \veclabel{{j}}{\Delta} }
\and        
        \quad{\text{\color{violet}M:\color{black}$\bot$L}} \;\;
        \inferrule
        {  }
        { \atomlabel{i}{\text{false}}, \veclabel{{k}}{\Gamma} \vdashu{\Psi}{\{i_{\mId}, \any{\vec{k}}, \any{\vec{j}}\} \cupm \Psi} \veclabel{{j}}{\Delta} }
    \end{mathpar}

\section{Metatheory}\label{app:metatheory}
  \begin{definition}[\textbf{Uniqueness}]
        Let $\veclabel{{\mathbf i}}{\Gamma} \vdashu{\Psi}{\Sigma} \veclabel{{\mathbf j}}{\Delta}$. A singleton label $\color{red}l \color{black}\in \color{red}\vec{\mathbf i}\color{black}, \color{red}\vec{\mathbf j}\color{black}$ is \emph{unique} iff $\color{red}l \color{black}\neq \color{red}l'\color{black} \; \forall \color{red}l'\color{black} \in \color{red}\vec{\mathbf i}, \vec{\mathbf j}\color{black} \setminus \color{red}l$.
  \end{definition}
\begin{definition}[\textbf{Freshness}]
        A singleton label $\color{red}l$ is \emph{fresh} for a valid sequent $\veclabel{{\mathbf i}}{\Gamma} \vdashu{\Psi}{\Sigma} \veclabel{{\mathbf j}}{\Delta}$ iff $\color{red}l \color{black}\notin \color{blue}\Psi\color{black}, \color{blue}\Sigma\color{black}, \color{red}\vec{\mathbf i}\color{black}, \color{red}\vec{\mathbf j}$.
  \end{definition}
  \begin{observation}[\textbf{Labeling schema}]
  The labels $\color{red}\vec{\mathbf i}\color{black}, \color{red}\vec{\mathbf j}$ in the conclusion of a proof $\veclabel{{\mathbf i}}{\Gamma} \vdashu{\Psi}{\Sigma} \veclabel{{\mathbf j}}{\Delta}$ are singleton and unique. The labels of a proof tree contain either singletons that appear in the conclusion of the proof or that are fresh.
  \end{observation}

\begin{lemma}[\textbf{Well-formed sets}]
    If $\veclabel{{\mathbf i}}\Gamma \vdashu{\Psi}{\Sigma} \veclabel{{\mathbf j}}\Delta$ is valid and the labels in $\color{blue}\Psi\color{black}$ are unique, then for all $\color{blue}\Psi$ each label in $\color{red}\vec{\mathbf i},\vec{\mathbf j}$ appears exactly once in $\color{blue}\Sigma$. 
\end{lemma}

\begin{proof}
    The proof is straightforward, following by structural induction on $\veclabel{{i}}\Gamma \vdashu{\Psi}{\Sigma} \veclabel{{j}}\Delta$. At each step, we note that label uniqueness with respect to sets is preserved by the operations on sets $\cupm$ and set difference.
\end{proof}

\begin{lemma}[\textbf{Composition of mutations}]
    If $\color{blue}A$ and $\color{blue}B$ are sets s.t. $\forall \color{red}l \color{black}\in \color{blue}A$ $\color{red}l \color{black}\notin \color{blue}B$
    , then $\color{violet}\color{violet}\mu_{A \cup B}\color{black}(\Gamma \vdash \Delta) \equiv \color{violet}{\color{violet}\mu_{A}}\color{black}(\color{violet}\color{violet}\mu_{B}\color{black}(\Gamma \vdash \Delta)) \equiv \color{violet}\color{violet}\mu_{B}\color{black}(\color{violet}{\color{violet}\mu_{A}}\color{black}(\Gamma \vdash \Delta))$.
\end{lemma}

\begin{proof}
    The proof proceeds by induction on the size of $A$. \\

    \textbf{Base case:} If $\left|A\right| = 0$, then $A = \emptyset$ and hence

    \[\mu_{A \cup B}(\Gamma \vdash \Delta) \equiv \mu_B(\Gamma \vdash \Delta) \equiv \mu_A(\mu_B(\Gamma \vdash \Delta)) \equiv \mu_B(\mu_A(\Gamma \vdash \Delta))\]

    \textbf{Inductive case:} Suppose that $\mu_{A' \cup B}(\Gamma \vdash \Delta) \equiv \mu_{A'}(\mu_{B}(\Gamma \vdash \Delta))$ for $\left|A'\right| = n$. We need to show that $\mu_{A \cup B}(\Gamma \vdash \Delta) \equiv \mu_A(\mu_B(\Gamma \vdash \Delta))$ for $A = l_\mathcal{M} \cup A'$ where $l_\mathcal{M} \notin A'$ (i.e. $\left|A\right| = n + 1$). By definition of $A$, we have
    \begin{align*}
        \mu_A(\mu_B(\Gamma \vdash \Delta)) &\equiv \mu_{l_\mathcal{M} \cup A'}(\mu_B(\Gamma \vdash \Delta)) \\
        &\equiv \mu_{l_\mathcal{M}}(\mu_A'(\mu_B(\Gamma \vdash \Delta)))  &\quad{(\text{by $A \cap B = \emptyset$, i.e., $l_\mathcal{M} \notin B$ for all $\mathcal{M}$})}\\
        &\equiv \mu_{l_\mathcal{M}}(\mu_{A'\cup B}(\Gamma \vdash \Delta))  &\quad{(\text{by IH})}
    \end{align*}

    Since $l_\mathcal{M} \notin A' \; \text{for all} \; \mathcal{M}$ and $A \cap B = \emptyset$, it follows that $l_\mathcal{M} \notin A' \cup B \; \text{for all} \; \mathcal{M}$. Therefore, we have
    \begin{align*}
        \mu_{l_\mathcal{M}}(\mu_{A'\cup B}(\Gamma \vdash \Delta)) &\equiv \mu_{(l_\mathcal{M} \cup A') \cup B}(\Gamma \vdash \Delta)\\
        &\equiv \mu_{A \cup B}(\Gamma \vdash \Delta) \; &\quad{(\text{by definition of $A$})}
    \end{align*}

    The argument for $\mu_{A \cup B}(\Gamma \vdash \Delta) \equiv \mu_B(\mu_A(\Gamma \vdash \Delta))$ follows a similar argument.
    $\square$
\end{proof}

\begin{customlem}{2}{\textbf{Monotonicity}}
    If $\mu_{\Sigma}(\Gamma \vdash \Delta)$ is valid for all $\mu_\Sigma$, then $\mu_{\Sigma \cupm \Omega}(\Gamma \vdash \Delta)$ is valid for all $\mu_{\Sigma \cupm \Omega}$.
    \end{customlem}
    
    \begin{proof}
    The proof proceeds by nested induction, first on the size of $\Omega$, and then on the depth of the proof tree. \\
    
    \textbf{Base case.} If $\left|\Omega\right| = 0$, then it must be that $\Omega = \emptyset$. So, 
    $\mu_{\Sigma \cupm \Omega}(\Gamma \vdash \Delta) \equiv \mu_{\Sigma}(\Gamma \vdash \Delta)$ (vacuously).
    Then the desired result follows by the assumption that $\mu_{\Sigma}(\Gamma \vdash \Delta)$ is valid for all choices of $\mu_\Sigma$. \\
    
    \textbf{Inductive step.} Assume that $\mu_{\Sigma}(\Gamma \vdash \Delta)$ is valid for all choices of $\mu_\Sigma$ implies 
    $\mu_{\Sigma \cupm \Omega'}(\Gamma \vdash \Delta)$ is valid for all $\Omega'$ such that $\left|\Omega'\right| = n$. We need to show that 
    $\mu_{\Sigma \cupm \Omega}(\Gamma \vdash \Delta)$ is valid for all $\Omega$ such that $\left|\Omega\right| = n + 1$.
    Fix an arbitrary $\Omega'$, and let $\Omega = \Omega' \cup \{l_\mathcal{M}\}$ for some $l_\mathcal{M}$. Since $l_\mathcal{M} \in \Omega$, the proof proceeds in two subcases (whether or not $l$ appears in $\Sigma$): \\
    
    \textit{\underline{Case 1: $l_\mathcal{M} \in \Omega$, $l_\mathcal{N} \in \Sigma$ ($\exists \; \mathcal{N}$).}} Suppose that $l_\mathcal{N} \in \Sigma$ for some $\mathcal{N}$, i.e. both $\Sigma$ and $\Omega$ mention label $l$. Letting $\Sigma = \Sigma' \cup \{l_{\mathcal{N}}\}$, it follows by definition of $\cupm$ that
    
    \[\Sigma \cupm \Omega = \{l_{\mathcal{M} \meet \mathcal{N}}\} \cup \left(\Sigma' \cupm \Omega'\right) \; \quad{(\dagger)}\]
    
    Since $l \notin \Omega'$, by uniqueness of labels, we have
    
    \[\left(\dagger\right) = \left(\{l_{\mathcal{M} \meet \mathcal{N}}\} \cup \Sigma'\right) \cupm \Omega'\]
    
    Since $\mathcal{M} \meet \mathcal{N} \subseteq \mathcal{N}$ and $\Sigma = \Sigma' \cup \{l_\mathcal{N}\}$, it must be that any choice of $\mu_{\{l_{\mathcal{M} \meet \mathcal{N}}\} \cup \Sigma'}$ is already included in the choices for $\mu_\Sigma$. Therefore, since $\mu_\Sigma(\Gamma \vdash \Delta)$ is valid for all $\mu_\Sigma$, it follows that $\mu_{\{l_{\mathcal{M} \meet \mathcal{N}}\} \cup \Sigma'}(\Gamma \vdash \Delta)$ is valid for all $\mu_{\{l_{\mathcal{M} \meet \mathcal{N}}\}} \cup \Sigma'$. 
    Then the desired result follows by the induction hypothesis.\\

    \textit{\underline{Case 2: $l_\mathcal{M} \in \Omega$, $l_\mathcal{N} \notin \Sigma$ (for all $\mathcal{N}$).}} Suppose that $l_\mathcal{N} \notin \Sigma$, i.e. $\Sigma$ has no constraints on $l$. Then by definition of $\cupm$, we have
    
    \begin{align*}
        \Sigma \cupm \Omega 
        &= \Sigma \cupm \left(\Omega' \cup \{l_\mathcal{M}\}\right) \\
        &= \{l_\mathcal{M}\} \cup (\Sigma \cupm \Omega')
    \end{align*}
    
    We need to show that $\mu_{\{l_\mathcal{M}\} \cup (\Sigma \cupm \Omega')}(\Gamma \vdash  \Delta)$ is valid. To this end, we need to show that $\mu_{\{l_\mathcal{M}\}}(\mu_{\Sigma \cupm \Omega'}(\Gamma \vdash  \Delta))$ is valid.
    The proof proceeds in two subcases (whether or not $l_\mathcal{M}$ labels an atom in $\Gamma \vdash \Delta$): \\
    
    \textit{\underline{Subcase 1: $l_\mathcal{M}$ does not label any atoms in $\Gamma \vdash \Delta$.}}  Then, by definition of $\mu$:
    
    \[\mu_{\{l_\mathcal{M}\} \cup (\Sigma \cupm \Omega')}(\Gamma \vdash \Delta) \equiv \mu_{\Sigma \cupm \Omega'}(\Gamma \vdash \Delta)\]
    
    By IH, $\mu_{\Sigma \cupm \Omega'}(\Gamma \vdash \Delta)$ is valid, and we conclude that $\mu_{\{l_\mathcal{M}\} \cup (\Sigma \cupm \Omega')}(\Gamma \vdash \Delta)$ is valid.\\
    
    \textit{\underline{Subcase 2: $l_\mathcal{M}$ labels some atom in $\Gamma \vdash \Delta$.}}  The proof proceeds by an inner induction on the depth of the proof tree. Writing $\mu_{\Sigma \cupm \Omega'}(\Gamma \vdash  \Delta)$ as $\Gamma' \vdash \Delta'$, we have the following:\\
    
    \textbf{Base case.}  If the depth of the proof tree is $1$, then it must be a leaf rule, i.e. M:id, M:auto, M:$\mathbb{R}$, M:$\top R$, M:$\bot L$.  \\
    
    \underline{Case M:id}  Suppose that $l_\mathcal{M}$ was generated by M:id, and let $\Gamma' = P, \Gamma_0$ and $\Delta' = P, \Delta_0$.
    
    \begin{mathpar}
        \quad{\text{\color{violet}M:\color{black}id}} \;\;
        \inferrule
        {  }
        { \veclabel{{i}}{P}, \veclabel{{k}}{\Gamma_0} \vdashu{\Psi}{\{\any{\fuse{\vec{i}}{\vec{j}}}, \any {\vec{k}}, \any {\vec{h}}\} \cupm \Psi } \veclabel{{j}}{P}, \veclabel{{h}}{\Delta_0} }
    \end{mathpar}
    
    Let $l_\mathcal{M} \in \Psi$ be an arbitrary label of some atom in $P, \Gamma_0 \vdash P, \Delta_0$ (i.e. $\vec{i}$, $\vec{j}$, $\vec{k}$, $\vec{h}$). Observe that this case holds vacuously because in this case $l_\mathcal{M} \in \Psi$ is a label of some atom in $P, \Gamma_0 \vdash P, \Delta_0$ and $l_\mathcal{N} \notin \{\any{\fuse{\vec{i}}{\vec{j}}}, \any {\vec{k}}, \any {\vec{h}}\} \cupm \Psi$ for any $\mathcal{N}$. \\
    
    
    \underline{Cases M:auto, M:$\mathbb{R}$, M:$\top R$, M:$\bot L$.} The proofs of these cases each follow arguments analogous to the one provided for M:id. The proof cases for M:auto, M:$\mathbb{R}$ rely on definition~\ref{defn:DA} which enforces that the set $DA(\vec{i},\vec{j})$ mentions every label in $\Gamma \vdash \Delta$. \\
    
    \textbf{Inductive step.}  Suppose that there exists a proof of $\Gamma'' \vdash \Delta''$ with depth $n$ for some $\Gamma'', \Delta''$. We need to show that there exists a proof of $\mu_{\{l_\mathcal{M}\}}(\Gamma' \vdash \Delta')$ with depth $n + 1$. By observation~\ref{obs:superset_output}, all other labels in $\Gamma' \vdash \Delta'$ already appear in $\Sigma \cupm \Omega'$. So there are only a select number of rules that could have generated $l_\mathcal{M}$, i.e. for which $l_\mathcal{M}$ appears in the output set of the conclusion but not of the premise. In other words, the only cases we need to consider here are the following: \\

    {\underline{Case M:CTR:}}  Suppose that $l_\mathcal{M}$ was generated by M:CTR.
    
    \begin{mathpar}
        \quad{\text{\color{violet}M:\color{black}CTR}} \;\;
        \inferrule
        { \vdashu{\Psi}{\Omega} \atomlabel{j}{e = k} }
        { \veclabel{{i}}{\Gamma} \vdashu{\Psi}{\Omega \cupm \{\any{\vec{i}}, \any{\vec{k}}\}} \atomlabel{j}{c(e) = c(k)}, \veclabel{{k}}{\Delta} }
    \end{mathpar}
    
    Here, we need to show that there exists a proof of $\mu_{l_\mathcal{M}}(\mu_{\Omega}(\Gamma \vdash c(e) = c(k), \Delta))$ for $l_\mathcal{M} \in \{\any{\vec{i}}, \any{\vec{k}}\}$. 
    Observe that in all cases, $l_{\mathcal{M}}$ does not label any atoms in $\vdash e = k$. Therefore, it follows by definition of $\mu$ that for all choices of $\mu_{l_\mathcal{M}}$:
    
    \[\left(\vdash e = k\right) \equiv \mu_{l_{\mathcal{M}}}(\vdash e = k)\]
    
    By the induction hypothesis, $\mu_\Omega(\Gamma \vdash c(e) = c(k), \Delta)$ is valid. Since $\mu_\Omega(\Gamma \vdash c(e) = c(k), \Delta)$ is valid, there exists a proof of it, and hence there exists a proof of $\mu_\Omega(\vdash e = k)$, i.e. $\mu_\Omega(\vdash e = k)$ is valid. Since $l_\mathcal{M}$ does not appear in $\vdashu{\Psi}{\Omega} \atomlabel{j}{e = k}$, there exists a proof of $\mu_{l_{\mathcal{M}}}(\mu_\Omega(\vdash e = k))$. Then by \dL rule CTR, there exists a proof of $\mu_{l_\mathcal{M}}(\mu_\Omega(\Gamma \vdash c(e) = c(k), \Delta))$. \\

    {\underline{Case M:CTL:}} This case follows in analogy to rule M:CTR. \\
    
    {\underline{Case M:\textit{W}R:}} Suppose that $l_\mathcal{M}$ was generated by M:\textit{W}R.
    
    \begin{mathpar}
        \quad{\text{M{:}\textit{W}R}} \;\;
        \inferrule
        { \vdashu{\Psi}{\Omega} \veclabel{{j}}{P} }
        { \veclabel{{i}}{\Gamma} \vdashu{\Psi}{\Omega \cupm \atoms{\any{\vec{i}}, \any{\vec{k}}}} \veclabel{{j}}{P}, \veclabel{{k}}{\Delta} }
    \end{mathpar}
    
    We need to show that there exists a proof of $\mu_{l_\mathcal{M}}(\mu_{\Omega}(\Gamma \vdash P, \Delta))$ for $l_\mathcal{M} \in \{\any{\vec{i}}, \any{\vec{k}}\}$. By the induction hypothesis, $\mu_{\Omega}(\Gamma \vdash P, \Delta)$ is valid, and hence there exists a proof of it. Then there exists a proof of the premise $\mu_{\Omega}(\vdash P)$, i.e. $\mu_{\Omega}(\vdash P)$ is valid.
    Since $\vec{i}, \vec{k}$ do not appear in $\vdashu{\Psi}{\Omega} \veclabel{{j}}{P}$, we have that $\mu_{l_\mathcal{M}}(\mu_{\Omega}(\vdash P))$ is valid for any choice of $l_\mathcal{M}$. Therefore, by application of the \dL rule \textit{W}R, $\mu_{l_\mathcal{M}}(\mu_{\Omega}(\Gamma \vdash P, \Delta))$ is valid. \\
    
    {\underline{Cases M:\textit{W}L, M:\textit{W}LR, M:WR, M:WL, M:GVR:}} These cases each follow arguments analogous to the one given for M:\textit{W}R. \\
    
    In all cases, there exists a proof of $\mu_{l_\mathcal{M}}(\mu_{\Omega}(\Gamma \vdash \Delta))$. Therefore, it must be valid. So it follows by lemma~\ref{lem:union_muts} that $\mu_{\{l_\mathcal{M}\} \cup (\Sigma \cupm \Omega')}(\Gamma \vdash  \Delta)$ is valid for all $\Omega'$. In general, we have shown that $\mu_{\Sigma \cupm \Omega}(\Gamma \vdash  \Delta)$ is valid for all $\Omega$.
    $\square$
    \end{proof}

\begin{customcor}{1}{\textbf{Monotonicity with freshness}} 
    If $\mu_{\Sigma}(\Gamma \vdash \Delta)$ is valid, then $\mu_{\Sigma \setminus {\vec{l}}}(\Gamma \vdash \Delta)$ is valid where $\vec{l}$ fresh for $\Gamma \vdash \Delta$.
    \end{customcor}
    
    \begin{proof}
        Since $\vec{l}$ is fresh, it follows by observation~\ref{obs:unique_labels} that $\vec{l}$ must not label any atom in $\Gamma \vdash \Delta$. Therefore, it follows by definition of $\mu$ that:
    
        \[\mu_{\Sigma}(\Gamma \vdash \Delta) \equiv \mu_{\Sigma \setminus {\vec{l}}}(\Gamma \vdash \Delta)\]
    
        Since $\mu_{\Sigma}(\Gamma \vdash \Delta)$ is valid by assumption, it follows that $\mu_{\Sigma \setminus {\vec{l}}}(\Gamma \vdash \Delta)$ is valid.
    $\square$
    \end{proof}

\begin{customthm}{1}{\textbf{Soundness of \uapc}}
    If $\veclabel{{i}}{\Gamma} \vdashu{\Psi}{\Sigma} \veclabel{{j}}{\Delta}$, then for all $\mu_\Sigma$, $\mu_{\Sigma}(\Gamma \vdash \Delta)$ is valid.
    \end{customthm}
    
    \begin{proof}
    
    The proof proceeds by structural induction over the usage-aware proof calculus.
    
    \paragraph{Case M:$\land$R. } Suppose the last rule in the proof tree is M:$\land$R:
    
    \begin{mathpar}
        \quad{\text{\color{violet}M:\color{black}$\land$R}} \;\;
        \inferrule
        { \veclabel{{k}}{\Gamma} \vdashu{\Psi}{\Sigma} \veclabel{{i}}{P}, \veclabel{{l}}{\Delta}
        \\ \veclabel{{k}}{\Gamma} \vdashu{\Psi}{\Omega} \veclabel{{j}}{Q}, \veclabel{{l}}{\Delta} }
        { \veclabel{{k}}{\Gamma} \vdashu{\Psi}{\Sigma \cupm \Omega} \veclabel{{i}}{P} \land \veclabel{{j}}{Q}, \veclabel{{l}}{\Delta} }
    \end{mathpar}
    
    Since there exists a proof of $\veclabel{{k}}{\Gamma} \vdashu{\Psi}{\Sigma \cupm \Omega} \veclabel{{i}}{P} \land \veclabel{{j}}{Q}$ with the last rule M:$\land$R, there must exist (smaller) proofs of $\veclabel{{k}}{\Gamma} \vdashu{\Psi}{\Sigma} \veclabel{{i}}{P}, \veclabel{{l}}{\Delta}$ and $\veclabel{{k}}{\Gamma} \vdashu{\Psi}{\Omega} \veclabel{{j}}{Q}, \veclabel{{l}}{\Delta}$. From the induction hypothesis applied to the proofs of these premises, we have:
    
    \begin{itemize}
        \item[(1)] $\mu_{\Sigma}(\Gamma \vdash P, \Delta)$ is valid, for any $\mu_{\Sigma}$
        \item[(2)] $\mu_{\Omega}(\Gamma \vdash Q, \Delta)$ is valid, for any $\mu_{\Omega}$
    \end{itemize}
    
    Since $\mu_{\Sigma}(\Gamma \vdash P, \Delta)$ is valid, and by lemma~\ref{lem: subset_muts} applied to (1), we have 
    
    \[\mu_{\Sigma \cupm \Omega}(\Gamma \vdash P, \Delta) \text{ is valid for a fixed arbitrary $\mu_{\Sigma \cupm \Omega}$.} \;\; (3)\] 
    

    Similarly, since $\mu_{\Omega}(\Gamma \vdash Q, \Delta)$ is valid, it follows by lemma~\ref{lem: subset_muts} applied to (2) that 
    
    \[\mu_{\Sigma \cupm \Omega}(\Gamma \vdash Q, \Delta) \text{ is valid for the same $\mu_{\Sigma \cupm \Omega}$. } \;\; (4)\]

    By the \dL $\land$R proof rule applied to (3) and (4), we have
    
    \[\text{$\mu_{\Sigma \cupm \Omega}(\Gamma \vdash P \land Q, \Delta)$ is valid.}\]
    
    Since our choice of $\mu_{\Sigma \cupm \Omega}$ was arbitrary, the desired result holds for \textit{any} $\mu_{\Sigma \cupm \Omega}$.
    
    \paragraph{Cases M:$\lor$L, M:$\leftrightarrow$R, M:$\to$L, M:$\leftrightarrow$L. } The proof cases for these rules each follow an argument analogous to the one provided for   M:$\land$R.
    
    \paragraph{Case M:cut. } Suppose the last rule in the proof tree is M:cut:
    
    \begin{mathpar}
        \quad{\text{\color{violet}M:\color{black}cut}^{\color{red}\vec{l}\color{black}}} \;\;
        \inferrule
        { \veclabel{{i}}{\Gamma}, \veclabel{{j}}{C} \vdashu{\Psi}{\Sigma} \veclabel{{k}}{\Delta}
        \\ 
        \veclabel{{i}}{\Gamma} \vdashu{\Psi}{\Omega} \veclabel{{k}}{\Delta}, \veclabel{{j}}{C}
        }
        { \veclabel{{i}}{\Gamma} \vdashu{\Psi}{(\Sigma \cupm \Omega) \setminus {\vec{l}}} \veclabel{{k}}{\Delta} }
        \quad{(\forall C_a \in \text{atoms}(C). \, \color{red}j_a \color{black}= \phifunc{}{C_a}{(\veclabel{{i}}{\Gamma}, \veclabel{{k}}{\Delta})}{\color{red}\vec{l}\color{black}})} 
    \end{mathpar}
    
    Since there exists a proof of $\veclabel{{i}}{\Gamma} \vdashu{\Psi}{(\Sigma \cupm \Omega) \setminus {\vec{l}}} \veclabel{{k}}{\Delta}$, there must exist (smaller) proofs of $\veclabel{{i}}{\Gamma}, \veclabel{{j}}{C} \vdashu{\Psi}{\Sigma} \veclabel{{k}}{\Delta}$ and $\veclabel{{i}}{\Gamma} \vdashu{\Psi}{\Omega} \veclabel{{k}}{\Delta}, \veclabel{{j}}{C}$. 
    So, by the induction hypothesis applied to the proofs of these premises, we have:
    
    \begin{itemize}
        \item $\mu_{\Sigma}(\Gamma, C \vdash \Delta)$ is valid, for any $\mu_{\Sigma}$.
        \item $\mu_{\Omega}(\Gamma \vdash \Delta, C)$ is valid, for any $\mu_{\Omega}$.
    \end{itemize}
    
    Now, we want to show that $\mu_{(\Sigma \cupm \Omega) \setminus {\vec{l}}}(\Gamma \vdash \Delta)$ is valid. Since $\mu_{\Sigma}(\Gamma, C \vdash \Delta)$ is valid, it follows by lemma~\ref{lem: subset_muts} that for a fixed arbitrary $\mu_{\Sigma \cupm \Omega}$:
    
    \[\mu_{\Sigma \cupm \Omega}(\Gamma, C \vdash \Delta) \; \text{ is valid}.\]
    
    From $\mu_{\Omega}(\Gamma \vdash \Delta, C)$, it follows by lemma~\ref{lem: subset_muts} that (for the same $\mu_{\Sigma \cupm \Omega}$):
    
    \[\mu_{\Sigma \cupm \Omega}(\Gamma \vdash \Delta, C) \; \text{ is valid}.\]
    
    Then, by the \dL proof rule cut (for the same $\mu_{\Sigma \cupm \Omega}$):
    
    \[\mu_{\Sigma \cupm \Omega}(\Gamma \vdash \Delta) \; \text{ is valid}.\]
    
    Since $\vec{l}$ is fresh for $\Gamma \vdash \Delta$, it follows by lemma~\ref{lem: minus_muts} that:
    
    \[\mu_{(\Sigma \cupm \Omega) \setminus {\vec{l}}}(\Gamma \vdash \Delta) \; \text{ is valid}.\]
    
    Since $\mu_{\Sigma \cupm \Omega}$ was arbitrary, the desired result holds for \textit{any} $\mu_{(\Sigma \cupm \Omega) \setminus {\vec{l}}}$.

    \paragraph{Cases M:MR, M:ML. } The proof case for this rule follows an argument analogous to the one provided for M:cut.

    \paragraph{Case M:cutR. } Suppose the last rule in the proof tree is M:cutR:
    
    \begin{mathpar}
        \quad{\text{\color{violet}M:\color{black}cutR}} \;\;
        \inferrule
        { \veclabel{{i}}{\Gamma} \vdashu{\Psi}{\Omega} \veclabel{{j}}{Q}, \veclabel{{k}}{\Delta}
        \\ \veclabel{{i}}{\Gamma} \vdashu{\Psi}{\Sigma} \veclabel{{j}}{Q} \to \veclabel{{l}}{P}, \veclabel{{k}}{\Delta}}
        { \veclabel{{i}}{\Gamma} \vdashu{\Psi}{(\Omega \cupm \Sigma) \setminus {\vec{j}}} \veclabel{{l}}{P}, \veclabel{{k}}{\Delta} }
        \quad{\text{($\vec{j}$ fresh)}}
    \end{mathpar}
    
    This case is similar to M:cut. Since there exists a proof of $\veclabel{{i}}{\Gamma} \vdashu{\Psi}{(\Omega \cupm \Sigma) \setminus {\vec{j}}} \veclabel{{l}}{P}, \veclabel{{k}}{\Delta}$, there must exist (smaller) proofs of 
    
    \begin{enumerate}
        \item[(1)] $\veclabel{{i}}{\Gamma} \vdashu{\Psi}{\Omega} \veclabel{{j}}{Q}, \veclabel{{k}}{\Delta}$ and
        \item[(2)] $\veclabel{{i}}{\Gamma} \vdashu{\Psi}{\Sigma} \veclabel{{j}}{Q} \to \veclabel{{l}}{P}, \veclabel{{k}}{\Delta}$. 
    \end{enumerate}

    By applying the induction hypothesis to the proof of $(1)$, we have
    
    \[\mu_{\Omega}({\Gamma} \vdash {Q}, {\Delta}) \;\; \text{is valid, for any $\mu_{\Omega}$.}\]
    
    By applying the induction hypothesis to the proof of $(2)$, we have
    
    \[\mu_{\Sigma}(\Gamma \vdash {Q} \to {P}, \Delta) \;\; {\text{is valid, for any $\mu_{\Sigma}$.}}\]
    
    Now, we want to show that $\mu_{(\Omega \cupm \Sigma) \setminus {\vec{j}}}(\Gamma \vdash P, \Delta)$ is valid. Since $\mu_{\Omega}({\Gamma} \vdash {Q}, {\Delta})$ is valid, it follows by lemma~\ref{lem: subset_muts} that for a fixed arbitrary $\mu_{\Omega \cupm \Sigma}$:
    
    \[\mu_{\Omega \cupm \Sigma}({\Gamma} \vdash {Q}, {\Delta}) \; \text{ is valid}.\]
    
    From $\mu_{\Sigma}({\Gamma} \vdash {Q} \to {P}, \Delta)$, it follows by lemma~\ref{lem: subset_muts} that (for the same $\mu_{\Omega \cupm \Sigma}$):
    
    \[\mu_{\Omega \cupm \Sigma}({\Gamma} \vdash {Q} \to {P}, \Delta) \; \text{ is valid}.\]
    
    Then, by the \dL proof rule cutR (for the same $\mu_{\Omega \cupm \Sigma}$):
    
    \[\mu_{\Omega \cupm \Sigma}(\Gamma \vdash P, \Delta) \; \text{ is valid}.\]
    
    Since $\vec{j}$ is fresh for $\Gamma \vdash \Delta$, it follows by lemma~\ref{lem: minus_muts} that:
    
    \[\mu_{(\Omega \cupm \Sigma) \setminus {\vec{j}}}(\Gamma \vdash P, \Delta) \; \text{ is valid}.\]
    
    Since $\mu_{\Omega \cupm \Sigma}$ was arbitrary, the desired result holds for \textit{any} $\mu_{(\Omega \cupm \Sigma) \setminus {\vec{j}}}$.

    \paragraph{Case M:cutL. } The proof case for this rule follows an argument analogous to the one provided for M:cutR.

    \paragraph{Case M:CER. } Suppose the last rule in the proof tree is M:CER:
    
    \begin{mathpar}
        \quad{\text{\color{violet}M:\color{black}CER}^{\color{red}\vec{n}\color{black}}} \;\;
        \inferrule
        { 
            \veclabel{{i}}{\Gamma} \vdashu{\Psi}{\Omega} \veclabel{ m}{C(\veclabel{j}{Q})}, \veclabel{{k}}{\Delta}
        \\ \vdashu{\Psi}{\Sigma} \veclabel{{j}}{Q} \leftrightarrow \veclabel{{l}}{P} 
        }
        { \veclabel{{i}}{\Gamma} \vdashu{\Psi}{(\Omega \cupm \Sigma) \setminus {\vec{l}}} \veclabel{ m}{C(\veclabel{{l}}{P})}, \veclabel{{k}}{\Delta} }
        \quad{(\forall Q_a \in \text{atoms}(Q). \, \color{red}j_a \color{black}= \phifunc{}{Q_a}{\veclabel{{l}}{P}}{\color{red}\vec{n}\color{black}})} 
    \end{mathpar}
    
    Since there exists a proof of $\veclabel{{i}}{\Gamma} \vdashu{\Psi}{(\Omega \cupm \Sigma) \setminus \vec{j}} \atomlabel{\vec m}{C(\veclabel{{l}}{Q_0})}, \veclabel{{k}}{\Delta}$, there must exist (smaller) proofs of 
    
    \begin{itemize}
        \item[(1)] $\veclabel{{i}}{\Gamma} \vdashu{\Psi}{\Omega} \veclabel{ m}{C(\veclabel{j}{Q})}, \veclabel{{k}}{\Delta}$, and
        \item[(2)] $\vdashu{\Psi}{\Sigma} \veclabel{{j}}{Q} \leftrightarrow \veclabel{{l}}{P}$
    \end{itemize}
    
    By the induction hypothesis applied to (1), we have that for all $\mu_\Omega$:
    
    \[\mu_\Omega(\Gamma \vdash C(P_0), \Delta) \; \text{is valid}.\]
    
    Likewise, applying the induction hypothesis to (2) yields that for all $\mu_\Sigma$:
    
    \[\mu_\Sigma( \vdash P_0 \leftrightarrow Q_0) \; \text{is valid}\]
    
    All together we have
    
    \begin{itemize}
        \item $\mu_\Sigma( \vdash P_0 \leftrightarrow Q_0)$ is valid for all $\mu_\Sigma$.
        \item $\mu_\Omega(\Gamma \vdash C(P_0), \Delta)$ is valid for all $\mu_\Omega$.
    \end{itemize}
    
    By application of lemma~\ref{lem: subset_muts}, we have for all $\mu_{\Omega \cupm \Sigma}$
    
    \begin{itemize}
        \item $\mu_{\Omega \cupm \Sigma}( \vdash P_0 \leftrightarrow Q_0)$ is valid
        \item $\mu_{\Omega \cupm \Sigma}(\Gamma \vdash C(P_0), \Delta)$ is valid
    \end{itemize}
    
    Fix an arbitrary $\mu_{\Omega \cupm \Sigma}$. Then by the \dL CER rule, we have:
    
    \[\mu_{\Omega \cupm \Sigma}(\Gamma \vdash C(Q_0), \Delta) \text{ is valid}.\]
    
    By freshness of $\vec{j}$, it follows by lemma~\ref{lem: minus_muts} that
    
    \[\mu_{(\Omega \cupm \Sigma) \setminus \vec{j}}(\Gamma \vdash C(Q_0), \Delta) \text{ is valid}.\]
    
    Since $\mu_{\Omega \cupm \Sigma}$ was arbitrary, the result holds for any $\mu_{(\Omega \cupm \Sigma) \setminus \vec{j}}$. \\

    \paragraph{Case M:CEL. } The proof case for this rule follows an argument analogous to the one provided for M:CER.

    \paragraph{Case M:CQR. } Suppose the last rule in the proof tree is M:CQR: 
    
    \begin{mathpar}
        \quad{\text{\color{violet}M:\color{black}CQR}} \;\;
        \inferrule
        { \veclabel{{i}}{\Gamma} \vdashu{\Psi}{\Omega} \veclabel{{j}}{p(k)}, \veclabel{{m}}{\Delta}
        \\ \vdashu{\Psi}{\Sigma} \atomlabel{{l}}{k = e}}
        { \veclabel{{i}}{\Gamma} \vdashu{\Psi}{(\Omega \cupm \Sigma) \setminus \atoms{l}} \veclabel{{j}}{p(e)}, \veclabel{{m}}{\Delta} }
        \quad{\text{(${l}$ fresh)}}
    \end{mathpar}
    
    Since there exists a proof of $\veclabel{{i}}{\Gamma} \vdashu{\Psi}{(\Omega \cupm \Sigma) \setminus \atoms{l}} \veclabel{{j}}{p(e)}, \veclabel{{m}}{\Delta}$, there must exist (smaller) proofs of
    
    \begin{itemize}
        \item[(1)] $\veclabel{{i}}{\Gamma} \vdashu{\Psi}{\Omega} \veclabel{{j}}{p(k)}, \veclabel{{m}}{\Delta}$, and
        \item[(2)] $\vdashu{\Psi}{\Sigma} \atomlabel{{l}}{k = e}$
    \end{itemize}
    
    The proof proceeds by an inner induction on the structure of $\vdashu{\Psi}{\Sigma} \atomlabel{{l}}{k = e}$:
    
    By application of the induction hypothesis, it follows that for any $\mu_\Sigma$, $\mu_{\Sigma}(k = e)$ is valid.
    By the induction hypothesis applied to (1), we have that for all $\mu_\Omega$:
    
    \[\mu_\Omega(\Gamma \vdash p(k), \Delta).\]
    
    All together we have
    
    \begin{itemize}
        \item $\mu_\Sigma( \vdash k = e)$ is valid for all $\mu_\Sigma$.
        \item $\mu_\Omega(\Gamma \vdash p(k), \Delta)$ is valid for all $\mu_\Omega$.
    \end{itemize}
    
    By application of lemma~\ref{lem: subset_muts}, it follows that
    
    \begin{itemize}
        \item $\mu_{(\Omega \cupm \Sigma)}( \vdash k = e)$ is valid for all $\mu_{(\Omega \cupm \Sigma)}$.
        \item $\mu_{(\Omega \cupm \Sigma)}(\Gamma \vdash p(k), \Delta)$ is valid for all $\mu_{(\Omega \cupm \Sigma)}$.
    \end{itemize}
    
    Fix an arbitrary $\mu_{(\Omega \cupm \Sigma)}$. Then by the \dL CQR rule, we have:
    
    \[\mu_{(\Omega \cupm \Sigma)}(\Gamma \vdash p(e), \Delta) \text{ is valid}.\]
    
    By the freshness of $l$, it follows by application of lemma~\ref{lem: minus_muts} that
    
    \[\mu_{(\Omega \cupm \Sigma) \setminus \atoms{l}}(\Gamma \vdash p(e), \Delta) \text{ is valid}.\]
    
    Since $\mu_{(\Omega \cupm \Sigma)}$ was arbitrary, the result holds for any $\mu_{(\Omega \cupm \Sigma) \setminus \atoms{l}}$. \\

    \paragraph{Case M:CQL. } The proof case for this rule follows an argument analogous to the one provided for M:CQR.

    \paragraph{Case M:$\neg$R. } Suppose the last rule in the proof tree is M:$\neg$R:
    
    \begin{mathpar}
        \quad{\text{\color{violet}M:\color{black}$\neg$R}} \;\;
        \inferrule
        { \veclabel{{k}}{\Gamma}, \veclabel{{i}}{P} \vdashu{\Psi}{\Sigma} \veclabel{{j}}{\Delta} }
        { \veclabel{{k}}{\Gamma} \vdashu{\Psi}{\Sigma} \neg \veclabel{{i}}{P}, \veclabel{{j}}{\Delta} }
    \end{mathpar}
    
    Since there exists a proof of $\veclabel{{k}}{\Gamma} \vdashu{\Psi}{\Sigma} \neg \veclabel{{i}}{P}, \veclabel{{j}}{\Delta}$, there must exist a (smaller) proof of $\veclabel{{k}}{\Gamma}, \veclabel{{i}}{P} \vdashu{\Psi}{\Sigma} \veclabel{{j}}{\Delta}$. From the induction hypothesis applied to the proof of this premise, we have:
    
    \[\mu_{\Sigma}(\Gamma, P \vdash \Delta) \; \text{ is valid}, \text{ for any $\mu_{\Sigma}$}\]
    
    Now observe that for a fixed arbitrary $\mu_{\Sigma}$ and by the \dL $\neg$R proof rule, $\mu_{\Sigma}(\Gamma \vdash \neg P, \Delta)$ is valid. Since our choice of $\mu_{\Sigma}$ was arbitrary, the desired result holds for \textit{any} $\mu_{\Sigma}$.

    \paragraph{Cases M:$\lor$R, M:$\neg$L, M:$\land$L, M:$\to$R, M:PR, M:PL, M:$\forall$R, M:$\forall$L, M:$\exists$R, M:$\exists$L, M:US, M:UR, M:BRR, M:BRL,  M:$\leftrightarrow$\textit{c}R, M:$\leftrightarrow$\textit{c}L, M:$\rightarrow2\leftrightarrow$, $DE_{\leftarrow}$. } The proof cases for these rules each follow an argument analogous to the one provided for M:$\neg$R.

    \paragraph{Case M:loop. } Suppose the last rule in the proof tree is M:loop:
    
    \begin{mathpar}
        \quad{\text{\color{violet}M:\color{black}loop}^{\color{red}\vec{l}\color{black}}} \;\;
        \inferrule
        { \veclabel{{k}}{\Gamma} \vdashu{\Psi}{\Sigma} \veclabel{{n}}{J}, \veclabel{{i}}{\Delta}
        \\ \veclabel{{n}}{J} \vdashu{\Psi}{\Omega} \veclabel{{j}}{P}
        \\ \veclabel{{n}}{J} \vdashu{\Psi}{\Theta} [\veclabel{{m}}{\alpha}] \veclabel{{n}}{J}
        }
        { \veclabel{{k}}{\Gamma} \vdashu{\Psi}{(\Sigma \cupm \Omega \cupm \Theta) \setminus {\vec{l}}} [\veclabel{{m}}{\alpha}^*]\veclabel{{j}}{P},\veclabel{{i}}{\Delta} }
        \quad{(\forall J_a \in \text{atoms}(J). \, \color{red}n_a \color{black}= \phifunc{}{J_a}{(\veclabel{{k}}{\Gamma}, \veclabel{{i}}{\Delta})}{\color{red}\vec{l}\color{black}})} 
    \end{mathpar}
    
    Since there exists a proof of $\veclabel{{k}}{\Gamma} \vdashu{\Psi}{(\Sigma \cupm \Omega \cupm \Theta) \setminus \vec{l}} [\veclabel{{m}}{\alpha}^*]\veclabel{{j}}{P},\veclabel{{i}}{\Delta}$, there must exist (smaller) proofs of 
    \begin{itemize}
        \item $\veclabel{{k}}{\Gamma} \vdashu{\Psi}{\Sigma} \veclabel{{n}}{J}, \veclabel{{i}}{\Delta}$ 
        \item $\veclabel{{n}}{J} \vdashu{\Psi}{\Omega} \veclabel{{j}}{P}$
        \item $\veclabel{{n}}{J} \vdashu{\Psi}{\Theta} [\veclabel{{m}}{\alpha}] \veclabel{{n}}{J}$
    \end{itemize}
    From the induction hypothesis applied to the proofs of each of these premises, we have:
    \begin{itemize}
        \item $\mu_\Sigma(\Gamma \vdash J, \Delta)$ is valid for any $\mu_\Sigma$
        \item $\mu_\Omega(J \vdash P)$ is valid for any $\mu_\Omega$
        \item $\mu_\Theta(J \vdash [\alpha] J)$ is valid for any $\mu_\Theta$
    \end{itemize}
    
    Now, fix arbitrary $\mu_\Sigma$, $\mu_\Omega$, and $\mu_\Theta$. Now we want to show $\mu_{(\Sigma \cupm \Omega \cupm \Theta) \setminus \vec{l}} (\Gamma \vdash [\alpha^*]P, \Delta)$ is valid.
    Since $\mu_\Sigma(\Gamma \vdash J, \Delta)$ is valid, it follows by lemma~\ref{lem: subset_muts} that:
    
    \[\mu_{\Sigma \cupm \Omega \cupm \Theta}(\Gamma \vdash J, \Delta) \; \text{ is valid}.\]
    
    Since $\mu_\Omega(J \vdash P)$ is valid, it follows by lemma~\ref{lem: subset_muts} that (for the same $\mu_{\Sigma \cupm \Omega \cupm \Theta}$):
    
    \[\mu_{\Sigma \cupm \Omega \cupm \Theta}(J \vdash P) \; \text{ is valid}.\]
    
    Since $\mu_\Theta(J \vdash [\alpha] J)$ is valid, it follows by lemma~\ref{lem: subset_muts} that (for the same $\mu_{\Sigma \cupm \Omega \cupm \Theta}$):
    
    \[\mu_{\Sigma \cupm \Omega \cupm \Theta}(J \vdash [\alpha] J) \; \text{ is valid}.\]
    
    Then, by the \dL loop proof rule (for the same $\mu_{\Sigma \cupm \Omega \cupm \Theta}$):
    
    \[\mu_{\Sigma \cupm \Omega \cupm \Theta}(\Gamma \vdash [\alpha^*]P,\Delta) \; \text{ is valid}.\]
    
    By the freshness of $\vec{l}$ and from lemma~\ref{lem: minus_muts}, we have:
    
    \[\mu_{(\Sigma \cupm \Omega \cupm \Theta) \setminus \vec{l}}(\Gamma \vdash [\alpha^*]P,\Delta) \; \text{ is valid}.\]
    
    Since $\mu_{\Sigma \cupm \Omega \cupm \Theta}$ was arbitrary, the desired result holds for \textit{any} $\mu_{(\Sigma \cupm \Omega \cupm \Theta) \setminus \vec{l}}$.

    \paragraph{Case M:WR. }
    
    Suppose the last rule in the proof tree is M:WR:
    
    \begin{mathpar}
        \quad{\text{\color{violet}M:\color{black}WR}} \;\;
        \inferrule
        { \veclabel{{i}}{\Gamma} \vdashu{\Psi}{\Sigma} \veclabel{{k}}{\Delta} }
        { \veclabel{{i}}{\Gamma} \vdashu{\Psi}{\Sigma \cupm \atoms{\any {\vec{j}}}} \veclabel{{j}}{P}, \veclabel{{k}}{\Delta} }
    \end{mathpar}
    
    Since there exists a proof of $\veclabel{{i}}{\Gamma} \vdashu{\Psi}{\Sigma \cupm \atoms{\any {\vec{j}}}} \veclabel{{j}}{P}, \veclabel{{k}}{\Delta}$, there must exist a (smaller) proof of $\veclabel{{i}}{\Gamma} \vdashu{\Psi}{\Sigma} \veclabel{{k}}{\Delta}$. From the induction hypothesis applied to the proof of this premise, we have:
    
    \[\mu_{\Sigma}(\Gamma \vdash \Delta) \; \text{ is valid}, \text{ for any $\mu_{\Sigma}$}.\]
    
    Now fix an arbitrary $\mu_{\Sigma}$.
    Then, the \dL WR rule applies and yields $\mu_{\Sigma}(\Gamma \vdash P, \Delta)$ is valid. 
    From here, it follows by lemma~\ref{lem: subset_muts} that
    
    \[\mu_{\Sigma \cupm \atoms{\any {\vec{j}}}}(\Gamma \vdash P, \Delta) \; \text{ is valid.}\]
    
    Since our choice of $\mu_{\Sigma}$ was arbitrary, the desired result holds for \textit{any} $\mu_{\Sigma \cupm \atoms{\any {\vec{j}}}}$.

    \paragraph{Cases M:WL, M:\textit{W}R, M:\textit{W}L, M:\textit{W}LR, M:CTR, M:CTL. } The proof cases for these rules follows arguments analogous to the one provided for M:WR.

     \paragraph{Cases M:$\top$R, M:$\bot$L. } The proof cases for these rules follow an argument analogous to the one provided for M:id.
    
    \paragraph{Cases M:$\left[?\right]$, M:$\left['\right]$, M:$\left[\cup\right]$, M:$\left[;\right]$, M:$\left[*\right]$, M:$\langle\cdot\rangle$, M:$K$, M:$I$, M:$V$, M:$[:=]_1$, M:$[:=]_2$, M:$[{:=}*]_1$, M:$[{:=}*]_2$, M:$\exists$, M:$DW$, M:$DI$, M:$DC$, M:$DG$, M:$DE_1$, M:$DE_2$, M:$DS$. }
    
    These axioms serve as leaf rules in the proof, and hence the arguments for these cases are analogous to the one given for M:id.

    \paragraph{Cases M:$c'$, M:$x'$, M:$+'$, M:$-'$, M:$\cdot'$, M:$/'$, M:$\circ'$. } These axioms serve as leaf rules in the proof, and hence the arguments for these cases are analogous to the one given for M:id.

    \paragraph{Case M:$\mathbb{R}$} Suppose the last rule in the proof tree is M:$\mathbb{R}$:
    
    \begin{mathpar}
        \quad{\text{\color{violet}M:\color{black}$\mathbb{R}$}} \;\;
        \inferrule
        {  }
        { \veclabel{{i}}{\Gamma} \vdashu{\Psi}{DA(\vec{i}, \vec{j}) \cupm \Psi} \veclabel{{j}}{\Delta} }
    \end{mathpar}
    
    Fix an arbitrary $\mu_{DA(\vec{i}, \vec{j})}$ and observe that the validity of $\mu_{DA(\vec{i}, \vec{j}) \cupm \Psi}(\Gamma \vdash \Delta)$ follows by lemma~\ref{lem: subset_muts}.

    \paragraph{Case M:auto} Suppose the last rule in the proof tree is M:auto:
    
    \begin{mathpar}
        \quad{\text{\color{violet}M:\color{black}auto}} \;\;
        \inferrule
        {  }
        { \veclabel{{i}}{\Gamma} \vdashu{\Psi}{DA(\vec{i}, \vec{j}) \cupm \Psi} \veclabel{{j}}{\Delta} }
    \end{mathpar}
    
    Fix an arbitrary $\mu_{DA(\vec{i}, \vec{j})}$ and observe that the validity of $\mu_{DA(\vec{i}, \vec{j}) \cupm \Psi}(\Gamma \vdash \Delta)$ follows by lemma~\ref{lem: subset_muts}.
    
    $\square$
    \end{proof}

    \begin{customthm}{2}{\textbf{Completeness of \uapc}}
    If $\Gamma \vdash \Delta$ is valid, then for all $\Psi$ there exists $\Sigma$ such that $\veclabel{{i}}{\Gamma} \vdashu{\Psi}{\Sigma} \veclabel{{j}}{\Delta}$.
    \end{customthm}
    
    \begin{proof}
        Since $\Gamma \vdash \Delta$ is valid, it has a proof in the \dL proof calculus. The argument for completeness then follows by structural induction on the proof of $\Gamma \vdash \Delta$. In each case, we observe that there is a matching (or set of matching) proof rules in the usage-aware proof calculus that each generate an output set $\Sigma$.
    
        \paragraph{Case $\land$R} Suppose the last rule in the proof tree is $\land$R:
    
        \begin{mathpar}
            \quad{\text{$\land$R}} \;\;
            \inferrule
            { {\Gamma} \vdash {P}
            \\ {\Gamma} \vdash {Q} }
            { {\Gamma} \vdash {P} \land {Q} }
        \end{mathpar}
    
        Let $\Psi$ be an arbitrary input set. Then by the induction hypothesis, there exist output sets $\Sigma$ and $\Omega$ such that $\veclabel{{k}}{\Gamma} \vdashu{\Psi}{\Sigma} \veclabel{{i}}{P}$ and $\veclabel{{k}}{\Gamma} \vdashu{\Psi}{\Omega} \veclabel{{j}}{Q}$, respectively. Therefore, it follows by the rule M:$\land$R that $\Sigma \cupm \Omega$ is an output set for $\veclabel{{k}}{\Gamma} \vdashu{\Psi}{\Sigma \cupm \Omega} \veclabel{{i}}{P} \land \veclabel{{j}}{Q}$.
        
        \begin{mathpar}
            \quad{\text{\color{violet}M:\color{black}$\land$R}} \;\;
            \inferrule
            { \veclabel{{k}}{\Gamma} \vdashu{\Psi}{\Sigma} \veclabel{{i}}{P}
            \\ \veclabel{{k}}{\Gamma} \vdashu{\Psi}{\Omega} \veclabel{{j}}{Q} }
            { \veclabel{{k}}{\Gamma} \vdashu{\Psi}{\Sigma \cupm \Omega} \veclabel{{i}}{P} \land \veclabel{{j}}{Q} }
        \end{mathpar}
    
        \paragraph{Cases cut, $\neg$R, $\lor$R, $\neg$L, $\land$L, $\lor$L, $\to$R, $\leftrightarrow$R, $\to$L, $\leftrightarrow$L, WR, WL, PR, PL, $\forall$R, $\forall$L, $\exists$R, $\exists$L, dW, dI, dC, dG, loop, MR, ML, GVR, iG, CER, CEL, CQR, CQL, CTL, US, UR, BRR, BRL, M:\textit{W}R, M:\textit{W}L, M:\textit{W}LR, M:cutR, M:cutL, M:$\leftrightarrow$cR, M:$\leftrightarrow$cL, M:$\rightarrow{2}\leftrightarrow$} The proof follows an argument analogous to the one presented for $\land$R.
    
        \paragraph{Case $[:=]$} Suppose the last rule in the proof tree is $[:=]$:
    
        \begin{mathpar}
            \quad{[:=]} \;\; \vdash [{x:=e}] {p(x)} \leftrightarrow {p(e)} \\
        \end{mathpar}
    
        Let $\Psi$ be an arbitrary input set. The proof proceeds in two subcases, where $x$ appears in the postcondition (i.e. $x \in FV(p(x))$) and where $x$ does not appear in the postcondition (i.e. $x \notin FV(p)$).
    
        \paragraph{Subcase $x \in FV(p(x))$} If $x \in FV(p(x))$, then it follows by M:$[:=]_1$ that $\{j_{\mId}, \vec{k}_{\mR}\} \cupm \Psi$ is an output set for $\vdashu{\Psi}{\{j_{\mId}, \vec{k}_{\mR}\} \cupm \Psi} [\atomlabel{j}{x:=e}] \;\veclabel{{k}}{p(x)} \leftrightarrow \veclabel{{k}}{p(e)}$.
    
        \begin{mathpar}
            \quad{\text{\color{violet}M:\color{black}$[:=]_1$}} \;\; \vdashu{\Psi}{\{j_{\mId}, \vec{k}_{\mR}\} \cupm \Psi} [\atomlabel{j}{x:=e}] \;\veclabel{{k}}{p(x)} \leftrightarrow \veclabel{{k}}{p(e)} \;\; \quad{x \in FV(p(x))}
        \end{mathpar}
    
        \paragraph{Subcase $x \notin FV(p)$} If $x \notin FV(p)$, then it follows by M:$[:=]_2$ that $\{j_{\mR}, \vec{k}_{\mR}\} \cupm \Psi$ is an output set for $\vdashu{\Psi}{\{j_{\mR}, \vec{k}_{\mR}\} \cupm \Psi} [\atomlabel{j}{x:=e}] \;\veclabel{{k}}{p} \leftrightarrow \veclabel{{k}}{p}$.
    
        \begin{mathpar}
            \quad{\text{\color{violet}M:\color{black}$[:=]_2$}} \;\; \vdashu{\Psi}{{\{j_{\mR}, \vec{k}_{\mR}\} \cupm \Psi}} [\atomlabel{j}{x:=e}] \;\veclabel{{k}}{p} \leftrightarrow \veclabel{{k}}{p} \;\; \quad{x \notin FV(p)}
        \end{mathpar}
    
        \paragraph{Cases $[{:}{*}]$, DE} The proof follows an argument analogous to the one presented for $[:=]$.
    
        \paragraph{Case id} Suppose the last rule in the proof tree is id:
    
        \begin{mathpar}
            \quad{\text{id}} \;\;
            \inferrule
            {  }
            { {P}, {\Gamma} \vdash {P}, {\Delta} }
        \end{mathpar} 
    
        Let $\Psi$ be an arbitrary input set. Therefore, it follows by the rule M:id that $\{\any{\fuse{\vec{i}}{\vec{j}}}, \any {\vec{k}}, \any {\vec{l}}\} \cupm \Psi$ is an output set for $\veclabel{{i}}{P}, \veclabel{{k}}{\Gamma} \vdashu{\Psi}{\{\any{\fuse{\vec{i}}{\vec{j}}}, \any {\vec{k}}, \any {\vec{l}}\} \cupm \Psi } \veclabel{{j}}{P}, \veclabel{{l}}{\Delta}$.
        
        \begin{mathpar}
            \quad{\text{\color{violet}M:\color{black}id}} \;\;
            \inferrule
            {  }
            { \veclabel{{i}}{P}, \veclabel{{k}}{\Gamma} \vdashu{\Psi}{\{\any{\fuse{\vec{i}}{\vec{j}}}, \any {\vec{k}}, \any {\vec{l}}\} \cupm \Psi } \veclabel{{j}}{P}, \veclabel{{l}}{\Delta} }
        \end{mathpar}

        \paragraph{Cases $\mathbb{R}$, $\bot$L, $\top$R, $[?]$, $[;]$, $\langle \cdot \rangle$, K, $[\cup]$, $[']$, $[*]$, I, V, DW, DI, DC, DG, $\forall i$, $\forall \rightarrow$, $V_\forall$, $\exists$, DS, $c'$, $x'$, $+'$, $-'$, $\cdot'$, $/'$, $\circ'$} The proof follows an argument analogous to the one presented for id.
    
        \paragraph{Case dW} We assume that all \dL proofs are given without use of derived rules and that wherever used, these derived rules are replaced with their derivations in the \dL proof calculus. Since the rule dW is derivable from the differential equation axiom DW, the completeness for this case follows from the cases given above.

        \paragraph{Cases dI, dC, dG}  The proof follows an argument analogous to the one presented for dW.
    $\square$    
    \end{proof}

\section{$\uapcplus$ Proof Calculus}\label{app:uapcpluscalculus}

In this section, we give the rules of the \uapcplus proof calculus. We present only the rules that change and note that all others are the same as those in \uapc. 

\subsection{$\uapcplus$ Propositional Sequent Calculus Rules}

    \begin{mathpar}
    \quad{\text{\color{violet}M$^+$:\color{black}$\land$R}} \;\;
    \inferrule
    { \veclabel{{k}}{\Gamma} \vdashu{\Psi}{\Sigma} \veclabel{{i}}{P}
    \\ \veclabel{{k}}{\Gamma} \vdashu{\Sigma}{\Omega} \veclabel{{j}}{Q} }
    { \veclabel{{k}}{\Gamma} \vdashu{\Psi}{\Omega} \veclabel{{i}}{P} \land \veclabel{{j}}{Q} }
\and
    \quad{\text{\color{violet}M$^+$:\color{black}cut}^{\color{red}\vec{l}\color{black}}} \;\;
        \inferrule
        { \veclabel{{i}}{\Gamma}, \veclabel{{j}}{C} \vdashu{\Psi}{\Sigma} \veclabel{{k}}{\Delta}
        \\ 
        \veclabel{{i}}{\Gamma} \vdashu{\Sigma}{\Omega} \veclabel{{k}}{\Delta}, \veclabel{{j}}{C}
        }
        { \veclabel{{i}}{\Gamma} \vdashu{\Psi}{\Omega \setminus {\vec{l}}} \veclabel{{k}}{\Delta} }
        \quad{(\forall C_a \in \text{atoms}(C). \, \color{red}j_a \color{black}= \phifunc{}{C_a}{(\veclabel{{i}}{\Gamma}, \veclabel{{k}}{\Delta})}{\color{red}\vec{l}\color{black}})} 
\and
    \quad{\text{\color{violet}M$^+$:\color{black}$\lor$L}} \;\;
    \inferrule
    { \veclabel{{i}}{P}, \veclabel{{k}}{\Gamma} \vdashu{\Psi}{\Sigma} \veclabel{{l}}{\Delta}
    \\ \veclabel{{j}}{Q}, \veclabel{{k}}{\Gamma} \vdashu{\Sigma}{\Omega} \veclabel{{l}}{\Delta} }
    { \veclabel{{i}}{P} \lor \veclabel{{j}}{Q}, \veclabel{{k}}{\Gamma} \vdashu{\Psi}{\Omega} \veclabel{{l}}{\Delta} }
\and
    \quad{\text{\color{violet}M$^+$:\color{black}$\leftrightarrow$R}} \;\;
    \inferrule
    { \veclabel{{k}}{\Gamma}, \veclabel{{i}}{P} \vdashu{\Psi}{\Sigma} \veclabel{{l}}{\Delta}, \veclabel{{j}}{Q}
    \\ \veclabel{{k}}{\Gamma}, \veclabel{{j}}{Q} \vdashu{\Sigma}{\Omega} \veclabel{{l}}{\Delta}, \veclabel{{i}}{P} }
    { \veclabel{{k}}{\Gamma} \vdashu{\Psi}{\Omega} \veclabel{{i}}{P} \leftrightarrow \veclabel{{j}}{Q}, \veclabel{{l}}{\Delta} }
\and    
    \quad{\text{\color{violet}M$^+$:\color{black}$\to$L}} \;\;
    \inferrule
    { \veclabel{{k}}{\Gamma} \vdashu{\Psi}{\Sigma} \veclabel{{l}}{\Delta}, \veclabel{{i}}{P}
    \\ \veclabel{{j}}{Q}, \veclabel{{k}}{\Gamma} \vdashu{\Sigma}{\Omega} \veclabel{{l}}{\Delta} }
    { \veclabel{{i}}{P} \to \veclabel{{j}}{Q}, \veclabel{{k}}{\Gamma} \vdashu{\Psi}{\Omega} \veclabel{{l}}{\Delta} }
\and    
    \quad{\text{\color{violet}M$^+$:\color{black}$\leftrightarrow$L}} \;\;
    \inferrule
    { \veclabel{{i}}{P} \land \veclabel{{j}}{Q}, \veclabel{{k}}{\Gamma} \vdashu{\Psi}{\Sigma} \veclabel{{l}}{\Delta}
    \\ \neg \veclabel{{i}}{P} \land \neg \veclabel{{j}}{Q}, \veclabel{{k}}{\Gamma} \vdashu{\Sigma}{\Omega} \veclabel{{l}}{\Delta} }
    { \veclabel{{i}}{P} \leftrightarrow \veclabel{{j}}{Q}, \veclabel{{k}}{\Gamma} \vdashu{\Psi}{\Omega} \veclabel{{l}}{\Delta} }
    \end{mathpar}

\subsection{\uapcplus Derived Rules}

In this section, we give the derived rules of \uapcplus. The rules M$^+$:cutR and M$^+$:cutL are similar to the rule M$^+$:cut, and all other rules in this section are identical to their counterparts in \uapc. 


\begin{mathpar}   
    \quad{\text{\color{violet}M$^+$:\color{black}cutR}^{\color{red}\vec{j}}} \;\;
    \inferrule
    { \veclabel{{i}}{\Gamma} \vdashu{\chi}{\Omega} \veclabel{{j}}{Q}, \veclabel{{k}}{\Delta}
    \\ \veclabel{{i}}{\Gamma} \vdashu{\Omega}{\Sigma} \veclabel{{j}}{Q} \to \veclabel{{l}}{P}, \veclabel{{k}}{\Delta}}
    { \veclabel{{i}}{\Gamma} \vdashu{\chi}{\Sigma \setminus {\vec{j}}} \veclabel{{l}}{P}, \veclabel{{k}}{\Delta} }
\and
    \quad{\text{\color{violet}M$^+$:\color{black}cutL}^{\color{red}\vec{i}}} \;\;
    \inferrule
    { \veclabel{{i}}{Q}, \veclabel{{j}}{\Gamma} \vdashu{\chi}{\Omega} \veclabel{{k}}{\Delta}
    \\ \veclabel{{j}}{\Gamma} \vdashu{\Omega}{\Sigma} \veclabel{{k}}{\Delta}, \veclabel{{l}}{P} \to \veclabel{{i}}{Q}}
    { \veclabel{{l}}{P}, \veclabel{{j}}{\Gamma} \vdashu{\chi}{\Sigma \setminus {\vec{i}}} \veclabel{{k}}{\Delta} }
\end{mathpar}

\subsection{\uapcplus Quantifier Sequent Calculus Proof Rules}

The quantifier sequent calculus proof rules of \uapcplus are identical to those of \uapc since each rule has only one premise.

    
    
    

\subsection{\uapcplus Sequent Calculus Proof Rules}

    \begin{mathpar}
        \centering
\and 
\quad{\text{\color{violet}M$^+$:\color{black}loop}^{\color{red}\vec{l}\color{black}}} \;\;
        \inferrule
        { \veclabel{{k}}{\Gamma} \vdashu{\Psi}{\Sigma} \veclabel{{n}}{J}, \veclabel{{i}}{\Delta}
        \\ \veclabel{{n}}{J} \vdashu{\Sigma}{\Omega} \veclabel{{j}}{P}
        \\ \veclabel{{n}}{J} \vdashu{\Omega}{\Theta} [\veclabel{{m}}{\alpha}] \veclabel{{n}}{J}
        }
        { \veclabel{{k}}{\Gamma} \vdashu{\Psi}{\Theta \setminus {\vec{l}}} [\veclabel{{m}}{\alpha}^*]\veclabel{{j}}{P},\veclabel{{i}}{\Delta} }
        \quad{(\forall J_a \in \text{atoms}(J). \, \color{red}n_a \color{black}= \phifunc{}{J_a}{(\veclabel{{k}}{\Gamma}, \veclabel{{i}}{\Delta})}{\color{red}\vec{l}\color{black}})} 
\and   
    \quad{\text{\color{violet}M$^+$:\color{black}MR}^{\color{red}\vec{k}\color{black}}} \;\;
    \inferrule
    { \veclabel{{i}}{\Gamma} \vdashu{\Psi}{\Sigma} \left[\veclabel{{j}}{\alpha}\right] \veclabel{{k}}{Q}, \veclabel{{l}}{\Delta} 
    \\ \veclabel{{k}}{Q} \vdashu{\Sigma}{\Omega} \veclabel{{m}}{P} }
    { \veclabel{{i}}{\Gamma} \vdashu{\Psi}{\Omega \setminus {\vec{k}}} \left[\veclabel{{j}}{\alpha}\right] \veclabel{{m}}{P}, \veclabel{{l}}{\Delta} }
\and    
    \quad{\text{\color{violet}M$^+$:\color{black}ML}^{\color{red}\vec{l}\color{black}}} \;\;
    \inferrule
    { \veclabel{{i}}{\Gamma}, \left[\veclabel{{j}}{\alpha}\right] \veclabel{{k}}{Q} \vdashu{\Psi}{\Sigma} \veclabel{{l}}{\Delta}
    \\ \veclabel{{m}}{P} \vdashu{\Sigma}{\Omega} \veclabel{{k}}{Q} }
    { \veclabel{{i}}{\Gamma}, \left[\veclabel{{j}}{\alpha}\right] \veclabel{{m}}{P} \vdashu{\Psi}{\Omega \setminus {\vec{k}}} \veclabel{{l}}{\Delta} }
\and 
        \quad{\text{\color{violet}M$^+$:\color{black}CER}^{\color{red}\vec{n}\color{black}}} \;\;
    \inferrule
    { \veclabel{{i}}{\Gamma} \vdashu{\Psi}{\Omega} \veclabel{ m}{C(\veclabel{j}{Q})}, \veclabel{{k}}{\Delta}
    \\ \vdashu{\Omega}{\Sigma} \veclabel{{j}}{Q} \leftrightarrow \veclabel{{l}}{P} }
    { \veclabel{{i}}{\Gamma} \vdashu{\Psi}{\Sigma \setminus {\vec{l}}} \veclabel{ m}{C(\veclabel{{l}}{P})}, \veclabel{{k}}{\Delta} }
    \quad{(\forall Q_a \in \text{atoms}(Q). \, \color{red}j_a \color{black}= \phifunc{}{Q_a}{\veclabel{{l}}{P}}{\color{red}\vec{n}\color{black}})} 
\and
    \quad{\text{\color{violet}M$^+$:\color{black}CEL}^{\color{red}\vec{n}\color{black}}} \;\;
    \inferrule
    { \veclabel{{i}}{\Gamma}, \veclabel{ m}{C(\veclabel{j}{Q})} \vdashu{\Psi}{\Omega} \veclabel{{k}}{\Delta}
    \\ \vdashu{\Omega}{\Sigma} \veclabel{{j}}{Q} \leftrightarrow \veclabel{{l}}{P} }
    { \veclabel{{i}}{\Gamma}, \veclabel{ m}{C(\veclabel{{l}}{P})} \vdashu{\Psi}{\Sigma \setminus {\vec{l}}} \veclabel{{k}}{\Delta} }
    \quad{(\forall Q_a \in \text{atoms}(Q). \, \color{red}j_a \color{black}= \phifunc{}{Q_a}{\veclabel{{l}}{P}}{\color{red}\vec{n}\color{black}})} 
\and
    \quad{\text{\color{violet}M$^+$:\color{black}CQR}^{\color{red}{l}\color{black}}} \;\;
    \inferrule
    { \veclabel{{i}}{\Gamma} \vdashu{\Psi}{\Omega} \veclabel{{j}}{p(k)}, \veclabel{{m}}{\Delta}
    \\ \vdashu{\Omega}{\Sigma} \atomlabel{{l}}{k = e}}
    { \veclabel{{i}}{\Gamma} \vdashu{\Psi}{\Sigma} \veclabel{{j}}{p(e)}, \veclabel{{m}}{\Delta} }
\and
    \quad{\text{\color{violet}M$^+$:\color{black}CQL}^{\color{red}{l}\color{black}}} \;\;
    \inferrule
    { \veclabel{{i}}{\Gamma}, \veclabel{{j}}{p(k)} \vdashu{\Psi}{\Omega} \veclabel{{m}}{\Delta}
    \\ \vdashu{\Omega}{\Sigma} \atomlabel{{l}}{k = e} }
    { \veclabel{{i}}{\Gamma}, \veclabel{{j}}{p(e)} \vdashu{\Psi}{\Sigma} \veclabel{{m}}{\Delta} }
    \end{mathpar}

\subsection{\uapcplus Axioms, Differential Axioms, and First-order Axioms}

The axioms are identical to those of \uapc since these are essentially leaf rules.

\subsection{Metatheory}

\begin{theorem}[\textbf{Soundness of \uapcplus}]
    If $\veclabel{{i}}{\Gamma} \vdashu{\Psi}{\Sigma} \veclabel{{j}}{\Delta}$, then for all $\mu_\Sigma$, $\mu_{\Sigma}(\Gamma \vdash \Delta)$ is valid.
\end{theorem}

\begin{sketch}
    The argument follows by lemma~\ref{lem: subset_muts} and theorem~\ref{thm:soundness}. $\square$
\end{sketch}

\begin{theorem}[\textbf{Completeness of \uapcplus}]
    If $\Gamma \vdash \Delta$ is valid, then for all $\Psi$ there exists $\Sigma$ such that $\veclabel{{i}}{\Gamma} \vdashu{\Psi}{\Sigma} \veclabel{{j}}{\Delta}$.
\end{theorem}

\begin{sketch}
    The argument follows by lemma~\ref{lem: subset_muts} and theorem~\ref{thm:completeness}. $\square$
\end{sketch}

\newpage

\subsection{Extension of \uapc to Proof Diagnostics}\label{apx:proof_diag}

In this section, we extend the \uapc proof calculus with the capability of tracking cut diagnostics. We accomplish this by retaining the fresh labels of cut atoms which we would normally discard in \uapc. We list only the rules that change (which are the ones that introduce fresh labels), and note that all other rules remain the same.

\begin{mathpar}
    \quad{\text{\color{violet}M$_\texttt{d}$:\color{black}loop}^{\color{red}\vec{l}\color{black}}} \;\;
    \inferrule
    { \veclabel{{k}}{\Gamma} \vdashu{\Psi}{\Sigma} \veclabel{{n}}{J}, \veclabel{{i}}{\Delta}
    \\ \veclabel{{n}}{J} \vdashu{\Psi}{\Omega} \veclabel{{j}}{P}
    \\ \veclabel{{n}}{J} \vdashu{\Psi}{\Theta} [\veclabel{{m}}{\alpha}] \veclabel{{n}}{J}
    }
    { \veclabel{{k}}{\Gamma} \vdashu{\Psi}{\Sigma \cupm \Omega \cupm \Theta} [\veclabel{{m}}{\alpha}^*]\veclabel{{j}}{P},\veclabel{{i}}{\Delta} }
    \quad{(\forall J_a \in \text{atoms}(J). \, \color{red}n_a \color{black}= \phifunc{}{J_a}{(\veclabel{{k}}{\Gamma}, \veclabel{{i}}{\Delta})}{\color{red}\vec{l}\color{black}})} 
    \and
    \quad{\text{\color{violet}M$_\texttt{d}$:\color{black}cut}^{\color{red}\vec{l}\color{black}}} \;\;
    \inferrule
    { \veclabel{{i}}{\Gamma}, \veclabel{{j}}{C} \vdashu{\Psi}{\Sigma} \veclabel{{k}}{\Delta}
    \\ 
    \veclabel{{i}}{\Gamma} \vdashu{\Psi}{\Omega} \veclabel{{k}}{\Delta}, \veclabel{{j}}{C}
    }
    { \veclabel{{i}}{\Gamma} \vdashu{\Psi}{\Sigma \cupm \Omega} \veclabel{{k}}{\Delta} }
    \quad{(\forall C_a \in \text{atoms}(C). \, \color{red}j_a \color{black}= \phifunc{}{C_a}{(\veclabel{{i}}{\Gamma}, \veclabel{{k}}{\Delta})}{\color{red}\vec{l}\color{black}})} 
    \and
    \quad{\text{\color{violet}M:\color{black}MR}^{\color{red}\vec{k}\color{black}}} \;\;
    \inferrule
    { \veclabel{{i}}{\Gamma} \vdashu{\Psi}{\Sigma} \left[\veclabel{{j}}{\alpha}\right] \veclabel{{k}}{Q}, \veclabel{{l}}{\Delta} 
    \\ \veclabel{{k}}{Q} \vdashu{\Psi}{\Omega} \veclabel{{m}}{P} }
    { \veclabel{{i}}{\Gamma} \vdashu{\Psi}{\Sigma \cupm \Omega} \left[\veclabel{{j}}{\alpha}\right] \veclabel{{m}}{P}, \veclabel{{l}}{\Delta} }
\and    
    \quad{\text{\color{violet}M$_\texttt{d}$:\color{black}ML}^{\color{red}\vec{l}\color{black}}} \;\;
    \inferrule
    { \veclabel{{i}}{\Gamma}, \left[\veclabel{{j}}{\alpha}\right] \veclabel{{k}}{Q} \vdashu{\Psi}{\Sigma} \veclabel{{l}}{\Delta}
    \\ \veclabel{{m}}{P} \vdashu{\Psi}{\Omega} \veclabel{{k}}{Q} }
    { \veclabel{{i}}{\Gamma}, \left[\veclabel{{j}}{\alpha}\right] \veclabel{{m}}{P} \vdashu{\Psi}{\Sigma \cupm \Omega} \veclabel{{l}}{\Delta} }
    %
    \and
    \quad{\text{\color{violet}M$_\texttt{d}$:\color{black}CER}^{\color{red}\vec{n}\color{black}}} \;\;
    \inferrule
    { \veclabel{{i}}{\Gamma} \vdashu{\Psi}{\Omega} \veclabel{ m}{C(\veclabel{j}{Q})}, \veclabel{{k}}{\Delta}
    \\ \vdashu{\Psi}{\Sigma} \veclabel{{j}}{Q} \leftrightarrow \veclabel{{l}}{P} }
    { \veclabel{{i}}{\Gamma} \vdashu{\Psi}{\Omega \cupm \Sigma} \veclabel{ m}{C(\veclabel{{l}}{P})}, \veclabel{{k}}{\Delta} }
    \quad{(\forall Q_a \in \text{atoms}(Q). \, \color{red}j_a \color{black}= \phifunc{}{Q_a}{\veclabel{{l}}{P}}{\color{red}\vec{n}\color{black}})} 
\and
    \quad{\text{\color{violet}M$_\texttt{d}$:\color{black}CEL}^{\color{red}\vec{n}\color{black}}} \;\;
    \inferrule
    { \veclabel{{i}}{\Gamma}, \veclabel{ m}{C(\veclabel{j}{Q})} \vdashu{\Psi}{\Omega} \veclabel{{k}}{\Delta}
    \\ \vdashu{\Psi}{\Sigma} \veclabel{{j}}{Q} \leftrightarrow \veclabel{{l}}{P} }
    { \veclabel{{i}}{\Gamma}, \veclabel{ m}{C(\veclabel{{l}}{P})} \vdashu{\Psi}{\Omega \cupm \Sigma} \veclabel{{k}}{\Delta} }
    \quad{(\forall Q_a \in \text{atoms}(Q). \, \color{red}j_a \color{black}= \phifunc{}{Q_a}{\veclabel{{l}}{P}}{\color{red}\vec{n}\color{black}})} 
    \and
    \quad{\text{\color{violet}M$_\texttt{d}$:\color{black}CQR}^{\color{red}\vec{l}\color{black}}} \;\;
    \inferrule
    { \veclabel{{i}}{\Gamma} \vdashu{\chi}{\Sigma} \veclabel{{j}}{p(k)}, \veclabel{{m}}{\Delta}
    \\ \vdashu{\chi}{\Omega} \atomlabel{{l}}{k = e} }
    { \veclabel{{i}}{\Gamma} \vdashu{\chi}{\Sigma} \veclabel{{j}}{p(e)}, \veclabel{{m}}{\Delta} }
    \quad{( l \text{ fresh} )}
    \and
    \quad{\text{\color{violet}M$_\texttt{d}$:\color{black}CQL}^{\color{red}\vec{l}\color{black}}} \;\;
    \inferrule
    { \veclabel{{i}}{\Gamma}, \veclabel{{j}}{p(k)} \vdashu{\chi}{\Omega} \veclabel{{m}}{\Delta}
    \\ \vdashu{\chi}{\Sigma} \atomlabel{{l}}{k = e}     
    }
    { \veclabel{{i}}{\Gamma}, \veclabel{{j}}{p(e)} \vdashu{\chi}{\Omega} \veclabel{{m}}{\Delta} }
    \quad{( l \text{ fresh})}
\end{mathpar}

\subsubsection{Metatheory}

The soundness and completeness theorems follow from \uapc, with slight modifications to the argument in the cases noted above.

\begin{theorem}[\textbf{Soundness}]\label{apx:thm:soundness_uapc_diag}
If $\veclabel{{i}}{\Gamma} \vdashu{\Psi}{\Sigma} \veclabel{{j}}{\Delta}$ is valid, then $\mu_{\color{blue}\Sigma}(\Gamma \vdash \Delta)$ is valid for all $\mu_{\color{blue}\Sigma}$.
\end{theorem}

\begin{proof}
    The proof is similar to the one for \uapc. We sketch an example case, the canonical one being \color{violet}M$_\texttt{d}$:\color{black}cut:

\paragraph{Case \color{violet}M$_\texttt{d}$:\color{black}cut. } Suppose the last rule in the proof tree is \color{violet}M$_\texttt{d}$:\color{black}cut:
    
    \begin{mathpar}
    \quad{\text{\color{violet}M$_\texttt{d}$:\color{black}cut}^{\color{red}\vec{l}\color{black}}} \;\;
    \inferrule
    { \veclabel{{i}}{\Gamma}, \veclabel{{j}}{C} \vdashu{\Psi}{\Sigma} \veclabel{{k}}{\Delta}
    \\ 
    \veclabel{{i}}{\Gamma} \vdashu{\Psi}{\Omega} \veclabel{{k}}{\Delta}, \veclabel{{j}}{C}
    }
    { \veclabel{{i}}{\Gamma} \vdashu{\Psi}{\Sigma \cupm \Omega} \veclabel{{k}}{\Delta} }
    \quad{(\forall C_a \in \text{atoms}(C). \, \color{red}j_a \color{black}= \phifunc{}{C_a}{(\veclabel{{i}}{\Gamma}, \veclabel{{k}}{\Delta})}{\color{red}\vec{l}\color{black}})}
    \end{mathpar}
    
    Since there exists a proof of $\veclabel{{i}}{\Gamma} \vdashu{\Psi}{\Sigma \cupm \Omega} \veclabel{{k}}{\Delta}$, there must exist (smaller) proofs of $\veclabel{{i}}{\Gamma}, \veclabel{{j}}{C} \vdashu{\Psi}{\Sigma} \veclabel{{k}}{\Delta}$ and $\veclabel{{i}}{\Gamma} \vdashu{\Psi}{\Omega} \veclabel{{k}}{\Delta}, \veclabel{{j}}{C}$. 
    So, by the induction hypothesis applied to the proofs of these premises, we have:
    
    \begin{itemize}
        \item $\mu_{\Sigma}(\Gamma, C \vdash \Delta)$ is valid, for any $\mu_{\Sigma}$.
        \item $\mu_{\Omega}(\Gamma \vdash \Delta, C)$ is valid, for any $\mu_{\Omega}$.
    \end{itemize}
    
    Now, we want to show that $\mu_{\Sigma \cupm \Omega}(\Gamma \vdash \Delta)$ is valid. Since $\mu_{\Sigma}(\Gamma, C \vdash \Delta)$ is valid, it follows by lemma~\ref{lem: subset_muts} that for a fixed arbitrary $\mu_{\Sigma \cupm \Omega}$:
    
    \[\mu_{\Sigma \cupm \Omega}(\Gamma, C \vdash \Delta) \; \text{ is valid}.\]
    
    From $\mu_{\Omega}(\Gamma \vdash \Delta, C)$, it follows by lemma~\ref{lem: subset_muts} that (for the same $\mu_{\Sigma \cupm \Omega}$):
    
    \[\mu_{\Sigma \cupm \Omega}(\Gamma \vdash \Delta, C) \; \text{ is valid}.\]
    
    Then, by the \dL proof rule cut (for the same $\mu_{\Sigma \cupm \Omega}$):
    
    \[\mu_{\Sigma \cupm \Omega}(\Gamma \vdash \Delta) \; \text{ is valid}.\]
    
    This is the desired result.
    
\end{proof}

\begin{theorem}[\textbf{Completeness}]\label{apx:thm:completeness_uapc_diag}
If $\Gamma \vdash \Delta$ is valid, then for all $\color{blue}\Psi$ there exists a $\color{blue}\Sigma$ such that $\veclabel{{i}}{\Gamma} \vdashu{\Psi}{\Sigma} \veclabel{{j}}{\Delta}$ is valid.
\end{theorem}

\begin{proof}
    The proof is similar to the one for \uapc. 
\end{proof}

\subsubsection{Diagnostics for the Parachute Example}

The need for such diagnostics arises in the parachute example in Fig.~\ref{fig:parachute_ex}. By inspecting the branch (3a) output $\color{blue}\Theta$, any mutation (including removal) is admissible for $\color{red}u_1$. Thus, $\color{red}u_1$, which was freshly cut into the proof via \color{violet}M$_\texttt{d}$:\color{black}loop, was unnecessary in this proof branch. In fact, this is the case overall if we were to look at the output sets of all three branches (1), (2), (3). In this extension, $\Xi$ indicates that the atom labeled by $\color{red}u_1$ was an unnecessary cut in the proof. Soundness guarantees that the removal of this atom maintains the proof structure. 

%% file: Sections/first_order_axioms.tex
\subsection{\uapc First-Order Axioms}

The first-order axioms provide mechanisms to handle quantifiers in a proof. When applied, the axiom $\forall i$ eliminates the universal quantifier from premise to conclusion by instantiating $x$ with $e$. The axiom propagates the binding $\forall x$ to each of the formulas $P$ and $Q$. The axiom $V_\forall$ wraps a formula that does not mention $x$ in the binding $\forall x$, and the axiom $\exists$ translates between universal and existential quantifiers.
\begin{align*}
    \quad{\text{\color{violet}M:\color{black}$\forall i$}} \;\; &\vdashu{\chi}{\any{\vec{j}} \cupm \chi} (\forall x \; \veclabel{{j}}{p(x)}) \to \veclabel{{j}}{p(e)} \\
    \quad{\text{\color{violet}M:\color{black}$\forall \to$}} \;\; &\vdashu{\chi}{\{\any{\vec{j}}, \any{\vec{k}}\} \cupm \chi} \forall x \; (\veclabel{{j}}{P} \to \veclabel{{k}}{Q}) \to (\forall x \; \veclabel{{j}}{P} \to \forall x \; \veclabel{{k}}{Q}) \\
    \quad{\text{\color{violet}M:\color{black}$V_\forall$}} \;\; &\vdashu{\chi}{\any{\vec{j}} \cupm \chi} \veclabel{{j}}{p} \to \forall x \; \veclabel{{j}}{p} \; \quad{(x \notin FV(p))} \\
    \quad{\text{\color{violet}M:\color{black}$\exists$}} \;\; &\vdashu{\chi}{\any{\vec{j}} \cupm \chi} \left(\lnot \forall x \lnot \veclabel{{j}}{P}\right) \leftrightarrow \exists x \; \veclabel{{j}}{P}
\end{align*}

%% file: Sections/diff_eq_axioms.tex
\subsection{\uapc Differential Equation Axioms}

In this section, we give the axioms that allow us to maneuver differential equations in a proof.

\begin{align*}
    \quad{\text{\color{violet}M:\color{black}DW}} \;\; &\vdashu{\chi}{\{\any{\fuse{\vec{k}}{\vec{m}}}, \any{j}, \any{\vec{l}}\} \cupm \chi} [\atomlabel{j}{x' = f(x)} \& \atomlabel{\fuse{\vec{k}}{\vec{m}}}{Q}]\veclabel{{l}}{P} \\
    &\leftrightarrow [\atomlabel{j}{x' = f(x)} \& \veclabel{{k}}{Q}](\veclabel{{m}}{Q} \to \veclabel{{l}}{P}) \\
    \quad{\text{\color{violet}M:\color{black}DI}} \;\; &\vdashu{\chi}{\{\any{\vec{j}}, \any{k}, \any{\vec{m}}, \any{\fuse{\vec l}{\vec{n}}}\} \cupm \chi} (\atomlabel{\vec m}{Q} \to [\atomlabel{k}{x' = f(x)} \& \veclabel{{j}}{Q}](\atomlabel{\fuse{\vec l}{\vec n}}{P})') \\
    &\rightarrow ([\atomlabel{k}{x' = f(x)} \& \veclabel{{j}}{Q}]\veclabel{{l}}{P} \leftrightarrow [?\atomlabel{\vec m}{Q}]\veclabel{{n}}{P}) \\
    \quad{\text{\color{violet}M:\color{black}DC}} \;\; &\vdashu{\chi}{\{\any{\fuse{i}{j}}, \any{\fuse{\vec{k}}{\vec{h}}}, \any{\vec{l}}, \vec{m}_{\mId}, \vec{n}_{\mId}\} \cupm \chi} [\atomlabel{\fuse{i}{j}}{x' = f(x)} \& \atomlabel{\fuse{\vec{k}}{\vec{h}}}{Q}]\veclabel{{l}}{C} \\
    &\rightarrow ([\atomlabel{j}{x' = f(x)} \& \veclabel{{k}}{Q}]\veclabel{{m}}{P} \leftrightarrow [\atomlabel{i}{x' = f(x)} \& \veclabel{{h}}{Q} \land \veclabel{{l}}{C}]\veclabel{{n}}{P})  \\
    \quad{\text{\color{violet}M:\color{black}DG}} \;\; &\vdashu{\chi}{\{m_{\mR}, \any{j}, \any{\vec{k}}, \any{\vec{l}}\} \cupm \chi} [\atomlabel{j}{x' = f(x)}, \atomlabel{m}{y' = a(x)y + b(x)} \& \veclabel{{k}}{Q}] \veclabel{{l}}{P} \\
    &\leftrightarrow [\atomlabel{j}{x' = f(x)} \& \veclabel{{k}}{Q}] \veclabel{{l}}{P} \\
    \quad{\text{\color{violet}M:\color{black}DE$_1$}} \;\; &\vdashu{\chi}{\{\any{\fuse{\vec{j}}{\vec{m}}}, \any{\vec{k}}, \any{\vec{l}}\} \cupm \chi} [\atomlabel{j}{x' = f(x)} \& \veclabel{{k}}{Q}][\atomlabel{m}{x' := f(x)}] \veclabel{{l}}{p(x)} \\
    &\leftrightarrow [\atomlabel{\fuse{\vec{j}}{\vec{m}}}{x' = f(x)} \& \veclabel{{k}}{Q}] \veclabel{{l}}{p(x)} \;\; &\quad{(x' \notin FV(p))} \\
    \quad{\text{\color{violet}M:\color{black}DE$_2$}} \;\; &\vdashu{\chi}{\{{\fuse{\vec{j}}{\vec{m}}}_{\mId}, \any{\vec{k}}, \any{\vec{l}}\} \cupm \chi} [\atomlabel{j}{x' = f(x)} \& \veclabel{{k}}{Q}][\atomlabel{m}{x' := f(x)}] \veclabel{{l}}{p(x)} \\
    &\leftrightarrow [\atomlabel{\fuse{\vec{j}}{\vec{m}}}{x' = f(x)} \& \veclabel{{k}}{Q}] \veclabel{{l}}{p(x)} \;\; &\quad{(x' \in FV(p(x)))} \\
\end{align*}

For example, the axioms M:DE$_1$ and M:DE$_2$ translate between differential equations and variable assignments. We define differential assignment in two axioms that differ depending on whether $x$ is mentioned in $FV(p)$. If $x \in FV(p(x))$, M:DE$_2$ enforces that the only admissible mutation for the differential equation and differential assignment is the identity mutation, i.e. these should remain untouched. Alternatively, if $x \notin FV(p)$, then the postcondition must be independent of the differential equation and differential assignment, so we allow them to take on any mutation. The reasoning behind these axioms is analogous to the discussion for axioms M:$[:=]_1$ and M:$[:=]_2$.

The axiom M:DW rephrases the differential equation by extracting the evolution domain constraint as an implication of the postcondition. If the evolution domain constraint is false, then the postcondition holds. No usage information can be discerned from this rule alone so the output set allows labels to take on any mutation. 

The axiom M:DI allows us to reason about differential invariants which can be helpful in proving properties about systems with differential equations. The test $[?Q]P$ holds if and only if the postcondition is valid after running the ODE. Similar to a loop invariant, the left-hand side of the implication asserts that the induction step can assume that the domain constraint holds. If the evolution domain constraint is initially false, then vacuously $[x' = f(x) \& Q](P)'$ holds.

The axiom M:DC is similar to the cut rule in that it cuts a formula $C$ into the evolution domain constraint and claims the result to be equivalent to the original formula provided the postcondition $C$ holds after running the ODE over $Q$. Since the original postcondition $P$ disappears moving from conclusion to premise in the proof, the proof calculus enforces that the labels on $\vec{m}$ and $\vec{n}$ remain unmutated to ensure validity. 

The axiom M:DG allows us to introduce a new differential equation, or differential ghost, that facilitates the proof. The variable $y$ should be new and not appear in the right-hand side or in $a(x)$ or in $b(x)$. 

%% file: Sections/differential_axioms.tex
\subsection{\uapc Differential Axioms} 

This section provides the differential axioms of \uapc. Analogous to the \uapc axioms, these axioms can be used in a proof with rules M:CQR and M:CQL. 
\begin{align*}
    \quad{\text{\color{violet}M:\color{black}DS}} \;\; &\vdashu{\chi}{\chi \cupm \{i_{\mR}, \vec{j}_{\mR}, \vec{k}_{\mR}, l_{\mId}, m_{\mId}\}} \left([\veclabel{i}x' = c() \& \veclabel{{j}}q(x)] \veclabel{{k}}p(x)\right) \\
    & \;\;\;\;\;\; \leftrightarrow \left(\forall \veclabel{l}{t \geq 0} ((\forall \veclabel{m}{0 \leq s \leq t} \; \veclabel{{j}}q(x + c()s)) \to [\veclabel{i}x:=x+c()t]\veclabel{{k}}p(x))\right) \\
    \quad{\text{\color{violet}M:\color{black}$c'$}} \;\; &\vdashu{\chi}{\chi \cupm \{l_{\mId}\}} \veclabel{l}{(c())' = 0} \\
    \quad{\text{\color{violet}M:\color{black}$x'$}} \;\; &\vdashu{\chi}{\chi \cupm \{l_{\mId}\}} \veclabel{l}(x)' = x' \\
    \quad{\text{\color{violet}M:\color{black}$+'$}} \;\; &\vdashu{\chi}{\chi \cupm \{l_{\mId}\}} \veclabel{l}(e+k)' = (e)'+(k)' \\
    \quad{\text{\color{violet}M:\color{black}$-'$}} \;\; &\vdashu{\chi}{\chi \cupm \{l_{\mId}\}} \veclabel{l}(e-k)' = (e)'-(k)' \\
    \quad{\text{\color{violet}M:\color{black}$\cdot'$}} \;\; &\vdashu{\chi}{\chi \cupm \{l_{\mId}\}} \veclabel{l}(e\cdot k)' = (e)'\cdot k + e\cdot(k)' \\
    \quad{\text{\color{violet}M:\color{black}$/'$}} \;\; &\vdashu{\chi}{\chi \cupm \{l_{\mId}\}} \veclabel{l}(e/k)' = ((e)'\cdot k - e\cdot(k)')/k^2 \\
    \quad{\text{\color{violet}M:\color{black}$\circ'$}} \;\; &\vdashu{\chi}{\chi \cupm \{i_{\mId}, j_{\mId}, k_{\mId}\}} [\veclabel{i}y:=g(x)][\veclabel{j}y':=1]\left(\veclabel{k}(f(g(x)))' = (f(y))' \cdot (g(x))'\right)
\end{align*}

%% file: Sections/derived_rules.tex
\subsection{\uapc Derived Rules}

These rules are derivable from the core usage-aware proof calculus. In proofs, it is often more efficient to apply these rules directly, so we extend \uapc to accommodate analysis of these rules. The rules M:\textit{W}R, M:\textit{W}L, M:\textit{W}LR are derived from the weakening rules above, and M:cutR and M:cutL are derived from M:cut. The rules M:$\leftrightarrow$cR and M:$\leftrightarrow$cL handle the symmetry of formula equivalence, and the rule M:${\rightarrow}{2}{\leftrightarrow}$ allows us to reason about equivalence in terms of implication by moving from premise to conclusion.
\begin{figure}[htp]
\begin{mathpar}
    \quad{\text{\color{violet}M:\color{black}\textit{W}R}} \;\;
    \inferrule
    { \vdashu{\chi}{\Omega} \veclabel{{j}}{P} }
    { \veclabel{{i}}{\Gamma} \vdashu{\chi}{\Omega \cupm \atoms{\any{\vec{i}}, \any{\vec{k}}}} \veclabel{{j}}{P}, \veclabel{{k}}{\Delta} }
\and    
    \quad{\text{\color{violet}M:\color{black}\textit{W}L}} \;\;
    \inferrule
    { \veclabel{{i}}{P} \vdashu{\chi}{\Omega} }
    { \veclabel{{i}}{P}, \veclabel{{j}}{\Gamma} \vdashu{\chi}{\Omega \cupm \atoms{\any{\vec{j}}, \any{\vec{k}}}} \veclabel{{k}}{\Delta} }
\and    
    \quad{\text{\color{violet}M:\color{black}\textit{W}LR}} \;\;
    \inferrule
    { \veclabel{{i}}{P} \vdashu{\chi}{\Omega} \veclabel{{k}}{Q} }
    { \veclabel{{i}}{P}, \veclabel{{j}}{\Gamma} \vdashu{\chi}{\Omega \cupm \atoms{\any{\vec{j}}, \any{\vec{l}}}} \veclabel{{k}}{Q}, \veclabel{{l}}{\Delta} }
\and    
    \quad{\text{\color{violet}M:\color{black}cutR}^{\color{red}\vec{j}}} \;\;
    \inferrule
    { \veclabel{{i}}{\Gamma} \vdashu{\chi}{\Omega} \veclabel{{j}}{Q}, \veclabel{{k}}{\Delta}
    \\ \veclabel{{i}}{\Gamma} \vdashu{\chi}{\Sigma} \veclabel{{j}}{Q} \to \veclabel{{l}}{P}, \veclabel{{k}}{\Delta}}
    { \veclabel{{i}}{\Gamma} \vdashu{\chi}{(\Omega \cupm \Sigma) \setminus {\vec{j}}} \veclabel{{l}}{P}, \veclabel{{k}}{\Delta} }
\and
    \quad{\text{\color{violet}M:\color{black}cutL}^{\color{red}\vec{i}}} \;\;
    \inferrule
    { \veclabel{{i}}{Q}, \veclabel{{j}}{\Gamma} \vdashu{\chi}{\Omega} \veclabel{{k}}{\Delta}
    \\ \veclabel{{j}}{\Gamma} \vdashu{\chi}{\Sigma} \veclabel{{k}}{\Delta}, \veclabel{{l}}{P} \to \veclabel{{i}}{Q}}
    { \veclabel{{l}}{P}, \veclabel{{j}}{\Gamma} \vdashu{\chi}{(\Omega \cupm \Sigma) \setminus {\vec{i}}} \veclabel{{k}}{\Delta} }
\and
    \quad{\text{\color{violet}M:\color{black}$\leftrightarrow$cR}} \;\;
    \inferrule
    { \veclabel{{i}}{\Gamma} \vdashu{\chi}{\Omega} \veclabel{{j}}{Q} \leftrightarrow \veclabel{{k}}{P}, \veclabel{{l}}{\Delta} }
    { \veclabel{{i}}{\Gamma} \vdashu{\chi}{\Omega} \veclabel{{k}}{P} \leftrightarrow \veclabel{{j}}{Q}, \veclabel{{l}}{\Delta} }
\and    
    \quad{\text{\color{violet}M:\color{black}$\leftrightarrow$cL}} \;\;
    \inferrule
    { \veclabel{{i}}{Q} \leftrightarrow \veclabel{{j}}{P}, \veclabel{{k}}{\Gamma} \vdashu{\chi}{\Omega} \veclabel{{l}}{\Delta} }
    { \veclabel{{j}}{P} \leftrightarrow \veclabel{{i}}{Q}, \veclabel{{k}}{\Gamma} \vdashu{\chi}{\Omega} \veclabel{{l}}{\Delta} }
\and    
    \quad{\text{\color{violet}M:\color{black}${\rightarrow}{2}{\leftrightarrow}$}} \;\;
    \inferrule
    { \veclabel{{i}}{\Gamma} \vdashu{\chi}{\Omega} \veclabel{{j}}{P} \leftrightarrow \veclabel{{k}}{Q}, \veclabel{{l}}{\Delta} }
    { \veclabel{{i}}{\Gamma} \vdashu{\chi}{\Omega} \veclabel{{j}}{P} \to \veclabel{{k}}{Q}, \veclabel{{l}}{\Delta} }
\end{mathpar}
\end{figure}

%% file: main.bbl

\begin{thebibliography}{39}


\ifx \showCODEN    \undefined \def \showCODEN     #1{\unskip}     \fi
\ifx \showDOI      \undefined \def \showDOI       #1{#1}\fi
\ifx \showISBNx    \undefined \def \showISBNx     #1{\unskip}     \fi
\ifx \showISBNxiii \undefined \def \showISBNxiii  #1{\unskip}     \fi
\ifx \showISSN     \undefined \def \showISSN      #1{\unskip}     \fi
\ifx \showLCCN     \undefined \def \showLCCN      #1{\unskip}     \fi
\ifx \shownote     \undefined \def \shownote      #1{#1}          \fi
\ifx \showarticletitle \undefined \def \showarticletitle #1{#1}   \fi
\ifx \showURL      \undefined \def \showURL       {\relax}        \fi
\providecommand\bibfield[2]{#2}
\providecommand\bibinfo[2]{#2}
\providecommand\natexlab[1]{#1}
\providecommand\showeprint[2][]{arXiv:#2}

\bibitem[che({[n.\,d.]})]%
        {cheatsheet}
 \bibinfo{year}{[n.\,d.]}\natexlab{}.
\newblock \bibinfo{title}{Keymaera X: Cheat Sheet}.
\newblock \bibinfo{howpublished}{https://keymaeraX.org/}.
\newblock


\bibitem[Alur(2015)]%
        {alur2015}
\bibfield{author}{\bibinfo{person}{Rajeev Alur}.}
  \bibinfo{year}{2015}\natexlab{}.
\newblock \bibinfo{booktitle}{\emph{Principles of Cyber-Physical Systems}}.
\newblock \bibinfo{publisher}{The MIT Press}.
\newblock
\showISBNx{0262029111}


\bibitem[Alur et~al\mbox{.}(1993)]%
        {alur1993}
\bibfield{author}{\bibinfo{person}{Rajeev Alur}, \bibinfo{person}{Costas
  Courcoubetis}, \bibinfo{person}{Thomas~A. Henzinger}, {and}
  \bibinfo{person}{Pei~Hsin Ho}.} \bibinfo{year}{1993}\natexlab{}.
\newblock \showarticletitle{Hybrid automata: An algorithmic approach to the
  specification and verification of hybrid systems}. In
  \bibinfo{booktitle}{\emph{Hybrid Systems}},
  \bibfield{editor}{\bibinfo{person}{Robert~L. Grossman}, \bibinfo{person}{Anil
  Nerode}, \bibinfo{person}{Anders~P. Ravn}, {and} \bibinfo{person}{Hans
  Rischel}} (Eds.). \bibinfo{publisher}{Springer Berlin Heidelberg},
  \bibinfo{pages}{209--229}.
\newblock
\showISBNx{978-3-540-48060-0}


\bibitem[Bartocci et~al\mbox{.}(2022)]%
        {BARTOCCI2022101254}
\bibfield{author}{\bibinfo{person}{Ezio Bartocci}, \bibinfo{person}{Roderick
  Bloem}, \bibinfo{person}{Benedikt Maderbacher}, \bibinfo{person}{Niveditha
  Manjunath}, {and} \bibinfo{person}{Dejan Ni{\v c}kovi{\'c}}.}
  \bibinfo{year}{2022}\natexlab{}.
\newblock \showarticletitle{Adaptive testing for specification coverage and
  refinement in {CPS} models}.
\newblock \bibinfo{journal}{\emph{Nonlinear Analysis: Hybrid Systems}}
  \bibinfo{volume}{46} (\bibinfo{year}{2022}), \bibinfo{pages}{101254}.
\newblock
\showISSN{1751-570X}
\urldef\tempurl%
\url{https://doi.org/10.1016/j.nahs.2022.101254}
\showDOI{\tempurl}


\bibitem[Bartocci et~al\mbox{.}(2023)]%
        {10132190}
\bibfield{author}{\bibinfo{person}{Ezio Bartocci}, \bibinfo{person}{Leonardo
  Mariani}, \bibinfo{person}{Dejan Ničković}, {and} \bibinfo{person}{Drishti
  Yadav}.} \bibinfo{year}{2023}\natexlab{}.
\newblock \showarticletitle{Property-Based Mutation Testing}. In
  \bibinfo{booktitle}{\emph{2023 IEEE Conference on Software Testing,
  Verification and Validation (ICST)}}. \bibinfo{pages}{222--233}.
\newblock
\urldef\tempurl%
\url{https://doi.org/10.1109/ICST57152.2023.00029}
\showDOI{\tempurl}


\bibitem[Bohrer et~al\mbox{.}(2018)]%
        {DBLP:conf/pldi/BohrerTMMP18}
\bibfield{author}{\bibinfo{person}{Rose Bohrer}, \bibinfo{person}{Yong~Kiam
  Tan}, \bibinfo{person}{Stefan Mitsch}, \bibinfo{person}{Magnus~O. Myreen},
  {and} \bibinfo{person}{Andr{\'{e}} Platzer}.}
  \bibinfo{year}{2018}\natexlab{}.
\newblock \showarticletitle{{VeriPhy}: Verified Controller Executables from
  Verified Cyber-Physical System Models}. In \bibinfo{booktitle}{\emph{PLDI}},
  \bibfield{editor}{\bibinfo{person}{Dan Grossman}} (Ed.).
  \bibinfo{publisher}{{ACM}}, \bibinfo{pages}{617--630}.
\newblock
\urldef\tempurl%
\url{https://doi.org/10.1145/3192366.3192406}
\showDOI{\tempurl}


\bibitem[Cervesato et~al\mbox{.}(1996)]%
        {10.1007/3-540-60983-0_5}
\bibfield{author}{\bibinfo{person}{Iliano Cervesato},
  \bibinfo{person}{Joshua~S. Hodas}, {and} \bibinfo{person}{Frank Pfenning}.}
  \bibinfo{year}{1996}\natexlab{}.
\newblock \showarticletitle{Efficient resource management for linear logic
  proof search}. In \bibinfo{booktitle}{\emph{Extensions of Logic
  Programming}}, \bibfield{editor}{\bibinfo{person}{Roy Dyckhoff},
  \bibinfo{person}{Heinrich Herre}, {and} \bibinfo{person}{Peter
  Schroeder-Heister}} (Eds.). \bibinfo{publisher}{Springer},
  \bibinfo{address}{Berlin, Heidelberg}, \bibinfo{pages}{67--81}.
\newblock
\showISBNx{978-3-540-49751-6}


\bibitem[Chirimar et~al\mbox{.}(1996)]%
        {chirimar_gunter_riecke_1996}
\bibfield{author}{\bibinfo{person}{Jawahar Chirimar}, \bibinfo{person}{Carl~A.
  Gunter}, {and} \bibinfo{person}{Jon~G. Riecke}.}
  \bibinfo{year}{1996}\natexlab{}.
\newblock \showarticletitle{Reference counting as a computational
  interpretation of linear logic}.
\newblock \bibinfo{journal}{\emph{Journal of Functional Programming}}
  \bibinfo{volume}{6}, \bibinfo{number}{2} (\bibinfo{year}{1996}),
  \bibinfo{pages}{195–244}.
\newblock
\urldef\tempurl%
\url{https://doi.org/10.1017/S0956796800001660}
\showDOI{\tempurl}


\bibitem[Chong et~al\mbox{.}(2023)]%
        {Chong2023}
\bibfield{author}{\bibinfo{person}{Stephen Chong}, \bibinfo{person}{Ruggero
  Lanotte}, \bibinfo{person}{Massimo Merro}, \bibinfo{person}{Simone Tini},
  {and} \bibinfo{person}{Jian Xiang}.} \bibinfo{year}{2023}\natexlab{}.
\newblock \showarticletitle{Quantitative Robustness Analysis of Sensor Attacks
  on Cyber-Physical Systems}. In \bibinfo{booktitle}{\emph{Proceedings of the
  26th ACM International Conference on Hybrid Systems: Computation and
  Control}} (San Antonio, TX, USA) \emph{(\bibinfo{series}{HSCC '23})}.
  \bibinfo{publisher}{Association for Computing Machinery},
  \bibinfo{address}{New York, NY, USA}, Article \bibinfo{articleno}{20},
  \bibinfo{numpages}{12}~pages.
\newblock
\showISBNx{9798400700330}
\urldef\tempurl%
\url{https://doi.org/10.1145/3575870.3587118}
\showDOI{\tempurl}


\bibitem[Cockett(2001)]%
        {COCKETT200188}
\bibfield{author}{\bibinfo{person}{Robin Cockett}.}
  \bibinfo{year}{2001}\natexlab{}.
\newblock \showarticletitle{Deforestation, program transformation, and
  cut-elimination}.
\newblock \bibinfo{journal}{\emph{Electronic Notes in Theoretical Computer
  Science}} \bibinfo{volume}{44}, \bibinfo{number}{1} (\bibinfo{year}{2001}),
  \bibinfo{pages}{88--127}.
\newblock
\showISSN{1571-0661}
\urldef\tempurl%
\url{https://doi.org/10.1016/S1571-0661(04)80904-6}
\showDOI{\tempurl}
\newblock
\shownote{CMCS 2001, Coalgebraic Methods in Computer Science (a Satellite Event
  of ETAPS 2001)}.


\bibitem[Cytron et~al\mbox{.}(1991)]%
        {Cytron1991}
\bibfield{author}{\bibinfo{person}{Ron Cytron}, \bibinfo{person}{Jeanne
  Ferrante}, \bibinfo{person}{Barry~K. Rosen}, \bibinfo{person}{Mark~N.
  Wegman}, {and} \bibinfo{person}{F.~Kenneth Zadeck}.}
  \bibinfo{year}{1991}\natexlab{}.
\newblock \showarticletitle{Efficiently Computing Static Single Assignment Form
  and the Control Dependence Graph}.
\newblock \bibinfo{journal}{\emph{ACM Trans. Program. Lang. Syst.}}
  \bibinfo{volume}{13}, \bibinfo{number}{4} (\bibinfo{date}{oct}
  \bibinfo{year}{1991}), \bibinfo{pages}{451–490}.
\newblock
\showISSN{0164-0925}
\urldef\tempurl%
\url{https://doi.org/10.1145/115372.115320}
\showDOI{\tempurl}


\bibitem[Davenport and Heintz(1988)]%
        {davenport88}
\bibfield{author}{\bibinfo{person}{James~H. Davenport} {and}
  \bibinfo{person}{Joos Heintz}.} \bibinfo{year}{1988}\natexlab{}.
\newblock \showarticletitle{Real Quantifier Elimination is Doubly Exponential}.
\newblock \bibinfo{journal}{\emph{J. Symb. Comput.}} \bibinfo{volume}{5},
  \bibinfo{number}{1/2} (\bibinfo{year}{1988}), \bibinfo{pages}{29--35}.
\newblock
\urldef\tempurl%
\url{https://doi.org/10.1016/S0747-7171(88)80004-X}
\showDOI{\tempurl}


\bibitem[Dokhanchi et~al\mbox{.}(2017)]%
        {Dokhanchi2017}
\bibfield{author}{\bibinfo{person}{Adel Dokhanchi}, \bibinfo{person}{Bardh
  Hoxha}, {and} \bibinfo{person}{Georgios Fainekos}.}
  \bibinfo{year}{2017}\natexlab{}.
\newblock \showarticletitle{Formal Requirement Debugging for Testing and
  Verification of Cyber-Physical Systems}.
\newblock \bibinfo{journal}{\emph{ACM Trans. Embed. Comput. Syst.}}
  \bibinfo{volume}{17}, \bibinfo{number}{2}, Article \bibinfo{articleno}{34}
  (\bibinfo{date}{dec} \bibinfo{year}{2017}), \bibinfo{numpages}{26}~pages.
\newblock
\showISSN{1539-9087}
\urldef\tempurl%
\url{https://doi.org/10.1145/3147451}
\showDOI{\tempurl}


\bibitem[Fulton et~al\mbox{.}(2017)]%
        {ITP2017}
\bibfield{author}{\bibinfo{person}{Nathan Fulton}, \bibinfo{person}{Stefan
  Mitsch}, \bibinfo{person}{Rose Bohrer}, {and} \bibinfo{person}{Andr{\'e}
  Platzer}.} \bibinfo{year}{2017}\natexlab{}.
\newblock \showarticletitle{Bellerophon: Tactical Theorem Proving for Hybrid
  Systems}. In \bibinfo{booktitle}{\emph{ITP}} \emph{(\bibinfo{series}{LNCS},
  Vol.~\bibinfo{volume}{10499})}, \bibfield{editor}{\bibinfo{person}{Mauricio
  Ayala-Rinc{\'o}n} {and} \bibinfo{person}{C{\'e}sar~A. Mu{\~n}oz}} (Eds.).
  \bibinfo{publisher}{Springer}, \bibinfo{pages}{207--224}.
\newblock
\showISBNx{978-3-319-66106-3}
\urldef\tempurl%
\url{https://doi.org/10.1007/978-3-319-66107-0\_14}
\showDOI{\tempurl}


\bibitem[Fulton and Platzer(2019)]%
        {fulton2019}
\bibfield{author}{\bibinfo{person}{Nathan Fulton} {and}
  \bibinfo{person}{Andr{\'{e}} Platzer}.} \bibinfo{year}{2019}\natexlab{}.
\newblock \showarticletitle{Verifiably Safe Off-Model Reinforcement Learning}.
  In \bibinfo{booktitle}{\emph{TACAS}} \emph{(\bibinfo{series}{LNCS},
  Vol.~\bibinfo{volume}{11427})}, \bibfield{editor}{\bibinfo{person}{Tomas
  Vojnar} {and} \bibinfo{person}{Lijun Zhang}} (Eds.).
  \bibinfo{publisher}{Springer}, \bibinfo{pages}{413--430}.
\newblock
\urldef\tempurl%
\url{https://doi.org/10.1007/978-3-030-17462-0\_28}
\showDOI{\tempurl}


\bibitem[Grebing and Ulbrich(2020)]%
        {grebing2020usability}
\bibfield{author}{\bibinfo{person}{Sarah Grebing} {and}
  \bibinfo{person}{Mattias Ulbrich}.} \bibinfo{year}{2020}\natexlab{}.
\newblock \showarticletitle{Usability recommendations for user guidance in
  deductive program verification}.
\newblock \bibinfo{journal}{\emph{Deductive Software Verification: Future
  Perspectives: Reflections on the Occasion of 20 Years of KeY}}
  (\bibinfo{year}{2020}), \bibinfo{pages}{261--284}.
\newblock


\bibitem[Henzinger(2000)]%
        {Henzinger2000}
\bibfield{author}{\bibinfo{person}{Thomas~A. Henzinger}.}
  \bibinfo{year}{2000}\natexlab{}.
\newblock \bibinfo{booktitle}{\emph{The Theory of Hybrid Automata}}.
\newblock \bibinfo{publisher}{Springer Berlin Heidelberg},
  \bibinfo{pages}{265--292}.
\newblock
\showISBNx{978-3-642-59615-5}
\urldef\tempurl%
\url{https://doi.org/10.1007/978-3-642-59615-5_13}
\showDOI{\tempurl}


\bibitem[Kabra et~al\mbox{.}(2024)]%
        {Kabra_2024}
\bibfield{author}{\bibinfo{person}{Aditi Kabra}, \bibinfo{person}{Jonathan
  Laurent}, \bibinfo{person}{Stefan Mitsch}, {and} \bibinfo{person}{André
  Platzer}.} \bibinfo{year}{2024}\natexlab{}.
\newblock \bibinfo{booktitle}{\emph{CESAR: Control Envelope Synthesis via
  Angelic Refinements}}.
\newblock \bibinfo{publisher}{Springer Nature Switzerland},
  \bibinfo{pages}{144–164}.
\newblock
\showISBNx{9783031572463}
\showISSN{1611-3349}
\urldef\tempurl%
\url{https://doi.org/10.1007/978-3-031-57246-3\_9}
\showDOI{\tempurl}


\bibitem[Kupferman and Vardi(2003)]%
        {Kupferman2003}
\bibfield{author}{\bibinfo{person}{Orna Kupferman} {and}
  \bibinfo{person}{Moshe~Y. Vardi}.} \bibinfo{year}{2003}\natexlab{}.
\newblock \showarticletitle{Vacuity detection in temporal model checking}.
\newblock \bibinfo{journal}{\emph{International Journal on Software Tools for
  Technology Transfer}} \bibinfo{volume}{4}, \bibinfo{number}{2}
  (\bibinfo{year}{2003}), \bibinfo{pages}{224--233}.
\newblock
\showISBNx{1433-2779}
\urldef\tempurl%
\url{https://doi.org/10.1007/s100090100062}
\showDOI{\tempurl}


\bibitem[Lee and Seshia(2015)]%
        {lee2015introduction}
\bibfield{author}{\bibinfo{person}{E.A. Lee} {and} \bibinfo{person}{S.A.
  Seshia}.} \bibinfo{year}{2015}\natexlab{}.
\newblock \bibinfo{booktitle}{\emph{Introduction to Embedded Systems: A
  Cyber-physical Systems Approach}}.
\newblock \bibinfo{publisher}{MIT Press}.
\newblock
\showISBNx{9781312427402}


\bibitem[Loos and Platzer(2016)]%
        {Loos2016}
\bibfield{author}{\bibinfo{person}{Sarah~M. Loos} {and}
  \bibinfo{person}{Andr{\'e} Platzer}.} \bibinfo{year}{2016}\natexlab{}.
\newblock \showarticletitle{Differential Refinement Logic}. In
  \bibinfo{booktitle}{\emph{LICS}}, \bibfield{editor}{\bibinfo{person}{Martin
  Grohe}, \bibinfo{person}{Eric Koskinen}, {and} \bibinfo{person}{Natarajan
  Shankar}} (Eds.). \bibinfo{publisher}{ACM}, \bibinfo{pages}{505--514}.
\newblock
\urldef\tempurl%
\url{https://doi.org/10.1145/2933575.2934555}
\showDOI{\tempurl}


\bibitem[Maler and Ni{\v c}kovi{\'c}(2013)]%
        {STL2013}
\bibfield{author}{\bibinfo{person}{Oded Maler} {and} \bibinfo{person}{Dejan
  Ni{\v c}kovi{\'c}}.} \bibinfo{year}{2013}\natexlab{}.
\newblock \showarticletitle{Monitoring properties of analog and mixed-signal
  circuits}.
\newblock \bibinfo{journal}{\emph{International Journal on Software Tools for
  Technology Transfer}} \bibinfo{volume}{15}, \bibinfo{number}{3}
  (\bibinfo{year}{2013}), \bibinfo{pages}{247--268}.
\newblock
\showISBNx{1433-2787}
\urldef\tempurl%
\url{https://doi.org/10.1007/s10009-012-0247-9}
\showDOI{\tempurl}


\bibitem[Mitsch and Platzer(2016)]%
        {DBLP:journals/fmsd/MitschP16}
\bibfield{author}{\bibinfo{person}{Stefan Mitsch} {and}
  \bibinfo{person}{Andr{\'e} Platzer}.} \bibinfo{year}{2016}\natexlab{}.
\newblock \showarticletitle{{ModelPlex}: Verified Runtime Validation of
  Verified Cyber-Physical System Models}.
\newblock \bibinfo{journal}{\emph{Form. Methods Syst. Des.}}
  \bibinfo{volume}{49}, \bibinfo{number}{1} (\bibinfo{year}{2016}),
  \bibinfo{pages}{33--74}.
\newblock
\showISSN{0925-9856}
\urldef\tempurl%
\url{https://doi.org/10.1007/s10703-016-0241-z}
\showDOI{\tempurl}
\newblock
\shownote{Special issue of selected papers from RV'14}.


\bibitem[Mitsch et~al\mbox{.}(2014)]%
        {Mitsch2014}
\bibfield{author}{\bibinfo{person}{Stefan Mitsch}, \bibinfo{person}{Jan-David
  Quesel}, {and} \bibinfo{person}{Andr{\'e} Platzer}.}
  \bibinfo{year}{2014}\natexlab{}.
\newblock \showarticletitle{Refactoring, Refinement, and Reasoning: A Logical
  Characterization for Hybrid Systems}. In \bibinfo{booktitle}{\emph{FM}},
  \bibfield{editor}{\bibinfo{person}{Cliff~B. Jones}, \bibinfo{person}{Pekka
  Pihlajasaari}, {and} \bibinfo{person}{Jun Sun}} (Eds.),
  Vol.~\bibinfo{volume}{8442}. \bibinfo{publisher}{Springer},
  \bibinfo{pages}{481--496}.
\newblock
\urldef\tempurl%
\url{https://doi.org/10.1007/978-3-319-06410-9_33}
\showDOI{\tempurl}


\bibitem[O'Hearn(2019)]%
        {10.1145/3371078}
\bibfield{author}{\bibinfo{person}{Peter~W. O'Hearn}.}
  \bibinfo{year}{2019}\natexlab{}.
\newblock \showarticletitle{Incorrectness Logic}.
\newblock \bibinfo{journal}{\emph{Proc. ACM Program. Lang.}}
  \bibinfo{volume}{4}, \bibinfo{number}{POPL}, Article \bibinfo{articleno}{10}
  (\bibinfo{date}{dec} \bibinfo{year}{2019}), \bibinfo{numpages}{32}~pages.
\newblock
\urldef\tempurl%
\url{https://doi.org/10.1145/3371078}
\showDOI{\tempurl}


\bibitem[Pek and Althoff(2020)]%
        {pek20}
\bibfield{author}{\bibinfo{person}{Christian Pek} {and}
  \bibinfo{person}{Matthias Althoff}.} \bibinfo{year}{2020}\natexlab{}.
\newblock \showarticletitle{Fail-Safe Motion Planning for Online Verification
  of Autonomous Vehicles Using Convex Optimization}.
\newblock \bibinfo{journal}{\emph{IEEE Transactions on Robotics}}
  \bibinfo{volume}{PP} (\bibinfo{date}{12} \bibinfo{year}{2020}).
\newblock
\urldef\tempurl%
\url{https://doi.org/10.1109/TRO.2020.3036624}
\showDOI{\tempurl}


\bibitem[Pfenning(2009)]%
        {pfenning2009}
\bibfield{author}{\bibinfo{person}{Frank Pfenning}.}
  \bibinfo{year}{2009}\natexlab{}.
\newblock \bibinfo{title}{Lecture Notes on Cut Elimination}.
\newblock
\newblock


\bibitem[Platzer(2008)]%
        {DBLP:journals/jar/Platzer08}
\bibfield{author}{\bibinfo{person}{Andr{\'e} Platzer}.}
  \bibinfo{year}{2008}\natexlab{}.
\newblock \showarticletitle{Differential Dynamic Logic for Hybrid Systems.}
\newblock \bibinfo{journal}{\emph{J. Autom. Reas.}} \bibinfo{volume}{41},
  \bibinfo{number}{2} (\bibinfo{year}{2008}), \bibinfo{pages}{143--189}.
\newblock
\showISSN{0168-7433}
\urldef\tempurl%
\url{https://doi.org/10.1007/s10817-008-9103-8}
\showDOI{\tempurl}


\bibitem[Platzer(2010)]%
        {Platzer10}
\bibfield{author}{\bibinfo{person}{Andr{\'e} Platzer}.}
  \bibinfo{year}{2010}\natexlab{}.
\newblock \bibinfo{booktitle}{\emph{Logical Analysis of Hybrid Systems: Proving
  Theorems for Complex Dynamics}}.
\newblock \bibinfo{publisher}{Springer}, \bibinfo{address}{Heidelberg}.
\newblock
\showISBNx{978-3-642-14508-7}
\urldef\tempurl%
\url{https://doi.org/10.1007/978-3-642-14509-4}
\showDOI{\tempurl}


\bibitem[Platzer(2012)]%
        {Platzer12b}
\bibfield{author}{\bibinfo{person}{Andr{\'e} Platzer}.}
  \bibinfo{year}{2012}\natexlab{}.
\newblock \showarticletitle{The Complete Proof Theory of Hybrid Systems}. In
  \bibinfo{booktitle}{\emph{LICS}}. \bibinfo{publisher}{IEEE},
  \bibinfo{pages}{541--550}.
\newblock
\showISBNx{978-1-4673-2263-8}
\urldef\tempurl%
\url{https://doi.org/10.1109/LICS.2012.64}
\showDOI{\tempurl}


\bibitem[Platzer(2015)]%
        {CADE2015}
\bibfield{author}{\bibinfo{person}{Andr{\'e} Platzer}.}
  \bibinfo{year}{2015}\natexlab{}.
\newblock \showarticletitle{A Uniform Substitution Calculus for Differential
  Dynamic Logic}. In \bibinfo{booktitle}{\emph{CADE}}
  \emph{(\bibinfo{series}{LNCS}, Vol.~\bibinfo{volume}{9195})},
  \bibfield{editor}{\bibinfo{person}{Amy~P. Felty} {and} \bibinfo{person}{Aart
  Middeldorp}} (Eds.). \bibinfo{publisher}{Springer},
  \bibinfo{pages}{467--481}.
\newblock
\urldef\tempurl%
\url{https://doi.org/10.1007/978-3-319-21401-6\_32}
\showDOI{\tempurl}
\showeprint{1503.01981}


\bibitem[Platzer(2017)]%
        {DBLP:journals/jar/Platzer17}
\bibfield{author}{\bibinfo{person}{Andr{\'e} Platzer}.}
  \bibinfo{year}{2017}\natexlab{}.
\newblock \showarticletitle{A Complete Uniform Substitution Calculus for
  Differential Dynamic Logic}.
\newblock \bibinfo{journal}{\emph{J. Autom. Reas.}} \bibinfo{volume}{59},
  \bibinfo{number}{2} (\bibinfo{year}{2017}), \bibinfo{pages}{219--265}.
\newblock
\urldef\tempurl%
\url{https://doi.org/10.1007/s10817-016-9385-1}
\showDOI{\tempurl}


\bibitem[Platzer(2018)]%
        {Platzer18}
\bibfield{author}{\bibinfo{person}{Andr{\'e} Platzer}.}
  \bibinfo{year}{2018}\natexlab{}.
\newblock \bibinfo{booktitle}{\emph{Logical Foundations of Cyber-Physical
  Systems}}.
\newblock \bibinfo{publisher}{Springer}, \bibinfo{address}{Cham}.
\newblock
\showISBNx{978-3-319-63587-3}
\urldef\tempurl%
\url{https://doi.org/10.1007/978-3-319-63588-0}
\showDOI{\tempurl}


\bibitem[Platzer and Quesel(2009)]%
        {platzer09}
\bibfield{author}{\bibinfo{person}{Andr{\'e} Platzer} {and}
  \bibinfo{person}{Jan-David Quesel}.} \bibinfo{year}{2009}\natexlab{}.
\newblock \showarticletitle{European Train Control System: A Case Study in
  Formal Verification}. In \bibinfo{booktitle}{\emph{Formal Methods and
  Software Engineering}}, \bibfield{editor}{\bibinfo{person}{Karin Breitman}
  {and} \bibinfo{person}{Ana Cavalcanti}} (Eds.). \bibinfo{publisher}{Springer
  Berlin Heidelberg}, \bibinfo{pages}{246--265}.
\newblock
\showISBNx{978-3-642-10373-5}


\bibitem[Platzer and Tan(2018)]%
        {DBLP:conf/lics/PlatzerT18}
\bibfield{author}{\bibinfo{person}{Andr{\'{e}} Platzer} {and}
  \bibinfo{person}{Yong~Kiam Tan}.} \bibinfo{year}{2018}\natexlab{}.
\newblock \showarticletitle{Differential Equation Axiomatization: The
  Impressive Power of Differential Ghosts}. In
  \bibinfo{booktitle}{\emph{LICS}}, \bibfield{editor}{\bibinfo{person}{Anuj
  Dawar} {and} \bibinfo{person}{Erich Gr{\"{a}}del}} (Eds.).
  \bibinfo{publisher}{ACM}, \bibinfo{address}{New York},
  \bibinfo{pages}{819--828}.
\newblock
\showISBNx{978-1-4503-5583-4}
\urldef\tempurl%
\url{https://doi.org/10.1145/3209108.3209147}
\showDOI{\tempurl}


\bibitem[Prebet and Platzer(2024)]%
        {Prebet2024}
\bibfield{author}{\bibinfo{person}{Enguerrand Prebet} {and}
  \bibinfo{person}{Andr{\'{e}} Platzer}.} \bibinfo{year}{2024}\natexlab{}.
\newblock \showarticletitle{Uniform Substitution for Differential Refinement
  Logic}. In \bibinfo{booktitle}{\emph{IJCAR}} \emph{(\bibinfo{series}{LNCS},
  Vol.~\bibinfo{volume}{14740})}, \bibfield{editor}{\bibinfo{person}{Christoph
  Benzm{\"{u}}ller}, \bibinfo{person}{Marijn~J. Heule}, {and}
  \bibinfo{person}{Renate~A. Schmidt}} (Eds.). \bibinfo{publisher}{Springer},
  \bibinfo{pages}{196--215}.
\newblock
\urldef\tempurl%
\url{https://doi.org/10.1007/978-3-031-63501-4_11}
\showDOI{\tempurl}


\bibitem[Reed(2009)]%
        {reed2009}
\bibfield{author}{\bibinfo{person}{Jason~C. Reed}.}
  \bibinfo{year}{2009}\natexlab{}.
\newblock \emph{\bibinfo{title}{A Hybrid Logical Framework}}.
\newblock PhD thesis. \bibinfo{school}{Carnegie Mellon University}.
\newblock


\bibitem[Selvaraj et~al\mbox{.}(2022)]%
        {selvaraj2022}
\bibfield{author}{\bibinfo{person}{Yuvaraj Selvaraj}, \bibinfo{person}{Jonas
  Krook}, \bibinfo{person}{Wolfgang Ahrendt}, {and} \bibinfo{person}{Martin
  Fabian}.} \bibinfo{year}{2022}\natexlab{}.
\newblock \showarticletitle{On How To Not Prove Faulty Controllers Safe In
  Differential Dynamic Logic}. In \bibinfo{booktitle}{\emph{International
  Conference on Formal Engineering Methods, ICFEM'22}} (Madrid, Spain)
  \emph{(\bibinfo{series}{LNCS})}. \bibinfo{publisher}{Springer-Verlag},
  \bibinfo{address}{Berlin, Heidelberg}, \bibinfo{pages}{281–297}.
\newblock
\showISBNx{978-3-031-17243-4}
\urldef\tempurl%
\url{https://doi.org/10.1007/978-3-031-17244-1\_17}
\showDOI{\tempurl}


\bibitem[Squires et~al\mbox{.}(2018)]%
        {squires18}
\bibfield{author}{\bibinfo{person}{Eric Squires}, \bibinfo{person}{Pietro
  Pierpaoli}, {and} \bibinfo{person}{Magnus Egerstedt}.}
  \bibinfo{year}{2018}\natexlab{}.
\newblock \showarticletitle{Constructive Barrier Certificates with Applications
  to Fixed-Wing Aircraft Collision Avoidance}. In
  \bibinfo{booktitle}{\emph{2018 IEEE Conference on Control Technology and
  Applications (CCTA)}}. \bibinfo{pages}{1656--1661}.
\newblock
\urldef\tempurl%
\url{https://doi.org/10.1109/CCTA.2018.8511342}
\showDOI{\tempurl}


\end{thebibliography}
